\documentclass{sfuthesis}
\usepackage[usenames,dvipsnames,table]{xcolor}
\usepackage{notoccite}

\title{\LARGE Data-Centric and Data-Aware Frameworks for Fundamentally Efficient Data Handling in\\Modern Computing Systems}
\author{Nastaran Hajinazar}
\previousdegrees{%
	M.Sc., Sharif University of Technology, 2011\\
	B.Sc., Shahid Chamran University, 2008}
\degree{Doctor of Philosophy}
\department{Department of Computing Science}
\faculty{Faculty of Applied Science}
\copyrightyear{2021}
\semester{Summer 2021}

\committee{
	\chair{Zhenman Fang}{Associate Member\\in School of Computing Science}
	\member{Onur Mutlu}{Supervisor \\ Professor of Computing Science \\ Department of Information Technology and Electrical Engineering \\
	ETH Zurich}
	\member{Arrvindh Shriraman}{Supervisor \\ Associate Professor, Computing Science}
	\member{Saugata Ghose}{Committee Member \\ Assistant Professor \\ Department of of Computer Science \\ University of Illinois at Urbana--Champaign}
	\member{Vivek Seshadri}{Committee Member \\ Senior Researcher \\ Microsoft Research India}
	\member{Alaa Alameldeen}{Examiner \\ Associate Professor, Computing Science}
	\member{Myoungsoo Jung}{External Examiner \\ Associate Professor \\ Department of Electrical Engineering \\ Korea Advanced Institute of Technology (KAIST)}
}

\usepackage{amsmath}                            
\usepackage{amssymb}                            
\usepackage{amsthm}                             
\usepackage{graphicx}                           
\usepackage[pdfborder={0 0 0}]{hyperref}        
% ...
\usepackage{cite}
\usepackage{todonotes}
\usepackage{soul}
\usepackage{listings}
\usepackage{fancyhdr}
\usepackage[normalem]{ulem}
\usepackage[keeplastbox]{flushend}
\usepackage[english]{babel}
\usepackage[utf8]{inputenc}
\usepackage{algorithm}
\usepackage{float}
\usepackage{wrapfig}
\usepackage[noend]{algpseudocode}
\usepackage{textcomp}
\usepackage{xspace}
\usepackage[binary-units, detect-all]{siunitx}
\usepackage{graphicx}
\usepackage{dblfloatfix} 
\usepackage{booktabs}
\usepackage{longfbox}
\usepackage{enumitem}
\usepackage{pifont}
\usepackage{microtype}
\usepackage[acronym,nonumberlist,nowarn]{glossaries}
    \newacronym{ADI}{ADI}{Alternating Direction Implicit}
\newacronym{AGU}{AGU}{Address Generation Unit}
\newacronym[longplural={Access History Tables}]{AHT}{AHT}{Access History Table}
\newacronym{AHT-R}{AHT-R}{Access History Table - Read}
\newacronym{AHT-W}{AHT-W}{Access History Table - Write-Back}
\newacronym{AIP}{AIP}{Access Interval Predictor}
\newacronym{AL}{AL}{Added Latency for column accesses}
\newacronym{ASIC}{ASIC}{Application Specific Integrated Circuit}
\newacronym{AVX}{AVX}{Advanced Vector Extensions}
\newacronym[longplural={Basic Block Vectors}]{BBV}{BBV}{Basic Block Vector}
\newacronym{BHT}{BHT}{Branch History Table}
\newacronym{HBM}{HBM}{Hybrid Bandwidth Memory}
\newacronym{DDR4}{DDR4}{Double-Data Rate 4}

\newacronym{NN}{NN}{Neural Network}
\newacronym{BTB}{BTB}{Branch Target Buffer}
\newacronym{CAS}{CAS}{Column Address Strobe}
\newacronym{AMAT}{AMAT}{Average Memory Access Time}
\newacronym{CCD}{CCD}{Column to Column Delay}
\newacronym{AI}{AI}{Arithmetic Intensity}
\newacronym[longplural={Columnar Database Systems}]{CDBS}{CDBS}{Columnar Database System}
\newacronym[longplural={Chip Multiprocessors}]{CMP}{CMP}{Chip Multiprocessor}
\newacronym{CMT}{CMT}{Chip Multithreading}
\newacronym[longplural={Coarse-Grain Reconfigurable Arrays}]{CGRA}{CGRA}{Coarse-Grain Reconfigurable Array}
\newacronym{C-RAM}{C-RAM}{Computational-RAM}
\newacronym{CWD}{CWD}{Column Write Delay}
\newacronym{DBI}{DBI}{Dynamic Binary Instrumentation}
\newacronym[longplural={Database Management Systems}]{DBMS}{DBMS}{Database Management System}
\newacronym{DDR}{DDR}{Double Data Rate}
\newacronym{DEWP}{DEWP}{Dead Line and Early Write-Back Predictor}
\newacronym{DRAM}{DRAM}{Dynamic Random Access Memory}
\newacronym{DSBP}{DSBP}{Dead Sub-Block Predictor}
\newacronym{DSM}{DSM}{Decomposed Storage Manage}
\newacronym{FAW}{FAW}{Four row Activation Window}
\newacronym{FP}{FP}{Floating-Point}
\newacronym{EDP}{EDP}{Energy-Delay Product}
\newacronym{FCFS}{FCFS}{First-Come First-Serve}
\newacronym{FIFO}{FIFO}{Fist In, First Out}
\newacronym{FSM}{FSM}{Finite State Machine}
\newacronym{FPGA}{FPGA}{Field-Programmable Gate Array}
\newacronym[longplural={Functional Units}]{FU}{FU}{Functional Unit}
\newacronym{GAg}{GAg}{Global Adaptive branch prediction using one Global PHT}
\newacronym{GAs}{GAs}{Global Adaptive branch prediction using per-Set PHT}
\newacronym{GCC}{GCC}{GNU Compiler Collection}
\newacronym{GEMS}{GEMS}{General Execution-driven Multiprocessor Simulator}
\newacronym[longplural={General Purpose Processors}]{GPP}{GPP}{General Purpose Processor}
\newacronym{HMC}{HMC}{Hybrid Memory Cube}
\newacronym{HIVE}{HIVE}{HMC Instruction Vector Extensions}
\newacronym{IATAC}{IATAC}{Inter-Access Time per Access Count}
\newacronym{ILP}{ILP}{Instruction Level Parallelism}
\newacronym{IPC}{IPC}{Instructions per Cycle}
\newacronym{ISA}{ISA}{Instruction Set Architecture}
\newacronym{IRAM}{IRAM}{Intelligent RAM}
\newacronym{KIPS}{KIPS}{Kilo Instructions per Second}
\newacronym{LDS}{LDS}{Linked Data Structure}
\newacronym{LFSR}{LFSR}{Linear Feedback Shift Register}
\newacronym{LFMR}{LFMR}{Last-to-First Miss-Ratio}
\newacronym{MLP}{MLP}{Memory-Level Parallelism}

\newacronym{LLC}{LLC}{last-level cache}
\newacronym{LRU}{LRU}{Least Recently Used}
\newacronym{LSU}{LSU}{Load Store Unit}
\newacronym{LTP}{LTP}{Last-Touch Predictor}
\newacronym{LvP}{LvP}{Live-time Predictor}
\newacronym{LWP}{LWP}{Last Write Predictor}
\newacronym{MARSS}{MARSS}{Micro Architectural and System Simulator}
\newacronym{McPAT}{McPAT}{Multi-core Power, Area, and Timing}
\newacronym{MIPS}{MIPS}{Microprocessor without Interlocked Pipeline Stages}
\newacronym{MOB}{MOB}{Memory Order Buffer}
\newacronym{MMU}{MMU}{Memory Management Unit}
\newacronym{MPKI}{MPKI}{Misses per Kilo-Instruction}
\newacronym{MSHR}{MSHR}{Miss-Status Handling Registers}
\newacronym{NAS}{NAS}{Numerical Aerodynamic Simulation}
\newacronym{NDP}{NDP}{Near-Data Processing}
\newacronym{NMP}{NMP}{Near Memory Processor}
\newacronym{NoC}{NoC}{Network-on-Chip}
\newacronym{NPB}{NPB}{NAS Parallel Benchmark}
\newacronym{NUCA}{NUCA}{Non-Uniform Cache Architecture}
\newacronym{NUMA}{NUMA}{Non-Uniform Memory Access}
\newacronym{OoO}{OoO}{Out-of-Order}
\newacronym{OpenMP}{OpenMP}{Open Multi-Processing}
\newacronym{OS}{OS}{operating system}
\newacronym{PAg}{PAg}{Per-address Adaptive branch prediction using one Global PHT}
\newacronym{PAs}{PAs}{Per-address Adaptive branch prediction using per-Set PHT}
\newacronym{PC}{PC}{Program Counter}
\newacronym[longplural={Pointer-Chasing Engines}]{PCE}{PCE}{Pointer-Chasing Engine}
\newacronym{PCM}{PCM}{Performance Counter Monitor}
\newacronym{PIM}{PIM}{Processing-in-Memory}
\newacronym[longplural={Pattern History Tables}]{PHT}{PHT}{Pattern History Table}
\newacronym{RAPL}{RAPL}{Running Average Power Limit}
\newacronym{RAS}{RAS}{Row Address Strobe}
\newacronym{RAT}{RAT}{Registers Alias Table}
\newacronym{RC}{RC}{Row Cycle}
\newacronym{RCD}{RCD}{RAS to CAS Delay}
\newacronym[longplural={Row-based Database Systems}]{RDBS}{RDBS}{Row-based Database System}
\newacronym{ROB}{ROB}{Reorder Buffer}
\newacronym{RRD}{RRD}{Row to Row activation Delay}
\newacronym{RP}{RP}{Row Precharge}
\newacronym{RTL}{RTL}{Register Transfer Level}
\newacronym{RTP}{RTP}{Read To Precharge}
\newacronym{RVU}{RVU}{Reconfigurable Vector Unit}

\newacronym{MIG}{MIG}{Majority-Inverter Graph}
\newacronym{AOIG}{AOIG}{AND/OR/Inverter Graph}
\newacronym{TRA}{TRA}{Triple-Row Activation}
\newacronym{AAP}{AAP}{ACTIVATION-ACTIVATION-PRECHARGE}
\newacronym{DCC}{DCC}{Dual-Contact Cell}

\newacronym{AP}{AP}{ACTIVATION-PRECHARGE}
\newacronym{SDP}{SDP}{Skewed Dead-Block Predictor}
\newacronym{SESC}{SESC}{Superescalar Simulator}
\newacronym{SSE}{SSE}{Streaming SIMD Extensions}
\newacronym{SFP}{SFP}{Spatial Footprint Predictor}
\newacronym{SiNUCA}{SiNUCA}{Simulator of Non-Uniform Cache Architectures}
\newacronym{SMT}{SMT}{Simultaneous Multi-Threading}
\newacronym{SIMD}{SIMD}{Single Instruction Multiple Data}
\newacronym{SPEC}{SPEC}{Standard Performance Evaluation Corporation}
\newacronym{SoC}{SoC}{System-on-Chip}
\newacronym{SPP}{SPP}{Spatial Pattern Predictor}
\newacronym{SRAM}{SRAM}{Static Random Access Memory}
\newacronym{SSOR}{SSOR}{Symmetric Successive Over-Relaxation}
\newacronym{SSV}{SSV}{Search Set Vector}
\newacronym{TLB}{TLB}{Translation Look-aside Buffer}
\newacronym{TLP}{TLP}{Thread Level Parallelism}
\newacronym[longplural={Through-Silicon Vias}]{TSV}{TSV}{Through-Silicon Via}
\newacronym{VWQ}{VWQ}{Virtual Write Queue}
\newacronym{WTR}{WTR}{Write Recovery time}
\newacronym{WR}{WR}{Write To Read delay time}

\newcommand{\sysfull}{Virtual Block Interface\xspace}
\newcommand{\sys}{VBI\xspace}

\newcommand{\xeightsix}{x86-64\xspace}

\newcommand{\requestvas}{\texttt{request\_vb}\xspace}

\newcommand{\enablevb}{{\texttt{enable\_vb}}\xspace}
\newcommand{\disablevb}{{\texttt{disable\_vb}}\xspace}

\newcommand{\attachvb}{{\texttt{attach}}\xspace}
\newcommand{\detachvb}{{\texttt{detach}}\xspace}

\newcommand{\clonevb}{{\texttt{clone\_vb}}\xspace}
\newcommand{\promotevb}{{\texttt{promote\_vb}}\xspace}

\newcommand{\cvt}{CVT\xspace}

\newcommand{\vit}{VB Info Table\xspace}
\newcommand{\vitshort}{VIT\xspace}

\newcommand{\mtlfull}{Memory Translation Layer\xspace}
\newcommand{\mtl}{MTL\xspace}

\usepackage{mwe}
\usepackage{mathtools}
    
    \DeclarePairedDelimiter\floor{\lfloor}{\rfloor}
\usepackage{multirow}
\usepackage[compact]{titlesec} 
\usepackage{subcaption}
\usepackage{cleveref}
\usepackage{tikz}
\usepackage{changepage}
\usepackage{textcomp}
\usepackage{hhline}
\usepackage[compatibility=false, skip=2pt]{caption}
\usepackage{hyperref}

\usetikzlibrary{decorations.pathreplacing,calc}
\newcommand{\tikzmark}[1]{\tikz[overlay,remember picture] \node (#1) {};}

\newcommand*{\AddNoteRed}[4]{%
    \begin{tikzpicture}[overlay, remember picture]
        \draw [decoration={brace,amplitude=0.5em},decorate,ultra thick,red]
            ($(#3)!(#1.north)!($(#3)-(0,1)$)$) --  
            ($(#3)!(#2.south)!($(#3)-(0,1)$)$)
                node [align=center, text width=2.5cm, pos=0.5, anchor=west] {#4};
    \end{tikzpicture}
}%
\newcommand*{\AddNoteBlue}[4]{%
    \begin{tikzpicture}[overlay, remember picture]
        \draw [decoration={brace,amplitude=0.5em},decorate,ultra thick,blue]
            ($(#3)!(#1.north)!($(#3)-(0,1)$)$) --  
            ($(#3)!(#2.south)!($(#3)-(0,1)$)$)
                node [align=center, text width=2.5cm, pos=0.5, anchor=west] {#4};
    \end{tikzpicture}
}%
\newcommand*{\AddNoteGreen}[4]{%
    \begin{tikzpicture}[overlay, remember picture]
        \draw [decoration={brace,amplitude=0.5em},decorate,ultra thick,black]
            ($(#3)!(#1.north)!($(#3)-(0,1)$)$) --  
            ($(#3)!(#2.south)!($(#3)-(0,1)$)$)
                node [align=center, text width=2.5cm, pos=0.5, anchor=west] {#4};
    \end{tikzpicture}
}%

\usepackage{xcolor}

\usepackage{tikz}
\usetikzlibrary{arrows}
\usetikzlibrary{patterns}
\usetikzlibrary{calc}
\usetikzlibrary{shapes}
\usetikzlibrary{positioning,fit}
\usepackage{float}
\usepackage{lipsum}

\newcommand{\nas}[1]{\textcolor{black}{#1}}

\newcommand{\incircle}[1]{\raisebox{0.5pt}{\protect%
  \tikz[baseline=(char.base)]{
  \node[shape=circle,draw,inner sep=0.5pt,minimum height=10pt, fill=black,text=white,font=\footnotesize\sffamily] (char) {#1};}%
}}%
\newcommand{\inhollowcircle}[1]{\raisebox{0.5pt}{\protect%
  \tikz[baseline=(char.base)]{
  \node[shape=circle,draw,inner sep=0.5pt,minimum height=10pt,font=\footnotesize\sffamily] (char) {#1};}%
}}%

\definecolor{amber}{rgb}{1.0, 0.49, 0.0}
\definecolor{darkgreen}{rgb}{0.0, 0.2, 0.13}
\definecolor{darkbyzantium}{rgb}{0.36, 0.22, 0.33}
\definecolor{darkseagreen}{rgb}{0.56, 0.74, 0.56}
\definecolor{darkspringgreen}{rgb}{0.09, 0.45, 0.27}
\definecolor{dollarbill}{rgb}{0.52, 0.73, 0.4}

\definecolor{darkcyan}{rgb}{0.0, 0.55, 0.55}
\definecolor{forestgreen}{rgb}{0.0, 0.27, 0.13}
\definecolor{azure}{rgb}{0.0, 0.5, 1.0}
\definecolor{amber}{rgb}{1.0, 0.49, 0.0}

% ...

\definecolor{dgreen}{rgb}{0.00, 0.75, 0.00}
\definecolor{dollarbill}{rgb}{0.52, 0.73, 0.4}
\definecolor{internationalorange}{rgb}{1.0, 0.31, 0.0}
\definecolor{islamicgreen}{rgb}{0.0, 0.56, 0.0}
\definecolor{amethyst}{rgb}{0.6, 0.4, 0.8}
\definecolor{mygreen}{rgb}{0,0.6,0}
\definecolor{mygray}{rgb}{0.5,0.5,0.5}
\definecolor{mymauve}{rgb}{0.58,0,0.82}
\definecolor{backcolour}{rgb}{0.95,0.95,0.92}
\definecolor{bluehl}{rgb}{0.8,0.874,1}
\definecolor{pinkhl}{rgb}{0.992156863,0.847058824,1}
\definecolor{greenhl}{rgb}{0.835,0.996,0.839}
\definecolor{yellowhl}{rgb}{0.996,0.957,0.8}
\newcommand\aap{\texttt{AAP}/\texttt{AP}\xspace}
\newcommand\aaps{\texttt{AAP}s/\texttt{AP}s\xspace}
\newcommand\uop{\textmu{}Op}
\newcommand\uprog{\textmu{}Program}
\newcommand\ureg{\textmu{}Register}
\newcommand\upc{\textmu{}PC}
\newcommand\uprogc{\textmu{}Program counter}

\newcommand{\revdel}[1]{}
\newcommand{\revdelrefa}[1]{}  
\newcommand{\revdelrefr}[1][0]{}  

\newcommand{\revonuriii}[1]{{#1}}
\newcommand{\omiii}[1]{\textcolor{black}{#1}} 
\newcommand{\omiv}[1]{\textcolor{black}{#1}} 
\newcommand{\omv}[1]{\textcolor{black}{#1}} 
\newcommand{\omvi}[1]{\textcolor{black}{#1}} 
\newcommand{\omvuii}[1]{\textcolor{black}{#1}} 
\newcommand{\omviii}[1]{\textcolor{black}{#1}}
\newcommand{\omix}[1]{\textcolor{black}{#1}}
\newcommand{\omdef}[1]{\textcolor{black}{#1}}
\newcommand{\omdefi}[1]{\textcolor{black}{#1}}
\newcommand{\omdefii}[1]{\textcolor{black}{#1}}

\newcommand{\geraldorevii}[1]{{#1}}
\newcommand{\onurt}[1]{\textcolor{black}{#1}}
\newcommand{\onurtt}[1]{\textcolor{black}{#1}}

\newcommand{\sgii}[1]{{#1}}
\newcommand{\nasrev}[1]{{#1}}
\newcommand{\om}[1]{{#1}}
\newcommand{\omi}[1]{{#1}} 
\newcommand{\omii}[1]{{#1}} 
 
\newcommand{\cutdel}[1]{}  
\newcommand{\sr}[1]{{#1}}

\newcommand{\cmr}[1]{\textcolor{black}{#1}} %

\newif\ifrevision
\revisionfalse

\ifrevision
     \newcommand\geraldorevi[1][0]{}
     
     \newcommand{\sgii}[1]{#1}
\else
    \newcommand{\geraldorevi}[1]{{#1}}
    \newcommand{\jgll}[1]{}
    
    \newcommand{\textfromsl}[1]{{\color{black}#1}}
    \newcommand{\gfrev}[1]{}
    
    \newcommand{\revonur}[1]{{#1}} 
    \newcommand{\revonurii}[1]{{#1}} 
    \newcommand{\revsgii}[1]{{#1}}
    \newcommand{\revonuri}[1]{{#1}} 
\fi

\ifrevision
     \newcommand\nasirevi[1][0]{}
\else
    \newcommand{\nasirevi}[1]{{#1}}
\fi

\newif\ifsubmission
\submissiontrue

\ifsubmission
    \newcommand\jgl[1]{}
    \newcommand\juang[1][0]{}
    \newcommand\juan[1][0]{}
    \newcommand\juangr[1][0]{}
    \newcommand\geraldo[1][0]{}
    \newcommand\revGeraldo[1][0]{}
    \newcommand\jr[1]{}
    \newcommand\gf[1]{}
    \newcommand\jds[1]{}
    \newcommand\joao[1][0]{}
    \newcommand\nasi[1][0]{}
    \newcommand\nass[1][0]{}
    \newcommand\nastaran[1][0]{}
    \newcommand\nasii[1][0]{}
    \newcommand\onur[1][0]{}
    \newcommand\minp[1][0]{}
    \newcommand\mpr[1][0]{}
    \newcommand\mpi[1][0]{}
    \newcommand{\sg}[1][0]{}
    \newcommand\new[1][0]{}
    \newcommand\juangg[1][0]{}
    \newcommand\edit[1][0]{}
    \newcommand\rev[1][0]{}
\else
    \newcommand{\jgl}[1]{[{\color{dgreen}JGL: #1}]}
    \newcommand{\juang}[1]{{\color{black}#1}}
    \newcommand{\juan}[1]{{\color{black}#1}}
    \newcommand{\juangr}[1]{\textcolor{black}{#1}}
    \newcommand{\juangg}[1]{{\color{dgreen}#1}}
    \newcommand{\geraldo}[1]{{\color{blue}#1}}
    \newcommand{\revGeraldo}[1]{\textcolor{blue}{#1}}
    \newcommand{\jr}[1]{[{\color{blue}Geraldo: #1}]}
    \newcommand{\gf}[1]{[{\color{blue}GF: #1}]}
    \newcommand{\jds}[1]{[{\color{black}JDS: #1}]}
    \newcommand{\joao}[1]{{\color{black}#1}}
    \newcommand{\nas}[1]{[{\color{teal}Nas: #1}]}
    \newcommand{\nasi}[1]{{\color{black}#1}}
    \newcommand{\nass}[1]{#1}
    \newcommand{\nastaran}[1]{\textcolor{black}{#1}}
    \newcommand{\nasii}[1]{{\color{VioletRed}#1}}
    
    \newcommand{\onur}[1]{\textcolor{Orange}{#1}}
    \newcommand{\minp}[1]{\textcolor{black}{#1}}
    \newcommand{\mpr}[1]{\textcolor{dgreen}{#1}}
    \newcommand{\mpi}[1]{{\textcolor{dgreen}#1}}
    \newcommand{\sg}[1]{\textcolor{MidnightBlue}{#1}}
    \newcommand{\new}[1]{\textcolor{black}{#1}}
    \newcommand{\edit}[1]{\textcolor{black}{#1}}
    \newcommand{\rev}[1]{\textcolor{teal}{#1}}
\fi
\newcommand{\revAMicro}[1]{\textcolor{black}{#1}}
\newcommand{\revBMicro}[1]{\textcolor{black}{#1}}

\newcommand{\revDMicro}[1]{\textcolor{black}{#1}}

\newcommand{\mech}{{SIMDRAM}\xspace} 
\newcommand\tempcommand[1]{\renewcommand{\arraystretch}{#1}}

\newcommand{\circled}[1]{\tikz[baseline=(char.base)]{\node[shape=circle,draw,inner sep=0pt,fill=black, text=white] (char) {#1};}}

\newcommand{\ignore}[1]{}

\algnewcommand\algorithmicforeach{\textbf{for each}}
\algdef{S}[FOR]{ForEach}[1]{\algorithmicforeach\ #1\ \algorithmicdo}
\algrenewcommand\algorithmicindent{0.5em}%

\lstdefinestyle{myC}{
  backgroundcolor=\color{backcolour},  
  basicstyle=\ttfamily\footnotesize,      
  breakatwhitespace=false,  
  breaklines=true,       
  captionpos=b,                   
  commentstyle=\color{mygreen},    
  deletekeywords={...},           
  escapechar=\%,
  xleftmargin=0pt,
  xrightmargin=0pt,
  aboveskip=\medskipamount,
  belowskip=\medskipamount,
  extendedchars=true,            
  keepspaces=true,               
  keywordstyle=\color{blue},      
  language=C++,              
  morekeywords={bbop_trsp_init, bbop_add, bbop_sub, bbop_greater, bbop_if_else, bbop_trsp_cpy, malloc, *,...},          
  numbers=left,                   
  numbersep=1pt,                   
  numberstyle=\tiny\color{mygray}, 
  rulecolor=\color{black},     
  showspaces=false,              
  showstringspaces=false,        
  showtabs=false,                 
  stepnumber=1,                    
  stringstyle=\color{mymauve},     
  tabsize=2,	                   
  title=\lstname                   
}

\newcommand{\HilightBlue}{\makebox[0pt][l]{\color{bluehl}\rule[-4pt]{\linewidth}{10pt}}}
\newcommand{\HilightPink}{\makebox[0pt][l]{\color{pinkhl}\rule[-4pt]{\linewidth}{10pt}}}
\newcommand{\HilightYellow}{\makebox[0pt][l]{\color{yellowhl}\rule[-4pt]{\linewidth}{10pt}}}
\newcommand{\HilightGreen}{\makebox[0pt][l]{\color{greenhl}\rule[-4pt]{\linewidth}{10pt}}}

\AtBeginDocument{\DeclareCaptionSubType{lstlisting}}
\crefname{sublstlisting}{listing}{listings}
\Crefname{sublstlisting}{Listing}{Listings}

 \titlespacing\section{0pt}{5pt plus 2pt minus 2pt}{0pt plus 2pt minus 2pt}
 \titlespacing\subsection{0pt}{5pt plus 2pt minus 2pt}{0pt plus 2pt minus 2pt}
 \titlespacing\subsubsection{0pt}{5pt plus 2pt minus 2pt}{0pt plus 2pt minus 2pt}

\definecolor{amber}{rgb}{1.0, 0.49, 0.0}
\definecolor{darkgreen}{rgb}{0.0, 0.2, 0.13}
\definecolor{darkbyzantium}{rgb}{0.36, 0.22, 0.33}
\definecolor{darkseagreen}{rgb}{0.56, 0.74, 0.56}
\definecolor{darkspringgreen}{rgb}{0.09, 0.45, 0.27}
\definecolor{dollarbill}{rgb}{0.52, 0.73, 0.4}

\definecolor{darkcyan}{rgb}{0.0, 0.55, 0.55}
\definecolor{forestgreen}{rgb}{0.0, 0.27, 0.13}
\definecolor{azure}{rgb}{0.0, 0.5, 1.0}

\newcommand{\sgdel}[1]{}

\newcommand{\rebuttdel}[1]{}

\newcommand{\cameradelete}[1]{}

\newcommand{\shepherddel}[1]{}
\newcommand{\sgaddi}[1]{}

\newcommand{\onurii}[1]{\textcolor{black}{#1}}
\newcommand{\onurthird}[1]{\textcolor{black}{#1}}

\newcommand{\onurv}[1]{\textcolor{black}{#1}}

\definecolor{amber}{rgb}{1.0, 0.49, 0.0}

\newcommand{\mpt}[1]{{\color{black}#1}}
\newcommand{\mhp}[1]{{\color{black}#1}} 
\newcommand{\pointer}[1]{{\color{black}#1}}
\newcommand{\onurvi}[1]{\textcolor{black}{#1}}
\newcommand{\onurvii}[1]{\textcolor{black}{#1}}
\newcommand{\onurx}[1]{\textcolor{black}{#1}}

\begin{document}

\frontmatter
\maketitle{}
\makecommittee{}


\begin{abstract}
\vspace{-20pt}
There is an explosive growth in the size of the input and/or intermediate data used and generated by modern and emerging applications. Unfortunately, modern computing systems are not capable of handling large amounts of data efficiently. Major concepts and components (e.g., the virtual memory system) and predominant execution models (e.g., the processor-centric execution model) used in almost all computing systems are designed without having \onurt{modern applications'} overwhelming data demand in mind. As a result, accessing, moving, and processing large amounts of data faces important challenges in today’s systems, making data a first-class concern and a prime performance and energy bottleneck in such systems. This thesis studies the root cause of inefficiency in modern computing systems when handling \onurt{modern applications'} data demand, and aims to fundamentally address such inefficiencies, with a focus on two directions. 

First, \onurt{we design a new} framework that \onurt{aids} the widespread adoption of processing-using-DRAM, a data-centric computation paradigm that improves the overall performance and efficiency of the system when computing large amounts of data by minimizing the cost of data movement and enabling computation where the data resides. 
To this end, we introduce SIMDRAM, an end-to-end processing-using-DRAM framework that (1) efficiently computes complex operations required by modern data intensive applications, and (2) provides the ability to implement new arbitrary operations as required, all in an in-DRAM massively-parallel \onurt{Single Instruction Multiple Data (SIMD)} substrate that requires minimal changes to the DRAM \onurt{(Dynamic Random Access Memory)} architecture. 

Second, \onurt{we design a new,} more scalable virtual memory framework that (1) eliminates the inefficiencies of the conventional virtual memory frameworks when handling the high memory demand in modern applications, and (2) is built from the ground up to understand, convey, and exploit data properties, to create opportunities for performance and efficiency improvements. To this end, we introduce the Virtual Block Interface (VBI), a novel virtual memory framework that (1) efficiently handles modern \onurt{applications'} high data demand, (2) conveys properties of different pieces of program data \onurt{(e.g., data structures)} to the hardware and exploits this knowledge for performance and efficiency optimizations, (3) better extracts performance from the wide variety of new system configurations that are designed to process large amounts of data \onurt{(e.g., hybrid memory systems)}, and (4) provides \onurt{all} the key features of the conventional virtual memory frameworks, \onurt{at low overhead}. 
\end{abstract}

 \begin{dedication}
 Dedicated to
 
 \qquad my beloved parents, Ezzatollah and Homa,
 
  \qquad my wonderful brothers, Samad and Siavash,
  
  \qquad my lovely husband, Amir,
  
   \qquad and my most precious son, Nick Ray.
 \end{dedication}

 \begin{acknowledgements}

Pursuing a PhD was a significant source of learning, inspiration and growth for me, to which many people have contributed in different ways. This is a humble attempt to thank them for their contribution and for helping me become who I am today.

First and foremost, I would like to express my deep and sincere gratitude to my advisor, Prof. Onur Mutlu, who believed in me, even when I did not. His unwavering support, despite the tough initial years, was my prime source of courage towards the completion of my PhD. Onur generously provided me with invaluable guidance, exceptional opportunities, incredible resources, and more importantly, extraordinary freedom to carry out my research. I also thank him for teaching me how to think critically, write thoroughly, speak clearly, and perform impactful research. His influence in shaping me certainly goes above and beyond this dissertation and extends to countless real-life lessons that I have learned from him.

I would like to thank my advisor, Prof. Arrvindh Shriraman, for all his help and support. I thank him for allowing me to find and follow the research direction that interested me. I also thank him for being open to my collaborations with students and researchers from other institutions.

I would also like to thank the members of my supervisory committee, Prof. Saugata Ghose and Dr. Vivek Seshadri. I thank Prof. Ghose for providing me with incredible technical and moral support and countless pieces of critical advice throughout my PhD journey. He helped me stay focused and navigate through my research with a clear and determined mind. My earnest thanks to Dr. Seshadri for teaching me how to find the right research problem to work on and how to perform quality research. I thank him for all his help and support through the times that I needed it the most.

During my PhD, I had the chance to work alongside many wonderful fellow graduate students at the SAFARI research group whom I am grateful to. My great friend, Giray Yaglikci was always amazingly kind and selfless in helping me and listening to me during the times when graduate school felt dark and lonely. I would like to thank him for all his support, as well as the hot cups of tea and delicious food that rescued me with when working late at school. Geraldo De Oliveira was my PIM guru, whose kind presence as well as his amazing home-made cakes and cookies brightened up my days at ETH. Minesh Patel has been a great friend with whom I have also enjoyed many research collaborations and real-life conversations. Juan Gómez Luna was my mentor and collaborator who assisted me in many ways. Kevin Hsieh was an extraordinary friend and mentor who was there to listen to me and support me when I was in need. Rachata Ausavarungnirun was a good friend who also taught me a great deal about how to conduct thorough simulations and evaluations. Damla Senol Cali was my sweet and kind friend and collaborator who was always there to support me and celebrate my achievements with me. Gagandeep Singh was a wonderful collaborator and friend who always amused me with his great sense of humor. I thank Can Firtina for his delightful friendship as well as the high-tech Tesla rides. I would also like to thank many other SAFARI members for their friendship and collaboration: Jeremie Kim, Mohammed Alser, Jawad Haj-Yahya, Lois Orosa, Jisung Park, Hasan Hassan, Tracy Ewen, and Christian Rossi.

I would also like to thank many wonderful people who made the long and at times tough journey of pursuing PhD easier for me, and I am lucky to call them my friends: Rajesh Rao, Sogol Barazandegan, Nasibeh Teimouri, Amir Shabani, and Ahmed Hamza.

And last, but not least, I would like to express my profound gratitude to my family for their unconditional and continued love and support. I am most grateful to my parents, Ezzatollah Hajinazar and Homa Golabian, for enabling me to pursue my passion for learning, and for being by my side at each and every step of this journey. I thank my mother for selflessly caring for me and for teaching me by example, the importance of hard work, perseverance, and integrity. I thank my father for always believing in me and for teaching me the important lesson that nothing is impossible if you fight for it. 
I thank my lovely brothers, Samad and Siavash Hajinazar, for their immense and unwavering love and support which has been a sublime source of motivation for me. I would like to give my heartfelt thanks to my wonderful husband, Amir Pourmand, for being my rock, my number one supporter, and my safe place. This thesis would have been impossible without him by my side. Finally, I thank my most precious son, Nick Ray, whose little kicks kept me motivated during the writing of this dissertation. Words will never be able to describe my love for him. He will always be the reason behind my smile and I am eternally proud and grateful to be his mother.

 \end{acknowledgements}

\addtoToC{Table of Contents}%
\tableofcontents%
\clearpage

\addtoToC{List of Tables}%
\listoftables
\clearpage

\addtoToC{List of Figures}%
\listoffigures
\clearpage

%
%

\mainmatter%

\chapter{Introduction}
\label{sec:chapter1}

Modern computing systems need to process increasingly large amounts of data. Many key applications and workloads of important and wide range of domains (e.g., data mining, machine learning, graph and text analytics, databases, augmented reality applications, and genome analysis) and their potential improvement depend on fast and efficient processing of large data volumes. With the advent of such applications, computing in modern systems is primarily bottlenecked by data. In other words, by how fast and efficient we are in accessing, moving and processing data. Unfortunately, \onurt{modern} computing systems do not handle (i.e., store, access, and process) data well. The large amount of input and intermediate data required by modern applications overwhelms many of the key components of the modern computing systems. As a result, data has become a prime bottleneck in today's computing systems, making it challenging to efficiently support important emerging applications with high data demand.

The importance of handling the large amount of data processed by modern applications in an efficient manner has inspired a large body of research in processor design, memory and storage architectures, and key system components. However, we argue that fundamentally efficient handling of the increasing data demand in modern applications requires a holistic rethinking of the key concepts and components used in modern computing systems.

\section{Motivation: Existing Computing Systems Are Designed Without Having \onurt{Modern Applications'} Data Demand in Mind}

Today's computing systems have two important characteristics that make it significantly challenging to efficiently handle large amounts of data:

\textbf{Characteristic 1: Processor-Centric Architectures.} Modern computing systems follow the \emph{processor-centric paradigm} in which computation is performed only in the processor \onurt{(or compute-centric accelerators)} and every piece of data needs to be transferred to/from main memory to enable the computation. The increasing prevalence and growing size of data in modern applications has made data movement between memory devices (e.g., DRAM) and the processor across bandwidth-limited memory channels a first-class performance and energy bottleneck. \onurt{For example, a} recent work~\cite{boroumand2018google} shows that the energy and performance costs of data movement across the memory hierarchy are significantly higher than that of computation, consuming more than 60\% of the total system energy, when executing four major commonly-used consumer workloads, including machine learning inference, video processing and playback, and web browsing. Furthermore, in a processor-centric configuration, every component in the system except the processor is designed to serve the processor by storing and accessing the data or moving it to the processor for computation. This leads to about 80-95\% of the chip area \onurt{to be} consumed by the components that are solely responsible for storing, accessing and moving the data to the processor~\cite{Mutlu2019Dac}. Spending the majority of the chip resources on elements that are not able to process data or understand and take advantage of the properties of the data is not the right mindset considering the advent of applications that require fast, efficient, and intelligent computation of significantly large volumes of data.

We conclude that the processor-centric paradigm as the predominant execution model used in almost all computing systems is designed without having \onurt{modern applications'} overwhelming data demand in mind, causing significant waste in terms of energy and performance by requiring \onurt{frequent} data movement across the entire system. This \onurt{causes} data \onurt{to become} a first-class concern and a prime performance and energy bottleneck in the system, which makes \onurt{it challenging to} efficiently support important emerging applications with high data demand in today’s \onurt{computing} systems.

\textbf{Characteristic 2: Data-Oblivious Policies.} In order to cater to the high and diverse memory requirements of modern applications, today's computing systems employ increasingly larger main memories~\cite{benjaminmicro2016, raoux2008, jedec, micron, lee2009, mutlu2013, shacham2010, lee2010phase, om1, om2, mutlu2020modern, onursuperfri} \onurx{and heterogeneous main memory architectures}~(e.g.,~\cite{salp, charm, diva-dram, tldram, lee2015decoupled, changhat, kchangphd, dongphd, benjaminmicro2016, chang.sigmetrics2016, Yoon2014, yoon2012, refree, raoux2008, het2, li2017utility, mezacal}). Efficiently exploiting the significantly larger main memory capacities and the increasing heterogeneity in the main memory architectures requires careful memory management that is conventionally performed using virtual memory. However, conventional virtual memory frameworks are designed without \onurt{considering} \onurt{modern} applications' overwhelmingly high memory demand, and \onurt{thus}, \onurx{without} considering the new larger more complex main memory designs. Therefore, continuing to adopt the conventional approach to virtual memory with the increasing capacity and heterogeneity in today's main memory architectures requires a lot of effort and often leads to important challenges and inefficiencies. Furthermore, in addition to the growth in the size of data that modern applications process, prior works~\cite{xmem, vijaykumar2018, yixin2014, eden19, vetter2012, vetter2013} show that different pieces of program data have different performance characteristics (latency/bandwidth/parallelism sensitivity), and other inherent properties (e.g., compressibility, persistence, approximability). As highlighted by recent works~\cite{xmem, vijaykumar2018, yixin2014, eden19}, conveying semantic information about application's data to the hardware that manages the physical memory resources can enable vastly more intelligent \emph{data-aware} management of the underlying hardware resources (e.g., better address translation, data mapping, migration, and scheduling decisions) and a host of new optimization opportunities.

\onurt{Unfortunately}, conventional virtual memory frameworks~\cite{denning1970, atlas, multics, wsdenning, fotheringham1961, kilburn1962, Talluri1995, multics2, multics3} as the key interface between the software stack and the hardware are not capable of conveying any insights regarding the properties and memory behaviour of different pieces of program data. Instead, programs are traditionally conveyed to the hardware in the form of ISA instructions and a set of memory accesses to virtual addresses. This semantic gap leads to hardware treating \emph{all data} as the same, thereby \onurt{being} unable to exploit data's semantics properties to employ more intelligent management or optimization policies. Accordingly, the management and optimization policies used in existing systems are \emph{data-oblivious} and mainly component-aware~\cite{omintelligent}, i.e., designed according to the characteristics of the system component as apposed to the properties of the data that it handles (e.g., tuning tile size to fit a specific cache size). By ignoring the valuable memory characteristics and semantic properties of application's data, each component of the system is required to \emph{predict} the application's data behaviour in order to optimize its policies. Such a strategy is quite challenging and often not very effective due to three main problems. First, each component in the system has a \emph{limited} and \emph{localized} view of the data and is not aware of the overall behaviour of the application. Therefore, its decisions may not be ideal when considering the big picture. Second, each component requires \emph{separate} resources for inferring and predicting the behaviour of the data. This leads to repeated overhead in every component that can be avoided using a unified \onurt{and} expressive interface that connects different layers of the computing stack. \onurt{Third}, the optimizations made by different components mainly \emph{react} to the behaviour of the data as the overall application's behaviour is not available \onurt{or predictable}. This makes it challenging to make timely optimization/management decisions.

Data-oblivious policies in modern systems are a direct result of how poor today’s systems are \onurx{at} exploiting the valuable properties of different pieces of application’s data, which results in ineffective policies and lost performance optimization opportunities that can be achieved by exploiting data properties to improve the computing systems policies. We \onurt{posit} that \onurx{conventional} virtual memory \onurx{frameworks}, as a critical component of the existing computing systems cannot efficiently \onurt{support} the high data demand and diversity in modern applications, as well as the diversity in today's system configurations that \onurx{have evolved} in response to the modern application's memory needs.




\section{Our Approach: Data-Centric and Data-Aware Architectures for Fundamentally Efficient Data Handling }

In this thesis, we argue that, moving forward, computing systems need to consider \onurt{large amounts of} data and the efficient computation of data as the ultimate priority of the system. In particular, modern computing systems should follow two main directions. \onurt{(1)~data-centric architectures, and (2)~data-aware architectures.}

\textbf{Data-Centric Architectures.} In contrast to the \onurt{dominant} processor-centric design \onurt{paradigm}, we believe that, in order to efficiently handle large amounts of data, modern computing systems need to be \emph{data-centric}, meaning that they should (1) minimize data movement, and (2) compute data in or near where the data resides. The data-centric approach to computing is highly effective as (1)~it improves performance by reducing/eliminating the need to move data to the processor for computation, and (2)~provides the ability to take advantage of the large internal bandwidth in the main memory to increase the efficiency of the computation. For example, we show that a processing-using-DRAM architecture that efficiently implements and computes complex operations in DRAM, and provides the ability to support new arbitrary operations can \onurt{significantly} improve the overall performance and efficiency of the system (Chapter~\ref{sec:data-centric}).


\textbf{Data-Aware Architectures.} In contrast to \onurt{the dominant} data-oblivious policies in existing systems, we believe that modern computing systems should enable \emph{data-aware} policies, by allowing the software to easily communicate properties and semantic information about \onurt{each} application’s \onurt{and system's} data to the hardware. A data-aware architecture (1) understands what it can do with and to each piece of data, and (2) makes use of different properties of data (e.g., compressibility, approximability, locality, sparsity, access semantics) to improve performance, efficiency and other metrics. For example, we show that a more scalable data-aware virtual memory framework that (1)~is fundamentally designed to handle large amounts of data more efficiently, and (2)~understands, conveys and exploits the properties of program's data to enable more intelligent memory management and optimizations, can significantly improve performance for both native execution and virtual machine environments, and significantly improve the effectiveness of heterogeneous main memory architectures (Chapter~\ref{sec:data-aware}).


\subsection{Thesis Statement}

This thesis, hence, provides evidence for the following thesis statement:

\indent \textbf{The performance and energy efficiency of computing systems can improve significantly when handling the increasingly large amounts of data in modern applications by employing data-centric and data-aware architectures that can (1) remove the overheads associated with data movement by processing data where it resides, (2)~efficiently \onurx{adapting to} the diversity in today's system configurations and memory architectures that are designed to process large amounts of data, and (3)~understand, convey, and exploit the characteristics of the data to make more intelligent memory management decisions.}

\section{Overview of Research}

In this thesis, we propose a \onurx{novel} data-centric processing-using-memory framework and a \onurx{novel} data-aware virtual memory framework, \onurt{which} we briefly describe next. \onurt{We also put these contributions in the context of relevant prior work in Sections~\ref{sec:sim} and \ref{sec:vib}. We provide detailed discussions of and comparisons to prior work in Chapters~\ref{sec:data-centric} and \ref{sec:data-aware}.}

\subsection{SIMDRAM: A Data-Centric Framework for Bit-Serial SIMD Processing using DRAM (Chapter~\ref{sec:data-centric})}
\label{sec:sim}


In order to provide processing capability in or near where data resides, many prior works have explored DRAM designs (as well as other memory technologies) that are capable of performing computation using memory~\cite{ataberk2021, pluto, wang2020figaro, nom2020, chang2016low, kim2019d, changhpca2018, paternpum, Chi2016, Shafiee2016, seshadri2017ambit, seshadri2019dram, li2017drisa, seshadri2013rowclone, seshadri2016processing, deng2018dracc, xin2020elp2im, song2018graphr, song2017pipelayer, gao2019computedram, eckert2018neural, aga2017compute, dualitycache, ali2019memory, angizi2019graphide, li2018scope, seshadri2017simple, seshadri2015fast, dlugosch2014efficient, subramaniyan2017parallel, angizi2018imce, pinatubo2016, he2020sparse, angizi2019redram, imani2019floatpim, angizi2018dima, vivekbook1, vivekphd, Orosa2021}. However, \onurt{these works} suffer from three major shortcoming. First, they support only basic operations and fall short on efficiently supporting more complex operations, which limits their applicability~\cite{seshadri2017ambit,angizi2019graphide, angizi2018imce, ali2019memory, pinatubo2016, gao2019computedram, xin2020elp2im, angizi2020pim, he2020sparse, angizi2019redram, deng2018dracc,imani2019floatpim,angizi2018dima, seshadri2017buddy}. Second, they support only a limited and specific set of operations, lacking the flexibility to support new operations and cater to the wide variety of applications that can potentially benefit from processing-using-DRAM~\cite{li2017drisa, deng2018dracc}. Third, they often require significant changes to the DRAM subarray, which makes them costly~\cite{li2017drisa, deng2018dracc}. \onurt{These shortcomings} highlight the need for a framework that aids the general adoption of processing-using-DRAM by efficiently implementing complex operations and providing the flexibility to support new desired operations, while requiring minimal changes to the DRAM architecture.

To this end, \onurt{this thesis introduces} SIMDRAM, a flexible general-purpose processing-using-DRAM framework that (1) enables the efficient implementation of complex operations, (2) provides a flexible mechanism to support the implementation of arbitrary user-defined operations, and (3) uses an in-DRAM massively-parallel SIMD substrate that requires minimal changes to the DRAM architecture. We build the in-DRAM substrate used in the SIMDRAM framework around two key techniques. The first key technique is \emph{vertical data layout} in DRAM. Prior works show that employing a vertical layout for the data in DRAM~\cite{batcher1982bit,shooman1960parallel, gao2019computedram,ali2019memory,eckert2018neural, dualitycache, cmhill, kahle1989connection,hillis1993cm, tucker1988architecture} eliminates the need for adding extra logic in DRAM to implement \onurx{the} bit-shift operation~\cite{deng2018dracc, li2017drisa} which is essential for many complex operations. Employing vertical data layout provides SIMDRAM with two key benefits: (1) implicit shift operation, and (2) massive parallelism, wherein each DRAM column operates as a SIMD lane by placing the source and destination operands of an operation on top of each other in the same DRAM column. The second key technique used in SIMDRAM substrate is \emph{majority-based computation}. As opposed to using basic logic operations such as AND/OR/NOT as building blocks to implement in-DRAM computation~\cite{seshadri2017ambit,gao2019computedram,seshadri2015fast,li2017drisa}, SIMDRAM uses logically complete set of majority (MAJ) and NOT operations to implement in-DRAM computation. Majority-based computation enables SIMDRAM to achieve higher performance, higher throughput, and lower energy consumption compared to using basic logical operations as building blocks for in-DRAM computation.

\onurt{The} SIMDRAM framework \onurt{we introduce} is composed of three main steps. The first step  of the framework builds an efficient MAJ/NOT representation of a desired operation from its AND/OR/NOT-based implementation. The second step allocates DRAM rows to the operation’s inputs and outputs and generates the required sequence of DRAM commands to execute the desired operation, which is called \uprog{}. The third step executes the \uprog{} to perform the operation. SIMDRAM uses a control unit in the memory controller that transparently issues the sequence of commands to DRAM, as dictated by the \uprog{}. 

We provide a detailed reference implementation of SIMDRAM \onurt{in this thesis}, including required hardware, programming, and ISA support, to (1) address key system integration challenges, and (2) allow programmers to define new operations without hardware changes. We demonstrate the generality of the SIMDRAM framework using 16 complex in-DRAM operations, and seven commonly-used real-world applications. We show that SIMDRAM is a promising processing-using-memory framework that (1) can ease the adoption of processing-using-DRAM architectures, and (2) improve the performance and efficiency of processing-using-DRAM architectures.

\onurt{The SIMDRAM framework is introduced, discussed and evaluated in detail in Chapter~\ref{sec:data-centric} of this thesis (as well as Appendixes~\ref{apdx:aoi-to-mig}, \ref{apdx:row-to-op}, \ref{apdx:op-class}, and \ref{appendix}). An earlier version of SIMDRAM was presented at \onurx{the} ASPLOS \onurx{2021} conference~\cite{hajinazar2021}.}


\subsection{The Virtual Block Interface: A Flexible Data-Aware Alternative to the Conventional Virtual Memory Framework (Chapter~\ref{sec:data-aware})}
\label{sec:vib}

Considering the key role that virtual memory has in the overall performance of the modern computing systems, a wide body of research (e.g., \cite{vm1,vm2,vm3,karakostas2015,vm5,vm6,pichai2014,vm8,mask,vm9,vm10,vm11,vm12,vm13,vm14,pham2014,vm16,pham2015,vm18,vm19,vm20,vm21,vm22,vm23,vm24,vm25,vm26,vm27,vm28,vm29,vm30,wood1986, teller1990, teller1988, black1989, vm32,vm33, ritchie1985, vm34,vm35,vm36,vm37,vm38,vm39,vm40,vm41,vm42,meswani15,het9,het10,sim14,het12,mitosis-asplos20,elastic-cuckoo-asplos20,meza2013, mondrian, page_overlays, tlbpref, multics, denning1970, wsdenning, atlas, Kavita1994, seshadri2015gather, mondrianthesis, sharedvm, Abhishek2010, Saulsbury2000, Artemiy2019, yale2020, latr, Xingbo2017, Licheng2013, Harsh2020, Zhulin2020, Nadav2017, Artemiy2021, Bogdan2010, caba2015, Lustig2013, ecoTLB, psTLB, Bhattacharjee2017}) propose mechanisms to alleviate the \onurt{various} overheads associated with it \onurt{(in some cases when handling the large data demand in modern applications)}. However, despite notable improvements, these solutions suffer from three major shortcomings. First, they are mainly designed based on specific system or workload characteristics and, thus, are applicable to only a limited set of problems or applications. Second, each solution requires specialized and not necessarily compatible changes to \onurx{either} the operating system \onurx{or} hardware \onurx{or both}. Therefore, implementing \onurx{a combination of} these proposals at the same time in a system is a daunting prospect. Third, \onurt{these proposals} do \onurx{\emph{not}} support understanding, conveying, and exploiting the properties of program data in order to enable more intelligent memory management decisions \onurt{(i.e., data-aware architectures)}. \onurt{These shortcomings} highlight the need for a holistic solution to efficiently support modern applications in today’s diverse system configurations, by (1) eliminating the inefficiencies of the conventional virtual memory framework when handling \onurx{modern applications'} large amount of data, (2) exploiting the properties of different pieces of program data to improve performance, efficiency and other metrics.

To this end, in this \onurt{thesis}, we \onurt{introduce} the Virtual Block Interface (VBI), a general-purpose alternative virtual memory framework that has three \onurx{major} properties. First, \onurt{VBI} is able to understand, convey, and exploit the properties of different pieces of program data to enable more intelligent management of main memory. Second, \onurt{VBI} efficiently and flexibly supports increasingly diverse system configurations that are employed today to process the high data demand in modern applications. Third, \onurt{VBI} provides the key features of the conventional virtual memory framework while eliminating its key inefficiencies when handling large \onurx{amounts} of data in modern applications. The key idea in VBI is to delegate the physical memory allocation and address translation to dedicated hardware in the memory controller. 

The VBI design is driven by three key guiding principles. First, programs should be allowed to choose the size of their virtual address space. This mitigates the translation overheads associated with unnecessarily large and fixed-sized virtual address spaces in current systems that results in increasingly \onurx{large} inefficiencies with the diverse memory requirements of modern applications. Second, address translation should be decoupled from memory protection, as they are logically separate. This enables opportunities to remove address translation from the critical path of an access protection check, and defer the address translation until physical memory must be accessed, thereby lowering the performance overheads of virtual memory when handling large amounts of data in modern applications. It also enables the flexibility of managing address \onurx{translation} and access protection using separate structures customized to their characteristics\onurx{,} which again helps with more efficient address translation mechanisms which reduce the overhead of processing large data volumes in modern applications. Third, software should be allowed to communicate semantic information about application data to the hardware. This helps hardware to exploit the rich properties of different pieces of data to manage the underlying memory resources more intelligently. 

VBI naturally enables a variety of important \onurx{optimizations} that improve overall system performance when handling the high memory demand in modern applications, including: (1)~enabling benefits akin to using virtually-indexed virtually-tagged (VIVT) caches (e.g., reduced address translation overhead), (2)~eliminating two-dimensional page table walks in virtual machine environments, (3)~delaying physical memory allocation until the first dirty last-level cache line eviction, and (4)~flexibly supporting different virtual-to-physical address translation structures for different memory regions. 

We demonstrate the benefits of VBI with two example use cases. First, we experimentally show that VBI significantly improves performance for both native execution and virtual machine environments. Second, we show that VBI significantly improves the effectiveness of heterogeneous main memory architectures. We demonstrate that VBI is a promising new virtual memory framework, that can enable several important optimizations, and increase the design flexibility for virtual memory to support efficient handling of data in modern computing systems.

\onurt{The VBI virtual memory framework is \onurx{introduced}, discussed, and evaluated in detail in Chapter~\ref{sec:data-aware} of this thesis. An earlier version of VBI was presented at the ISCA 2020 conference~\cite{hajinazar2020}.}

\section{Contributions}

\onurtt{To our knowledge, this thesis is the first to propose and study new frameworks for fundamentally-efficient data handling in modern computing systems with a focus on two key directions at the same time, i.e., data-centric and data-aware architectures. In this thesis, we make two major contributions.}

\begin{itemize}
    \item \onurt{We present SIMDRAM, an end-to-end processing-using-DRAM framework that aids the widespread adoption of processing-\onurx{using}-DRAM, a data-centric computation paradigm that improves the overall performance and efficiency of the system when computing \onurx{on} large amounts of data by minimizing the cost of data movement and enabling computation where the data resides. To this end, SIMDRAM (1)~efficiently computes complex operations required by modern data intensive applications, (2)~provides the ability to implement new arbitrary operations as required, and (3)~uses an in-DRAM massively-parallel SIMD substrate that requires minimal changes to the DRAM architecture. We provide a detailed reference implementation of SIMDRAM, including required changes to applications, ISA, and hardware. We demonstrate the effectiveness and generality of the SIMDRAM framework \onurx{at} improving the system performance and efficiency using a wide range of complex operations and commonly-used real-world applications.}
     
    \item \onurt{We introduce VBI, a \onurx{novel} data-aware scalable general-purpose virtual memory framework that enables efficient handling of large amounts of data in modern applications by (1)~efficiently understanding, conveying, and exploiting the properties of different pieces of program data to enable more intelligent management of main memory, (2)~efficiently and flexibly supporting increasingly diverse system configurations and memory architectures that are employed today to process the high data demand in modern applications, and (3)~providing the key features of the conventional virtual memory framework while eliminating its key inefficiencies when handling large \onurx{amounts} of data in modern applications. We provide a detailed reference implementation of VBI, including required changes to applications, system software, ISA, and hardware. We demonstrate the effectiveness of VBI \onurx{at} significantly improving the overall system performance in native and virtualized environments. We also show that VBI significantly improves the effectiveness of heterogeneous main memory architectures.}
    
\onurtt{Other contributions of this thesis are listed in Chapters~\ref{sec:data-centric} and \ref{sec:data-aware}. We specifically direct the reader to the final subsections in each chapter for a concise summary and contributions of each chapter, i.e., Sections~\ref{sec:conclusion_simd} and \ref{sec:conclusion_vbi} in this thesis. }

\end{itemize}

\chapter{SIMDRAM}
\label{sec:data-centric}

As discussed in Chapter~\ref{sec:chapter1}, the increasing prevalence and growing size of data in modern applications
has led to high \omiv{energy and latency} costs for computation in traditional computer architectures.
Moving large \om{amounts} of data between memory (e.g., DRAM) and the CPU across 
bandwidth-limited memory channels can consume more than
60\% of the total energy in modern systems~\cite{mutlu2019, boroumand2018google}. 
To mitigate \omiv{such} costs, \om{the} \emph{processing-in-memory} (PIM) \om{paradigm moves} computation closer to where the 
data resides, reducing (and in some cases eliminating) the
need to move data between memory and the processor.

There are two main approaches to PIM~\cite{ghoseibm2019, mutlu2020modern}:
(1)~processing-near-memory, where PIM logic is added to the same die as memory or to the logic layer of 3D-stacked memory~\cite{deoliveira2021, lee2016simultaneous, ahn2015scalable, nai2017graphpim, boroumand2018google, lazypim, top-pim, gao2016hrl, kim2018grim, drumond2017mondrian, santos2017operand, NIM, PEI, gao2017tetris, Kim2016, gu2016leveraging, HBM, HMC2, boroumand2019conda, hsieh2016transparent, cali2020genasm,Sparse_MM_LiM, NDC_Micro_2014, farmahini2015nda,loh2013processing, pattnaik2016scheduling,akin2016data, hsieh2016accelerating, babarinsa2015jafar, lee2015bssync, devaux2019true, ghose2018enabling, deoliveira2021optane, giannoula2021SynCron, mohammedprimer, kim2018grimarx, lazypimarx, napel, Guo2014, Gwangsun2017, Zhiyu2017, Dai2018, Mingxing2018, Huang2020, Youwei2019, Lloyd2015, Lloyd2018, Gokhale2015, Nair2015, Sura2015, Balasubramonian2014, Polynesia, Mensa, Hashemi2016, hashemi2016continuous, asghari2016chameleon, MEMSYS_MVX, Huangfu2019, nero, Maciej2021arx, Maciej2021, Gagandeep2021}; and (2)~processing-using-memory, which makes use of the operational principles of the memory cells themselves to perform computation by enabling interactions between cells~\cite{ataberk2021, pluto, wang2020figaro, nom2020, chang2016low, kim2019d, changhpca2018, paternpum, Chi2016, Shafiee2016, seshadri2017ambit, seshadri2019dram, li2017drisa, seshadri2013rowclone, seshadri2016processing, deng2018dracc, xin2020elp2im, song2018graphr, song2017pipelayer, gao2019computedram, eckert2018neural, aga2017compute, dualitycache, ali2019memory, angizi2019graphide, li2018scope, seshadri2017simple, seshadri2015fast, dlugosch2014efficient, subramaniyan2017parallel, angizi2018imce, pinatubo2016, he2020sparse, angizi2019redram, imani2019floatpim, angizi2018dima, vivekbook1, vivekphd, Maciej2021arx, Maciej2021}. Since processing-using-memory operates directly in the memory \om{arrays}, it benefits from the large internal bandwidth and parallelism available inside the memory arrays, which processing-near-memory solutions \omiv{cannot take advantage of}.

A common approach for processing-using-memory architectures is to make use of bulk bitwise computation. Many widely-used data-intensive applications (e.g., databases, neural networks, graph analytics) heavily rely on a broad set of simple (e.g., AND, OR, XOR) and complex (e.g., equality check, multiplication, addition) bitwise operations. Ambit~\cite{seshadri2017ambit, seshadri2015fast}, an in-DRAM processing-using-memory accelerator, was the first work to propose exploiting DRAM's analog operation\om{al principles} to perform bulk bitwise AND, OR, and NOT logic operations. Inspired by Ambit, many prior works have explored DRAM (as well as NVM) designs that are capable of performing in-memory bitwise operations~\omiv{\cite{angizi2019graphide, angizi2018imce, ali2019memory, pinatubo2016, gao2019computedram, xin2020elp2im, angizi2020pim, he2020sparse, angizi2019redram, deng2018dracc,imani2019floatpim,angizi2018dima, vivekphd, seshadri2017ambit, seshadri2017buddy, seshadri2015fast, seshadri2013rowclone, seshadri2017simple, seshadri2019dram, seshadri2016processing, vivekbook1}}. However, a major shortcoming prevents these proposals from becoming widely applicable: 
they support only basic operations (e.g., Boolean operations, addition) and fall short on flexibly \om{and easily} supporting new and more 
complex operations. Some prior works propose processing-using-DRAM designs that support more complex operations~\cite{li2017drisa, deng2018dracc}. However, such designs (1)~require significant changes to the DRAM subarray, and (2)~support only a limited and specific set of operations, lacking the flexibility to support new operations and cater to the wide variety of applications that can potentially benefit from in-memory computation. 

\textbf{Our goal} in this work is to design a framework that aids the adoption of processing-using-DRAM by efficiently implementing complex operations and providing the flexibility to support new desired operations. To this end, we propose SIMDRAM, an end-to-end processing-using-DRAM framework that provides the programming interface, the ISA, and the hardware support for (1)~efficiently computing \emph{complex} operations, and (2)~providing the ability to implement \emph{arbitrary} operations as required, all in an in-DRAM massively-parallel SIMD substrate.
At its core, we build the SIMDRAM framework around a DRAM substrate that enables two previously-proposed techniques:
(1)~vertical data layout in DRAM, and 
(2)~majority-based logic for computation.

\textbf{Vertical Data Layout.} Supporting bit-shift operations is essential for implementing complex computations, such as addition or multiplication. Prior works show that employing a vertical layout~\omiv{\cite{batcher1982bit,shooman1960parallel, gao2019computedram,ali2019memory,eckert2018neural, dualitycache, cmhill, kahle1989connection,hillis1993cm, tucker1988architecture}} for the data in DRAM,  such that all bits of an operand are placed in a single DRAM column (i.e., in a single bitline), eliminates the need for adding extra logic in DRAM to implement shifting~\cite{deng2018dracc, li2017drisa}. Accordingly, SIMDRAM supports efficient bit-shift operations by storing operands in a vertical fashion in DRAM. This provides SIMDRAM with two key benefits. First, a bit-shift operation can be performed by simply copying a DRAM row into another row (using RowClone~\cite{seshadri2013rowclone}, LISA~\cite{chang2016low}, NoM~\cite{nom2020} or FIGARO~\cite{wang2020figaro}). For example, SIMDRAM can perform a left-shift-by-one operation by copying the data in DRAM row $j$ to DRAM row \textit{j+1}. (Note that while SIMDRAM supports bit shifting, we can optimize many applications to avoid the need for explicit shift operations, by simply changing the row indices of the SIMDRAM commands that read the shifted data). Second, SIMDRAM enables massive parallelism, wherein each DRAM column operates as a SIMD lane by placing the source and destination operands of an operation on top of each other in the same DRAM column.

\textbf{Majority-Based Computation.} Prior works use majority operations to implement basic logical operations~\cite{seshadri2017ambit,gao2019computedram,seshadri2015fast,li2017drisa} (e.g., AND, OR) or addition~\cite{angizi2019graphide,ali2019memory,gaillardon2016programmable,li2017drisa,deng2018dracc,gao2019computedram}. These basic operations are then used as basic building blocks to implement the target in-DRAM computation. SIMDRAM extends the use of the majority operation by directly using \omiv{only} the logically complete set of majority (MAJ) and NOT operations to implement in-DRAM computation. Doing so enables SIMDRAM to achieve higher performance, throughput, and reduced energy consumption compared to using basic logical operations as building blocks for in-DRAM computation. We find that a computation typically requires fewer DRAM commands using MAJ and NOT than using basic logical operations AND, OR, and NOT.

\nasirevi{To aid the adoption of processing-using-DRAM by flexibly supporting new and more complex operations, SIMDRAM addresses \revonurii{two key} challenges: (1)~how to synthesize new arbitrary in-DRAM operations, and (2)~how to exploit an optimized implementation and control flow for such \revonurii{newly-added operations} \revonurii{while} taking into account key limitations of in-DRAM processing (e.g., DRAM operations \revonurii{that destroy input data}, limited number of DRAM rows that are capable of \revonurii{processing-using-DRAM}, and the need to avoid costly in-DRAM copies).}
As a result, \nasii{SIMDRAM is the first end-to-end framework for \revonurii{processing-using-DRAM}. SIMDRAM provides (1)~\onur{a\omiv{n} \omiv{effective algorithm}} to generate an efficient MAJ/NOT-based implementation of \revonurii{a given desired operation}; (2)~\omiv{\onur{\omiv{an} algorithm to} appropriately allocate DRAM rows \omiii{to} the operands of the operation and an algorithm to map the computation \nasirevi{to an efficient sequence of DRAM commands to execute \emph{any} MAJ-based computation;}} 
and (3)~the \revonurii{programming} interface, ISA support and hardware components required to }\revonurii{(i)~}\nasirevi{compute any new user-\omiv{defined} in-DRAM operation without hardware modifications\revonurii{, and (ii)~}program the memory controller for issuing DRAM commands to the corresponding DRAM rows and correctly performing the computation.}
\textfromsl{\omiv{Such} end-to-end support enables SIMDRAM as a holistic approach that facilitates the adoption of processing-using-DRAM \nasirevi{through (1) enabling the flexibility to support new in-DRAM operations by providing the user with a simplified interface to add desired operations, and (2) eliminating the need for adding extra logic to DRAM}.}

\textfromsl{The SIMDRAM framework efficiently supports a wide range of operations of different types. In this work, we demonstrate the functionality of the SIMDRAM framework using an example set of \omiv{16} operations including  (1)~\emph{N}-input logic operations (e.g., AND/OR/XOR of more than 2 input bits); (2)~relational operations (e.g., equality/inequality check, greater than, maximum, minimum); (3)~arithmetic operations (e.g., addition, subtraction, multiplication, division); (4)~predication (e.g., if-then-else); and (5)~other complex operations such as bitcount and ReLU~\cite{goodfellow2016deep}. The SIMDRAM framework is not limited to these \omiv{16} operations, and can enable processing-using-DRAM for other existing and future operations. 
\geraldorevi{SIMDRAM is well-suited to application classes that (i) \omiv{are SIMD-friendly, (ii)} have a regular access pattern, and (\omiv{iii}) are memory bound. Such applications are common in domains such as database analytics, \omiv{high-performance computing}, \omiv{image processing,} and machine learning.}}

We compare the benefits of SIMDRAM to different state-of-the-art computing platforms (CPU, GPU, and the Ambit~\cite{seshadri2017ambit} in-DRAM computing mechanism). We comprehensively evaluate SIMDRAM's reliability, area overhead, throughput, and energy efficiency. We leverage the SIMDRAM framework to accelerate seven application kernels from machine learning, databases, and image processing (VGG-13~\cite{simonyan2014very}, VGG-16~\cite{simonyan2014very}, LeNET~\cite{lecun2015lenet}, kNN~\cite{lee1991handwritten}, TPC-H~\cite{tpch}, BitWeaving~\cite{li2013bitweaving}, \omii{b}rightness~\cite{gonzales2002digital}). 
\omiv{Using a single DRAM bank, SIMDRAM provides
(1)~2.0$\times$ the throughput and 2.6$\times$ the energy efficiency of Ambit~\cite{seshadri2017ambit}, averaged across the 16 implemented operations; and
(2)~2.5$\times$ the performance of Ambit, averaged across the seven application kernels.}
\omiv{Compared to a CPU and a high-end GPU, SIMDRAM using 16 DRAM banks provides (1)~257$\times$ and 31$\times$ the energy efficiency, and \omi{88}$\times$ and \omi{5.8}$\times$ the throughput of the CPU and GPU, respectively, averaged across the 16 operations; and (2)}~\omiv{21$\times$ and 2.1$\times$ the performance of the CPU and GPU, respectively, averaged across the \omv{seven} application kernels.}  \omiv{SIMDRAM incurs no additional area overhead on top of Ambit\omiii{, and a total area overhead of \sgii{only 0.2\% in a high-end CPU.}}} We also evaluate the reliability of SIMDRAM under different degrees of manufacturing process variation, and observe that it guarantees correct operation as the DRAM process technology node scales down to smaller sizes. 

\section{Background}
\label{background}
%
%
%
\om{W}e \om{first} briefly explain the architecture of \geraldorevii{a typical} DRAM \geraldorevii{chip}. \nasrev{Next, we describe prior processing-using-DRAM works that \mech builds on top of (RowClone~\cite{seshadri2013rowclone} and Ambit~\om{\cite{seshadri2017ambit, seshadri2015fast,seshadri2019dram}})}\revdel{We then} \om{and} \nasii{explain the implications of majority-based computation.} \onurt{For an even more detailed operation of DRAM, we refer the reader to many prior works~\cite{kchangphd, chang2014improving, chang2016low, changsig2017, diva-dram, charge-dram, softmc, salp, ramulator, kim2014flipping, lee2016simultaneous, diva-dram, tldram, lee2015decoupled, lee2015adaptive, patel19, seshadri2017ambit, seshadri2013rowclone, seshadri2017simple, vivekphd, patel20, dongphd, avatar, chang.sigmetrics2016, changhpca2018, hassancrow, seshadri2015gather, Jamie2013, Khan2016, Khan2014, Liu2012, Patel2017, Yoonguphd, Khan2017, Khan2017cal, Yaohua2018, kim2019d, kim2018solar}.}
%

\subsection{DRAM {Basics}}
\label{sec:dram-basics}

\begin{figure}[ht]
    \centering
    \includegraphics[width=0.95\linewidth]{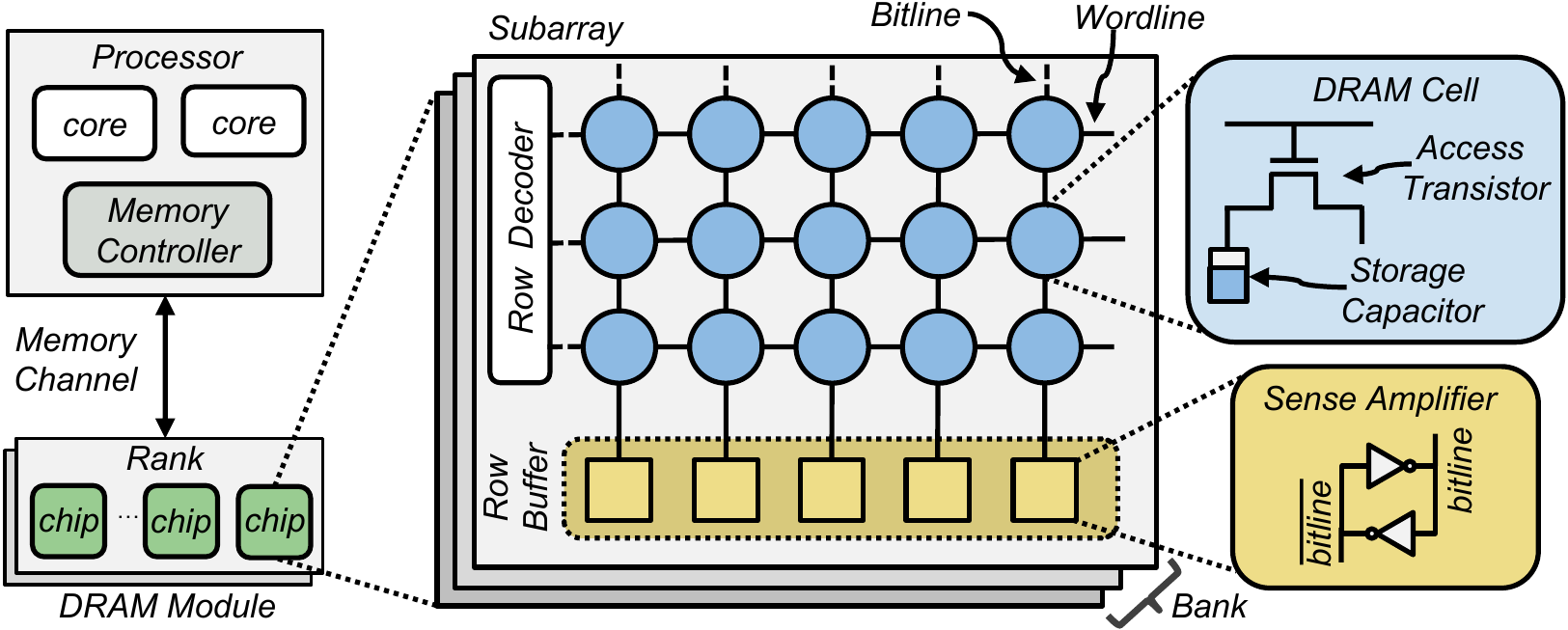}
    \caption{\geraldorevii{High-level overview of DRAM organization.}}
    \label{fig_subarray_dram}
\end{figure}

A DRAM \revdel{module}\om{system} comprises a hierarchy of \revdel{2D }components, \geraldorevii{as \cmr{Figure}~\ref{fig_subarray_dram} shows,} starting with \emph{channels} at the highest level. A channel is\revdel{ further} subdivided into \emph{ranks}, and a rank is subdivided into multiple \emph{banks} \om{(e.g., 8-16).}\revdel{, which are e}\om{~Each bank is} composed of \om{multiple (e.g., 64-128)} 2D arrays of cells known as \emph{subarrays}. Cells within a subarray are organized into \geraldorevii{multiple} \emph{rows} \geraldorevii{(e.g., 512-1024)} and \geraldorevii{multiple} \emph{columns}\revdel{,} \geraldorevii{(e.g., 2-8~kB)}~\cite{lee2015adaptive, diva-dram, kim2018solar}. A cell consists of an \emph{access transistor} and a \emph{storage capacitor} that encodes a single bit of data using its voltage level. The source \nasrev{node\om{s} of the} access transistors of all the cells in the same column connect the cells' storage capacitor\om{s} to the same \emph{bitline}. Similarly, the gate \nasrev{node\om{s} of} the access transistor\om{s} of all the cells in the same row connect the\revdel{ir} \om{cells'} \revdel{storage capacitor\om{s}}\geraldorevii{access transistors} to the same \emph{wordline}. 

\omi{When a wordline is asserted, all cells along the wordline are connected to their corresponding bitlines, which perturbs the voltage of each bitline depending on the value stored in each cell's capacitor. A two-terminal \emph{sense amplifier} connected to each bitline senses the voltage difference between the bitline (connected to one terminal) and a reference voltage (typically  $\frac{1}{2}V_{DD}$; connected to the other terminal) and amplifies it to a CMOS-readable value. In doing so, the sense amplifier terminal connected to the reference voltage is amplified to the \emph{opposite} (i.e., \emph{negated}) value, which is shown as the $\overline{\mbox{bitline}}$ terminal in \cmr{Figure}~\ref{fig_subarray_dram}.}
The set of sense amplifiers in each subarray forms a logical \emph{row buffer}, which \omi{maintains the sensed data for as long as the row is \emph{open} (i.e., the wordline continues to be asserted).}
\nasii{\sgii{A read or write} operation \nasii{in DRAM} includes \emph{three} steps:
\begin{enumerate}[noitemsep,topsep=0pt,parsep=0pt,partopsep=0pt,labelindent=0pt,itemindent=0pt,leftmargin=*]
    \item \nasrev{\texttt{ACTIVATE}}. The \emph{wordline} of the target row is asserted, which connects all cells along the row to their respective bitlines. Each \omi{bitline shares charge with its corresponding cell capacitor, and the resulting bitline voltage shift is sensed and amplified by the bitline's sense amplifier. Once the sense amplifiers \omiii{finish amplification}, 
    the row buffer contains the values originally stored within the cells along the asserted wordline.}
    \item \texttt{\sgii{\om{RD/WR}}}. The memory controller then \revdel{performs}\geraldorevii{issues} read or write commands to columns within the activated row \om{\omi{(i.e., the data within the row buffer)}}.
    \item \nasrev{\texttt{PRECHARGE.}} \revdel{\nasrev{In this step, s}\om{S}ense amplifiers are disabled, and each bitline is restored to its quiescent state (e.g., typically $\frac{1}{2}V_{DD}$).} The capacitor is disconnected from the bitline by disabling the wordline, \omi{and the bitline voltage is restored to its quiescent state (e.g., typically $\frac{1}{2}V_{DD}$)}.  
\end{enumerate}
}

\subsection{Processing-using-DRAM}
\label{sec:in-dram}

\subsubsection{In-DRAM Row Copy.} 
\label{sec_rowclone}

~RowClone~\cite{seshadri2013rowclone} is a\revdel{n in-DRAM} mechanism that exploits the vast internal DRAM bandwidth to \nasrev{efficiently} copy rows \revdel{within a subarray}\om{inside DRAM} without CPU intervention. RowClone enables copying \juang{a source} row~$A$ to \juang{a destination} row~$B$ \om{in the same subarray} by issuing two consecutive \texttt{ACTIVATE}
commands to these two rows, \juang{followed by a \texttt{PRECHARGE} command}. 
\juang{This \om{command} sequence is called \texttt{AAP}~\om{\cite{seshadri2017ambit}}.} \juang{The first \texttt{ACTIVATE} command copies the contents of the source row $A$ into the row buffer. \geraldorevii{The second \texttt{ACTIVATE} command connects the cells in the destination row~$B$ to the bitlines.} \revdel{When the second \texttt{ACTIVATE} command is issued,} 
\omi{Because the sense amplifiers \omiii{have already sensed \omiv{and amplified} the source data } 
by the time row~$B$ is activated, the data (i.e., voltage level) in each cell of row~$B$ is overwritten by the data stored in the row buffer (i.e., row~$A$'s data).}}
\omi{Recent work~\cite{gao2019computedram} experimentally demonstrates the feasibility of executing in-DRAM row copy operations in unmodified off-the-shelf DRAM chips.}


\subsubsection{In-DRAM Bitwise Operations.} 
\label{sec_ambit_logic} 

Ambit~\om{\cite{seshadri2017ambit, seshadri2015fast, seshadri2019dram}} shows that simultaneously activating \emph{three} DRAM \juang{rows} \juang{(\omi{via a DRAM operation} called \emph{Triple Row Activation, TRA})} can be used to perform \om{bitwise} Boolean \omi{AND, OR, and NOT operations} on the values contained within the cells \omi{of the three rows}. \juang{\omi{When activating three rows}, three cells connected to each bitline share charge simultaneously and contribute to the perturbation of the bitline. 
\omi{Upon sensing the perturbation, the sense amplifier amplifies the bitline voltage to}
\geraldorevii{$V_{DD}$ or 0 if at least two of the capacitors of the three DRAM cells are charged or discharged, respectively.}} \juang{As \omi{such},} a TRA results in a Boolean \emph{majority operation} ($MAJ$) \juang{among the three DRAM cells on each bitline}. A majority operation MAJ \omi{outputs a 1 (0)} only if more than half of \omi{its} inputs are \omi{1 (0)}. \juang{In terms of AND ($\cdot$) and OR (+) operations, a 3-input majority operation can be expressed as \texttt{MAJ(A, B, C) = A $\cdot$ B + A $\cdot$ C + B $\cdot$ C.}}

\omi{Ambit implements MAJ by introducing a custom row decoder (discussed in \cmr{\Cref{subarray}}) that can perform a TRA by simultaneously addressing three wordlines. To use this decoder, Ambit defines a new command sequence called \texttt{AP}, which issues (1)~a TRA to compute the MAJ of three rows, followed by (2)~a \texttt{PRECHARGE} to close all three rows.\footnote{\omi{Although the `\texttt{A}' in \texttt{AP} refers to a TRA operation instead of a conventional \texttt{ACTIVATE} command, we use this terminology to remain consistent with the Ambit paper~\cite{seshadri2017ambit}\omiii{, since an \texttt{ACTIVATE} command can be internally translated to a TRA operation by the DRAM chip~\cite{seshadri2017ambit}}.}}}
Ambit uses \omi{\texttt{AP} command sequences} to implement Boolean \om{AND} and \om{OR} operations by simply setting one of the inputs (e.g., $C$) to \omi{1} or \omi{0}. The AND operation is computed by setting $C$ to 0 (i.e., \texttt{MAJ(A, B, 0) = A AND B})\revdel{, while t}\om{. T}he OR operation is computed by setting $C$ to 1 (i.e., \texttt{MAJ(A, B, 1) = A OR B}).

\omi{To achieve functional completeness alongside AND and OR operations, Ambit implements NOT operations by exploiting the differential design of DRAM sense amplifiers. As \cmr{\Cref{sec:dram-basics}} explains, the sense amplifier already generates the complement of the sensed value as part of the activation process ($\overline{\mbox{bitline}}$ in \cmr{Figure}~\ref{fig_subarray_dram}). Therefore, Ambit simply forwards $\overline{\mbox{bitline}}$ to a special DRAM row in the subarray that consists of DRAM cells with \emph{two} access transistors, called \emph{dual-contact cells} (DCCs). Each access transistor is connected to one side of the sense amplifier and is controlled by a separate wordline (\emph{d-wordline} or \emph{n-wordline}). By activating \omiii{either} the d-wordline or the n-wordline, the row of DCCs can \omiii{provide} the true or negated value stored in the \omiii{row's} cells, respectively.}



\subsubsection{Majority-Based Computation.}
\label{sec_maj-based}


\juangr{~Activating multiple rows simultaneously reduces the reliability of the value read by the sense amplifiers \mpi{due to manufacturing process variation, which} introduces non-uniformities in circuit-level electrical characteristics (e.g., \omi{variation in} cell capacitance \omi{level\omiii{s}})~\om{\cite{seshadri2017ambit}}. This effect worsens with (1) an increased  number of simultaneously activated rows, and (2) more advanced technology nodes with smaller sizes. Accordingly, \geraldorevii{although}\revdel{while} \geraldorevii{processing-using-DRAM} can potentially support majority operations with more than \nasrev{three} inputs 
(as proposed by prior works~\cite{ali2019memory, angizi2019graphide,pinatubo2016}) \mpi{our realization of 
\geraldorevii{processing-using-DRAM} uses} the minimum number of inputs required for a majority operation ($N$=3) to maintain \nasrev{the reliability of the} computation. In \cmr{\Cref{sec_reliability}}, we \om{demonstrate via SPICE simulations}\revdel{show} that using 3-input MAJ \nasrev{operations} \revdel{improves}\om{provides} \revdel{the}\om{higher} reliability \revdel{of \gfrev{in-DRAM operations?}\mech }compared to designs with more than \nasrev{three} inputs per MAJ \nasrev{operation}.} Using 3-input MAJ, \revdel{\mech}\geraldorevii{a processing-using-DRAM substrate} does not require modifications to the subarray organization (\cmr{Figure}~\ref{fig_subarray}) beyond the ones proposed by Ambit \omi{(\cmr{\Cref{subarray}})}. 
\mpi{\om{R}ecent work~\cite{gao2019computedram} experimentally demonstrates the feasibility of executing MAJ operations by activating three rows \nasrev{in \om{unmodified off-the-shelf}\revdel{commodity DRAM} \omi{DRAM} chips}.}

\section{SIMDRAM Overview}
\label{mechanism}


\sgii{SIMDRAM is a processing-using-DRAM framework whose goal is to (1)~enable the efficient implementation of complex operations and (2)~provide a flexible mechanism to support the implementation of arbitrary user-defined operations.}
\nasii{\revdel{In this section, w}\om{W}e present the subarray organization in SIMDRAM, describe an overview of the SIMDRAM framework, and explain how \sgii{to integrate SIMDRAM} into \sgii{a} system.}

\subsection{Subarray Organization}
\label{subarray}

\begin{figure}[ht]
    \centering
    \includegraphics[width=0.8\linewidth]{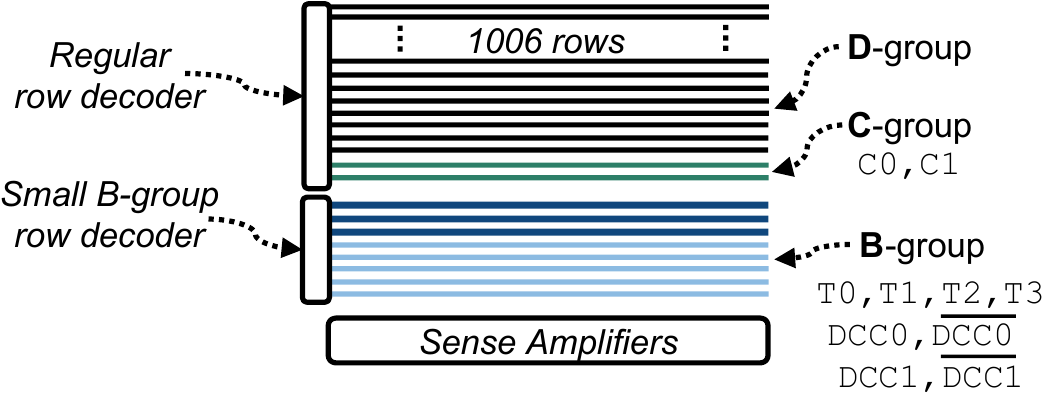}
    \caption{\geraldorevi{\mech subarray organization~\cite{seshadri2017ambit}. \revdel{Adapted from~\cite{seshadri2017ambit}.}}}
    \label{fig_subarray}
\end{figure}

\sgii{In order to perform processing-using-DRAM, SIMDRAM makes use of a subarray organization that incorporates additional functionality to perform logic primitives (i.e., MAJ and NOT). This subarray organization is \emph{identical} to Ambit's~\cite{seshadri2017ambit} and is similar to DRISA's~\cite{li2017drisa}.}
\textfromsl{\cmr{Figure}~\revdelrefr{\ref{fig_subarray}}\revdelrefa{\hl{1}} illustrates the internal organization of a subarray in \mech, which \juang{resembles a conventional DRAM subarray}. \juang{\mech requires \om{only} minimal modifications to the DRAM subarray (namely, a small row decoder that can activate three rows simultaneously) to enable computation}. \om{Like Ambit~\cite{seshadri2017ambit},} \juang{\mech divides DRAM rows} into \emph{three groups}: \sgii{the} \textbf{D}ata group (D-group), \sgii{the} \textbf{C}ontrol group (C-group) and \sgii{the} \textbf{B}itwise group (B-group).}

The D-group contains regular rows that store \revdel{users'}\om{program or system} data. The \sgii{C-group} \juang{consists of two constant} \sgii{rows, called C0 and C1,} that \juang{contain} \om{all-}0 and \om{all-}1 \juang{values}, respectively. \geraldo{These rows are used (1)~\sgii{as} initial input values for a given \sgii{SIMDRAM} \geraldorevii{operation}
(e.g., the initial carry-in bit in a full \omi{addition}), or (2)~to perform operations that naturally require\revdel{s} AND/OR operations (e.g., AND/OR reductions)}. The D-group and the C-group are connected to the regular row decoder, which selects a single row \juang{at a time}.

The \sgii{B-group} contains six \sgii{regular rows, called T0, T1, T2, and T3;} and two rows of dual-contact cells \sgii{(see \cmr{\Cref{sec_ambit_logic}}), 
whose d-wordlines are called DCC0 and DCC1, and whose n-wordlines are called $\overline{\mbox{DCC0}}$ and $\overline{\mbox{DCC1}}$, respectively}.
\sgii{The B-group} rows\om{, called \emph{compute rows}, are} designated to perform bitwise operations. They are all connected to a special row decoder that can simultaneously activate three rows using a single address 
(i.e., perform a TRA) 

\geraldorevi{\revonur{Using} a typical subarray size of 1024 rows~\revdelrefr{\cite{chang2014improving, salp, kim2018solar,tldram,kim2019d}}\revdelrefa{\hl{[19,50,53,57,64]}}, \mech splits the row addressing into 1006 \sgii{D-group rows, 2 C-group rows, and 16 B-group rows}.}

\subsection{Framework Overview}
\label{overview}

\begin{figure}[t!]
\begin{subfigure}{\linewidth}
    \centering
    \includegraphics[width=\textwidth]{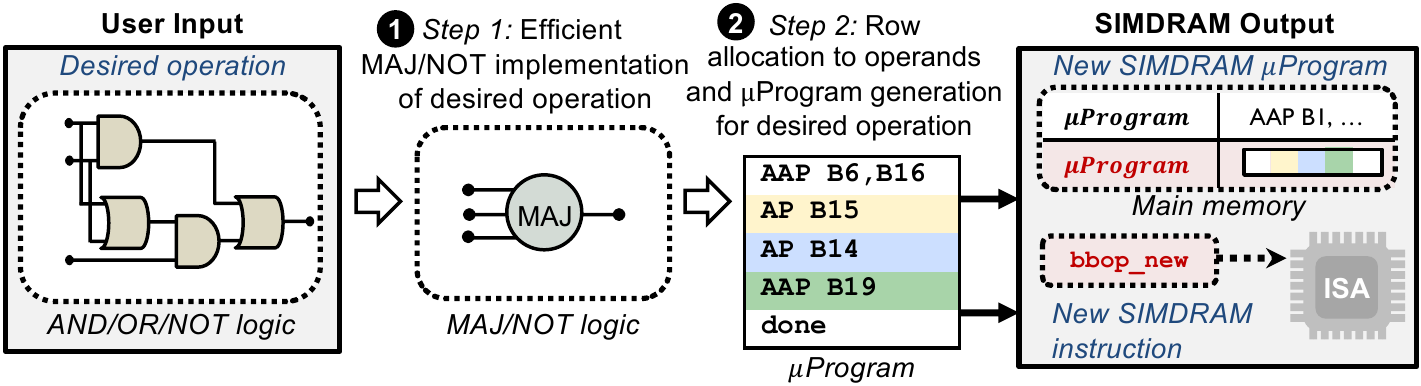}%
    \caption{\geraldorevi{SIMDRAM Framework: Steps 1 and 2}}
    \label{fig_framework_1_2}
\end{subfigure}
\par\bigskip 
\begin{subfigure}{\linewidth}
  \centering
    \includegraphics[width=\textwidth]{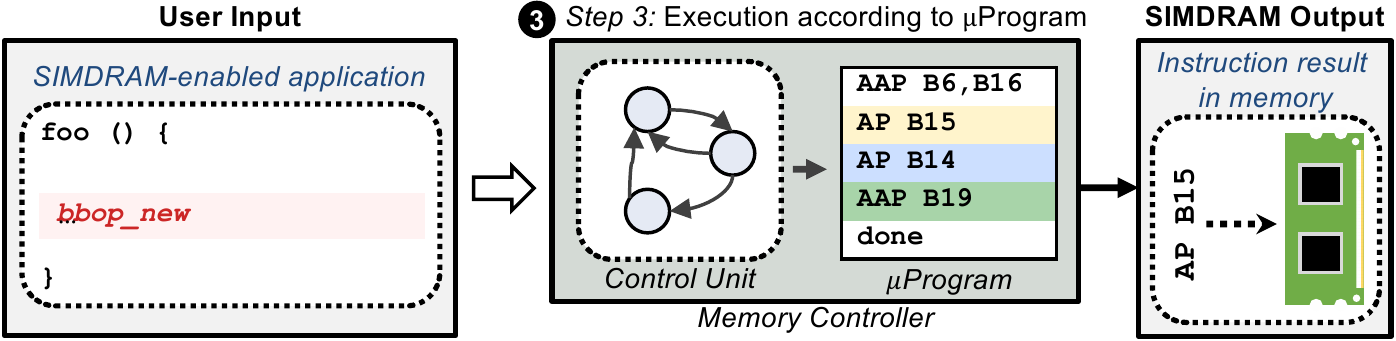}%
    \caption{\geraldorevi{SIMDRAM Framework: Step 3}}
    \label{fig_framework_3}
\end{subfigure}
\caption{\geraldorevi{Overview of the \mech framework.}}
\label{fig:figoverview}
\end{figure}

\sgii{SIMDRAM \omi{is} an end-to-end framework that provides the user with the ability to implement an \emph{arbitrary} operation in DRAM using the \aap command sequences.
The framework comprises}
three key steps, \geraldorevi{which \sgii{are illustrated in \cmr{Figure}~\ref{fig:figoverview}.}
\geraldo{The first two steps \sgii{of the framework} \sg{give} \omi{the} user the 
\juangg{ability} to efficiently implement 
any desired \revonuri{operation} in DRAM, while the third step controls the execution flow of the in-DRAM computation transparently from the user. 
\revonuri{We} briefly describe these steps \sgii{below, and discuss each step in detail in \cmr{\Cref{sec:operations}}}.}}

\om{T}he first step \geraldorevi{(\ding{182} in \cmr{Figure}~\ref{fig_framework_1_2}; \sgii{\cmr{\Cref{sec:framework:step1}}})} \sgii{builds an efficient MAJ/NOT representation of \revonuri{a given} desired operation 
\omi{from its AND/OR/NOT-based implementation}.  Specifically, this step} \om{takes as input a desired operation and} use\om{s} logic optimization to minimize the number of \sgii{logic primitives}
(and, therefore, the \om{computation} latency) required to perform \revdel{a specific}\om{the} operation. \sg{Accordingly, for a desired \om{operation input into the \mech framework by the user}, \jgll{Weird writing.} the first step 
derive\geraldorevii{s} its \emph{optimized} MAJ/NOT-based implementation.  }

The second step (\geraldorevi{\ding{183} in \cmr{Figure}~\ref{fig_framework_1_2}; \sgii{\cmr{\Cref{sec:framework:step2}}}}) 
\omi{allocates DRAM rows \omiii{to}} 
\sgii{the operation's inputs and outputs and generates the required sequence of DRAM commands to execute the desired operation. Specifically, this step}
translates the 
MAJ/NOT-based implementation \nasrev{of the operation} 
\sg{into} \geraldorevii{\aaps}. 
This step 
\sgii{involves}
(1)~\omi{allocating the designated compute rows in DRAM \omiii{to} the operands}, and (2)
\juangg{~determining} the \geraldorevii{optimized} sequence of \geraldorevii{\aaps} that are required to perform the \geraldorevii{operation}. \juangg{While doing so, SIMDRAM minimizes} the number of \geraldorevii{\aaps}\revdel{AAPs/TRAs} required for a specific operation. \geraldorevii{\cmr{This step's output} is a \uprog{}, i.e., the optimized sequence of \aaps that \omi{is stored in main memory and} will be used to execute the  operation \sgii{at} runtime. }

\sg{The third step \geraldorevi{(\ding{184} in \cmr{Figure}~\ref{fig_framework_3}; \sgii{\cmr{\Cref{sec:framework:step3}}})} 
\sgii{executes the \uprog{} to perform the operation.
\omiii{Specifically, when a user program encounters a \emph{bbop} instruction (\cmr{\Cref{sec:bbop}}) associated with a SIMDRAM operation, the \emph{bbop} instruction triggers the execution of the SIMDRAM operation by performing its \uprog{} in the memory controller.} }
SIMDRAM uses a \emph{control unit} in the memory controller that transparently \sgii{issues} the sequence of \geraldorevii{\aaps}\revdel{AAPs/TRAs} 
\sgii{to DRAM, as dictated by the \uprog{}}.}
\sgii{Once the \uprog{} is complete, the result of the operation is held in DRAM.}

\subsection{Integrating SIMDRAM in a System}
\label{sec:integ-overview}

\sg{As we discuss earlier, SIMDRAM operates on data using a vertical layout. \geraldorevii{\cmr{Figure}~\ref{fig:datalayout} illustrates how data is organized within a DRAM subarray when employing a horizontal data layout (\cmr{Figure}~\ref{fig:datalayout}a) and a vertical data layout (\cmr{Figure}~\ref{fig:datalayout}b). We assume that each data element is \sgii{four bits} wide, and \sgii{that} there are four data elements (each one represented by a different color). In a conventional horizontal data layout, data elements are stored in different DRAM rows, \sgii{with the contents of each data element ordered} from the most significant bit to the least significant bit (or vice versa) \sgii{in a single row}. In contrast, in a vertical data layout, \sgii{the} DRAM row holds \sgii{only} the $i$-th bit of \sgii{\emph{multiple}} data elements \sgii{(where the number of elements is determined by the bit width of the row)}. Therefore, when activating a single DRAM row in a vertical data layout organization, a \emph{single} bit of \sgii{data from each} data \sgii{element is} read at once, which enables \cmr{in\omvuii{-}DRAM} bit-serial parallel computation~\omiv{\cite{batcher1982bit,shooman1960parallel,seshadri2019dram,li2017drisa,ali2019memory,gu2016leveraging}}.} 

\begin{figure}[ht]
    \centering
    \includegraphics[width=0.8\linewidth]{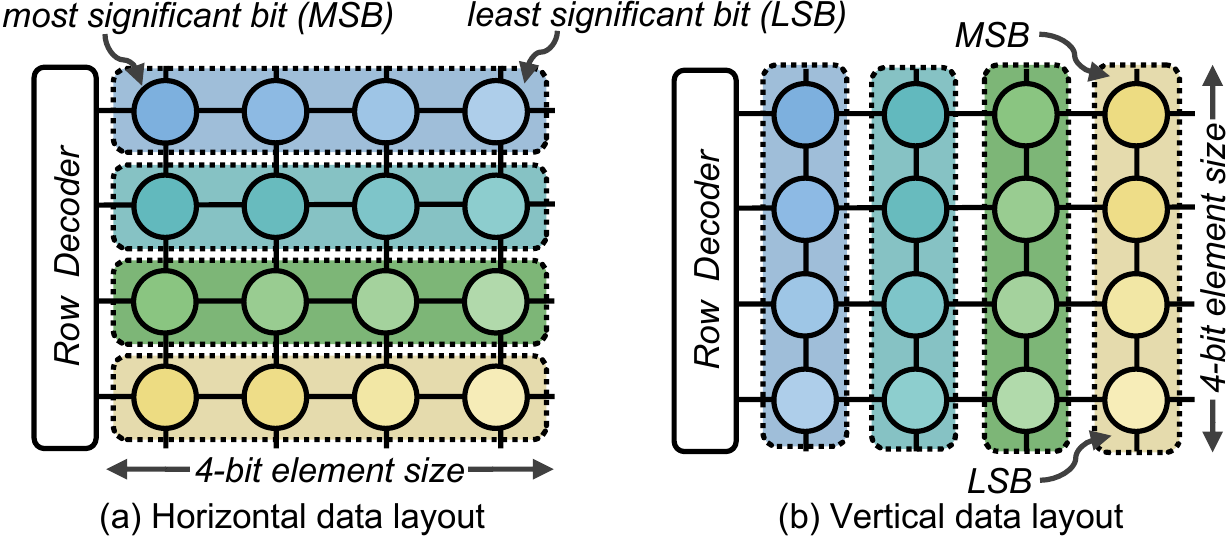}
    \caption{\geraldorevi{Data layout\omi{: horizontal vs.\ vertical}}.}
    \label{fig:datalayout}
\end{figure}

To maintain compatibility with traditional system software, we store regular data in the conventional horizontal layout and provide hardware support (explained in \cmr{\Cref{sec:transposing}}) to transpose horizontally\omi{-}laid\omi{-}out data into \omi{the} vertical layout for in-DRAM computation. To simplify program integration, we provide ISA extensions that
expose SIMDRAM operations to the programmer (\cmr{\Cref{sec:bbop}}).}


\section{SIMDRAM Framework}
\label{sec:operations}


\cmr{\omi{\omvuii{\omiii{We describe}
the three steps of the \mech framework introduced in \cmr{\Cref{overview}}, using the full addition operation as a running example.}}}

\subsection{Step 1: Efficient MAJ/NOT Implementation}
\label{sec:aoi-mi}
\label{sec:framework:step1}

\nasrev{SIMDRAM implements in-DRAM computation using the logically\omi{-}complete set of MAJ and NOT logic primitives, which requires fewer \aap \omiii{command sequences} to perform a given operation when compared to using AND/OR/NOT. As a result, the goal of the first step in the \mech framework is to build an optimized MAJ/NOT implementation of a given operation that executes the operation using as few \aap \omiii{command sequences} as possible, thus \omi{minimizing the operation's latency}. To this end, Step 1 transforms an AND/OR/NOT representation of a given operation to an optimized MAJ/NOT representation using \sgii{a transformation process} formalized by prior work~\cite{epflmaj}.}

\nasrev{\sgii{The transformation process uses a graph-based representation of the logic \sgii{primitives}, called \sgii{an} \textit{AND--OR--Inverter Graph\revdel{s}} (AOIG) \sgii{for AND/OR/NOT logic, and a} \textit{Majority--Inverter Graph\revdel{s}} (MIG) \sgii{for MAJ/NOT logic}. An AOIG is a logic representation structure in the form of a directed acyclic graph \omi{where} each node represents an AND or OR \sgii{logic primitive}. Each edge in an AOIG represents an input/output dependency between nodes. The incoming edges to a node represent input operands of the node and the outgoing edge of a node represent\omi{s} the output of the node. The edges in \sgii{an} AOIG can be either regular or complemented (\sgii{which represents \omi{an inverted} input operand; \omi{\omiii{denoted by}}} \omiii{a bubble on the} edge).
The direction of the edges follows the natural direction of computation from inputs to outputs. 
Similarly, a MIG is a directed acyclic graph in which each node represents a three-input MAJ \sgii{logic primitive}, and each regular/complemented edge represents one input or output to the MAJ \sgii{primitive} that the node represents.
The transformation process consists of two parts that operate on an input AOIG.}}

\sgii{The first part of the transformation process naively substitutes AND/OR primitives with MAJ primitives.
Each two-input AND or OR primitive is simply replaced with a three-input MAJ primitive, where one of the inputs is tied to logic 0 or logic 1, respectively.
This naive substitution yields \omi{a} MIG that \emph{correctly} replicates the functionality of the input AOIG, but the MIG is \omi{likely} \emph{inefficient}.}

\begin{figure}[ht]
    \centering
    \includegraphics[width=1\linewidth]{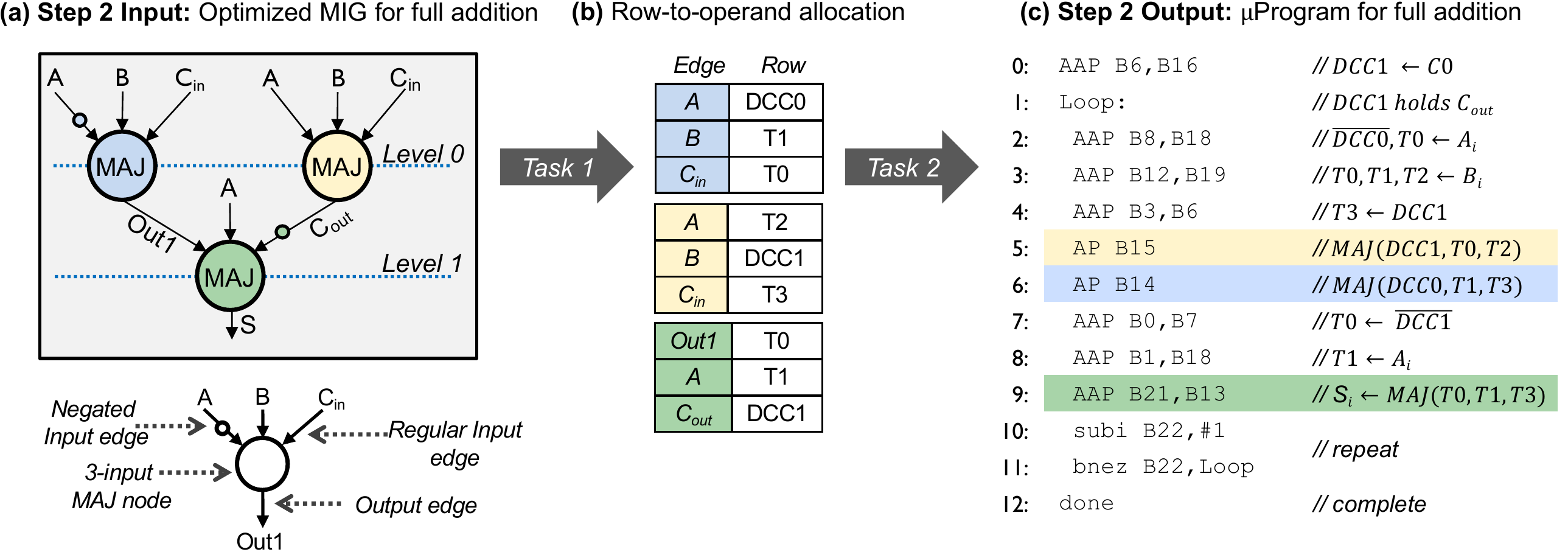}
    \caption{\sgii{(a)~\omiv{Optimized \cmr{MIG}}\cmr{;} (b)~\omi{row-to-operand allocation}\cmr{;} (c)~\uprog{}} for full \omi{addition.}}
    \label{fig_output_mapping}
\end{figure}

\sgii{The second part of the transformation process takes the inefficient MIG and uses a greedy algorithm 
\omi{(\omdef{see Appendix~\ref{apdx:aoi-to-mig})}} to apply a series of transformations that identifies how to consolidate multiple MAJ primitives into a smaller number of MAJ primitives with identical functionality. This yields a smaller MIG, which in turn requires fewer logic primitives to perform the same operation that the unoptimized MIG (and, thus, the input AOIG) perform\omi{s}. \cmr{Figure}~\ref{fig_output_mapping}a shows the optimized MIG produced by the transformation process for a full addition operation.}

\subsection{Step 2: \sgii{\uprog{} Generation}}
\label{sec:mi-aap}
\label{sec:framework:step2}


\nasrev{\sgii{Each \mech operation is stored as a \emph{\uprog{}}, which consists of a series of microarchitectural operations (\uop{}s) that \mech uses to execute the \mech operation in DRAM.
The goal of the second step is to take the optimized MIG generated in Step~1 and generate a \uprog{} that executes the 
\omi{\mech operation that the MIG represents}.
To this end, \omi{as shown in \cmr{Figure}~\ref{fig_output_mapping}}, the second step of the framework performs two key tasks on the optimized MIG:}
(1)~\emph{\omi{allocating DRAM rows \omiii{to} the operands}}, \sgii{which assigns each input operand (i.e., an incoming edge) of each MAJ node in the MIG to a DRAM row \omi{(\cmr{Figure}~\ref{fig_output_mapping}b)};} and (2)~\emph{generating the \uprog{}}, \sgii{which creates the \omiii{series} of \uop{}s that perform the MAJ and NOT logic primitives (i.e., nodes) in the MIG, while maintaining} the correct flow of the computation \omi{(\cmr{Figure}~\ref{fig_output_mapping}c)}. In this section, we first describe the \uop{}s used in \mech (\cmr{\Cref{sec:framework:step2:uops}}). \sgii{Second}, we explain the process of \omi{allocating DRAM rows \omiii{to}} the input operands of the MAJ nodes in the MIG to DRAM rows (\cmr{\Cref{sec:framework:step2:mapping}}).  \sgii{Third, we explain the process of generating} the \uprog{} (\cmr{\Cref{sec:framework:step2:generating}}).}

\subsubsection{\textbf{\mech \uop{}s.}}
\label{sec:framework:step2:uops}

\begin{figure}[!ht]
    \centering
    \includegraphics[width=0.9\linewidth]{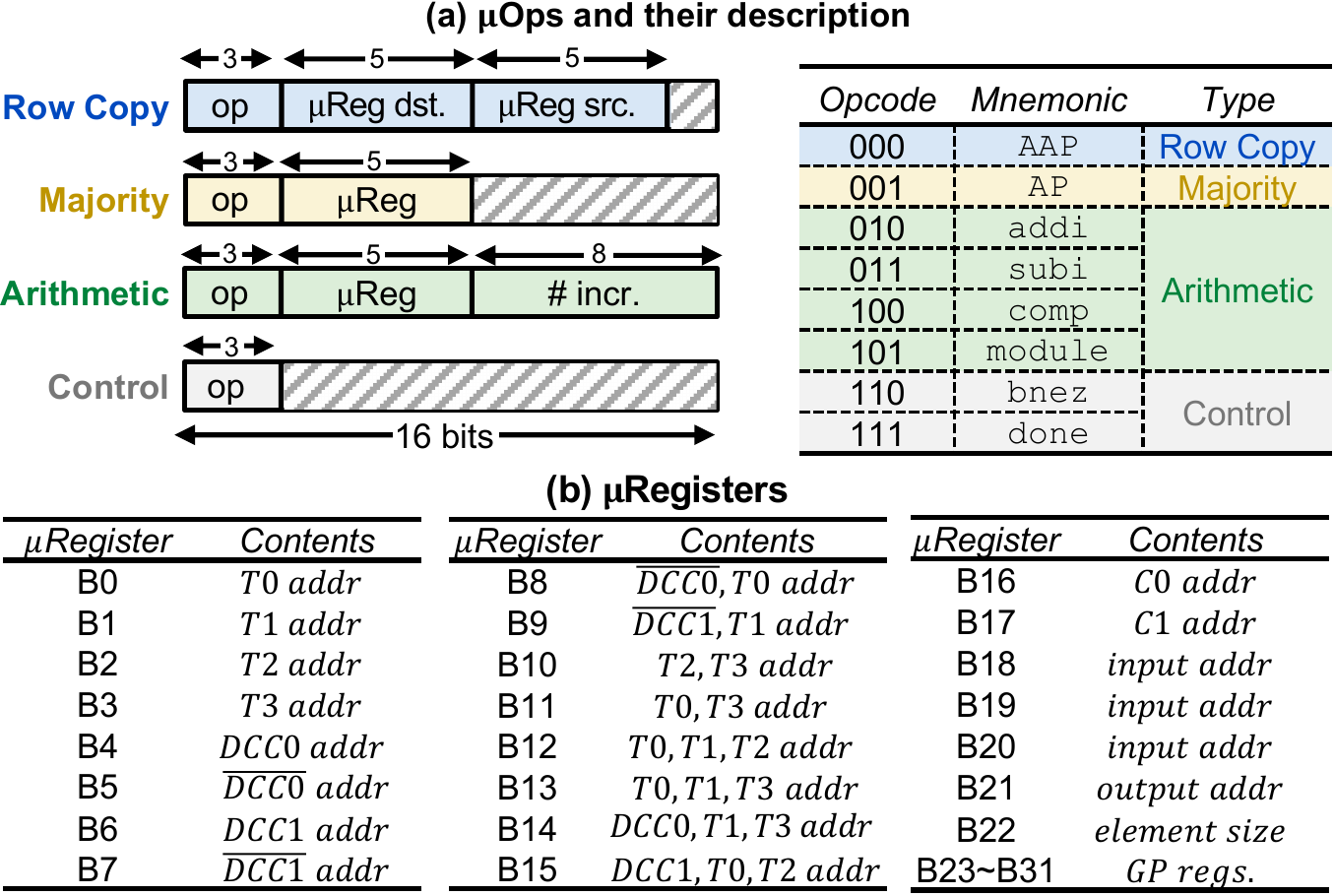}%
    \caption{\sgii{\uop{}s and \ureg{}s in \mech.} 
    \jgll{Keep this in mind when improving the figure: A MAJ can be executed with an AP or with an AAP -- AP if the destination row is among the source rows. AAP if not.}}
    \label{fig_opcodes}
\end{figure}

\nasrev{~\cmr{Figure}~\ref{fig_opcodes}a shows the set of \uop{}s \sgii{that we implement in \mech. Each \uop{} is either
(1)~a \emph{command sequence} that is issued by \mech to a subarray to perform a portion of the in-DRAM computation, or
(2)~a \emph{control} operation that is used by the \mech control unit (see \cmr{\Cref{sec:framework:step3}}) to manage the execution of the \mech operation.}
We further break down the command sequence \uop{}s into one of three types:
(1)~\textit{row copy}, \sgii{a \uop{} that} performs in-DRAM copy from a source memory address to a destination memory address using an \texttt{AAP} \omi{command sequence};
(2)~\textit{majority}, \sgii{a \uop{} that performs a majority logic primitive on three DRAM rows using an \texttt{AP} \omi{command sequence} (i.e., it performs a TRA); and} 
(3)~\textit{arithmetic}, \sgii{four \uop{}s that} \omi{perform simple arithmetic operations \omi{on \mech control unit registers} required to control the execution of the operation (\texttt{addi}, \texttt{subi}, \texttt{comp}, \texttt{module})}.}
\omi{We provide two control operation \uop{}s to support loops and termination in the \mech control flow (\texttt{bnez}, \texttt{done}).}

\nasrev{\sgii{During \uprog{} generation, the \mech framework converts the MIG into a series of \uop{}s.  \cmr{Note that MIG represents a 1-bit\omi{-wide} computation of an operation.}} \sgii{\cmr{Thus, to} implement a multi-bit\omi{-wide} \mech operation, the framework} needs to repeat the \omiii{series} of the \uop{}s that implement the MIG $n$ times, 
where $n$ is the number of bits in the operands of the \sgii{\mech} operation. To this end, SIMDRAM uses the arithmetic and control \uop{}s to 
repeat the 1-bit\omi{-wide} computation $n$ times\omi{, transparently to the programmer}}.

\nasrev{To support the execution of \uop{}s, SIMDRAM utilizes a set of \sgii{\emph{\ureg{}s}} (\cmr{Figure}~\ref{fig_opcodes}b) located in the \omi{\mech} control unit \omi{(\cmr{\Cref{sec:mc}})}. \sgii{The framework uses \ureg{}s}
(1)~to store the memory addresses of DRAM rows in the B-group and C-group (\cmr{Figure}~\ref{subarray}) of the subarray (\sgii{\ureg{}s B0--B17}),
(2)~to store the memory addresses of input and output rows for the computation (\sgii{\ureg{}s B18--B22}), and
(3)~as general-purpose registers during the execution of arithmetic and control operations (\sgii{\ureg{}s B23--B31}).
\jgll{
We also need to justify why 32 registers and why 32-bit.}}

\subsubsection{Task 1: \omi{Allocating DRAM Rows \omiii{to} the Operands}.}
\label{sec:framework:step2:mapping}

\nasrev{~The goal of this task is to \omi{allocate DRAM rows \omiii{to}} the  
input operands (i.e., incoming edges) of each MAJ node in the operation's MIG, \sgii{such that we} minimize the total number of \uop{}s \cmr{needed} to compute the operation. To this end, we present a new \omi{allocation} algorithm inspired by the linear scan register allocation algorithm~\cite{poletto1999}. However, unlike register allocation algorithms, our \omi{allocation algorithm} considers two extra constraints that are specific to processing-using-DRAM:
(1)~performing MAJ in DRAM has destructive behavior, i.e., a TRA overwrites the original values of the three input rows with the MAJ output; and 
(2)~the number of compute rows (i.e., B-group in \cmr{Figure}~\ref{fig_subarray}) that are designated to perform bitwise operations is limited 
(\sgii{\omi{there are only} six compute rows in \omi{each} subarray, as discussed in \cmr{\Cref{subarray}}}).}

\nasrev{\cutdel{Algorithm~\ref{alg_mapping} describes the \emph{mapping algorithm} used in \mech to map the input operands of each MAJ node in the operation's MIG to DRAM rows. The algorithm takes the optimized MIG of the operation as input, and generates an optimized mapping 
of the input operands of each MAJ node in the MIG to DRAM rows as the output. The \mech \emph{mapping algorithm} assumes that (1)~C0 and C1 rows in the C-group in the subarray (\cmr{Figure}~\ref{fig_subarray}) contain all-0 and all-1 values, respectively; (2)~the input operands of the SIMDRAM operation are already stored in separate rows of the D-group in the subarray using vertical layout \sgii{(\cmr{\Cref{sec:integ-overview}})}, before the computation of the operation starts; and (3)~the final output of the MIG is also stored in a D-group row after the computation of the entire MIG ends.}} 

\omi{The \mech \emph{\omdefi{row-to-operand} allocation algorithm} receives the operation's MIG as input. The algorithm assumes that the input operands of the operation are already stored in separate rows of the D-group in the subarray using vertical layout \sgii{(\cmr{\Cref{sec:integ-overview}})}, before the computation of the operation starts. The algorithm then does a topological traversal starting with the leftmost MAJ node \omiii{at} the highest level of the MIG (e.g., level 0 in \cmr{Figure}~\ref{fig_output_mapping}a), 
\omi{allocating compute rows \omiii{to}} the input operands of each MAJ node in the current level of the MIG, before moving to the next lower level of the graph. The algorithm finishes once \omi{DRAM rows are allocated \omiii{to}} all the input operands of all the MAJ nodes in the MIG. \cmr{Figure}~\ref{fig_output_mapping}b shows these \omi{allocations} as the output of Task 1 for the full addition example. \sgii{The resulting} \omi{row-to-operand allocation} is then used in the second task in step two (\cmr{\Cref{sec:framework:step2:generating}}) to generate the \omiii{series} of \uop{}s to compute the operation that the MIG represents. \omdef{We describe our \omdefi{row-to-operand} \omi{allocation} algorithm in Appendix~\ref{apdx:row-to-op}}. 
}




\subsubsection{Task 2: Generating \omi{a \uprog{}}.}
\label{sec:framework:step2:generating}
~\geraldorevii{\sgii{The goal of this task is} to \sgii{use the MIG and the DRAM row \omi{allocations} from Task~1 to generate the \uop{}s of the \uprog{} for our \mech operation.}
\sgii{To this end, Task~2}
(1)~translates the MIG into \sgii{a \omiii{series} of} row copy and majority \omiii{\uop{}s} \omi{(i.e., \aaps)}, 
(2)~optimizes \sgii{the \omiii{series of \uop{}s to reduce the number of \aaps}}, and 
(3)~\sgii{generalizes} the \sgii{one-bit bit-serial operation described by the MIG into an} $n$-bit operation by utilizing SIMDRAM's arithmetic and control \uop{}s.}

\omi{\textbf{\emph{Translating the MIG into \omiii{a Series of Row Copy and Majority \uop{}s.}}}} The \omi{allocation} produced \omi{during Task~1}
dictates how \omi{DRAM rows are allocated \omiii{to}} each edge in the MIG \sgii{during the \uprog{}}. With this information, \sgii{the framework} can generate the appropriate \omiii{series} of row copies and majority \omiii{\uop{}s} to reflect the MIG's computation in DRAM. To do so, we \sgii{traverse} the input MIG in topological order. For \emph{each} node, we \emph{first} assign row copy \omiii{\uop{}s} (using the \texttt{AAP} \omiii{command sequence}) to the node's edges. Then, we assign a majority \omiii{\uop{}} (using the \texttt{AP} \omiii{command sequence}) to execute the current MAJ node, following the DRAM row \omi{allocation} assigned to each edge of the node.  The \sgii{framework repeats this procedure} for all the nodes in the MIG. To illustrate, we assume that 
\omi{the SIMDRAM \omi{allocation} algorithm}
\omi{allocates DRAM rows DCC0, T1, and T0 \omiii{to} edges A, B, and C$_{in}$, respectively, \omi{of the blue node} in the full \omi{addition} MIG (\omi{\cmr{Figure}~\ref{fig_output_mapping}a}).} 
Then, when \omviii{visiting} this node, we generate the following \omiii{series} of \uop{}s:

\begin{center}
\tempcommand{.8}
\begin{tabular}{ll}
\texttt{AAP DCC0, A};      & // DCC0 $\leftarrow$ A      \\
\texttt{AAP T1, B};        & // T1 $\leftarrow$ B        \\
\texttt{AAP T0, C$_{in}$}; & // T0 $\leftarrow$ C$_{in}$ \\
\texttt{AP $\overline{\mbox{DCC0}}$, T1, T0}   & // MAJ(NOT(A), B, C$_{in}$)
\end{tabular}
\end{center}

\textbf{\emph{\omiii{Optimizing the Series of \uop{}s.}}} 
\geraldorevii{After traversing all of the nodes in the MIG and generating the appropriate \omiii{series} of \uop{}s, we optimize the \omiii{series} of \uop{}s by \sgii{coalescing \aap \omiii{command sequences}, which we can do in one of two cases}. 

\noindent\sgii{\textbf{Case 1:}} we can coalesce a \omiii{series} of row copy \sgii{\omiii{\uop{}s} if all of the \omiii{\uop{}s} have the same \ureg{} source as an input}. For example, consider a \omiii{series} of two \omiii{\texttt{AAP}s} that copy data array $A$ into rows T2 and T3. We can coalesce this \omiii{series} of \omiii{\texttt{AAP}s} into a single \texttt{AAP} issued to the wordline address stored in \ureg{}~B10 (see \cmr{Figure}~\ref{fig_opcodes}a). This wordline address leverages the special row decoder in the B-group \omi{(which is part of the Ambit subarray structure~\cite{seshadri2017ambit})} to activate \sgii{multiple DRAM rows in the group} \omvuii{at once} with a single activation command.
\sgii{For our example, activating \ureg{}~B10 allows the \texttt{AAP} command \omiii{sequence} to copy array $A$ into both rows T2 and T3 at once.}

\noindent\sgii{\textbf{Case 2:}} we can coalesce an \texttt{AP} command \omiii{sequence} (i.e., a majority \omiii{\uop{}}) followed by an \texttt{AAP} \sgii{command sequence (i.e., \omiii{a} row copy \omiii{\uop{}}) when the destination of the \texttt{AAP} is one of the rows used by the \texttt{AP}.
For example, consider an \texttt{AP} that performs a MAJ logic primitive on DRAM rows T0, T1, and T2 (storing the result in all three rows), followed by an \texttt{AAP} that copies \ureg{}~B12 (which refers to rows T0, T1, and T2) to row~T3.  \omiii{The \texttt{AP} followed by the \texttt{AAP}} puts the majority value in all four rows (T0, T1, T2, T3).
The two command \omiii{sequences} can be coalesced into a single \texttt{AAP} (\texttt{AAP} T3, B12), as the first ACTIVATE would automatically perform the majority on rows T0, T1, and T2 by activating all three rows simultaneously. The second ACTIVATE then copies the value from those rows into T3.}}

\omi{\textbf{\emph{Generalizing the Bit-Serial Operation into an $n$-bit Operation\omiii{.}}}} 
\sgii{Once all potential \uop{} coalescing is complete, the framework now has an optimized 1-bit version of the computation.}
We generalize \sgii{this 1-bit \uop{} \omiii{series}}
into a loop body that repeats $n$ times to implement an $n$-bit operation. We leverage the arithmetic and control \uop{}s available in SIMDRAM to orchestrate the $n$-bit computation. \omi{Data produced by the computation of one bit that \omviii{needs} to be used for computation of the next bit (e.g., the carry bit in full addition) is kept in a B-group row across the two computations, allowing for bit-to-bit data transfer without the need for dedicated shifting circuitry.}

\geraldorevii{The final \omiii{series} of  \uop{}s produced after this step is then packed into a  \uprog{} and stored in DRAM for future use.\footnote{\label{footnote:upgram}\sgii{In our example implementation of \mech, a \uprog{} has a maximum size of 128~bytes, as this is enough to store the largest \uprog{} generated in our evaluations (the division operation, which requires 56 \uop{}s, each two bytes wide, resulting in a total \uprog{} size of 112~bytes.)}} \cmr{Figure}~\ref{fig_output_mapping}c shows the final \uprog{} produced at the end of Step 2 for the full \omi{addition} operation. The figure shows the optimized \omiii{series} of \uop{}s that generates the 1-bit implementation of the full \omi{addition} (lines 2--9), and the arithmetic and control \uop{}s included to enable the $n$-bit implementation of the operation (lines 10--11).}


\omi{\textbf{\emph{Benefits of the \uprog{} Abstraction\omiii{.}}}}
The \uprog{} abstraction \sgii{that we use to store SIMDRAM operations} provides three main advantages to the framework. First, it allows SIMDRAM to minimize the total number of new \omi{CPU} \omiii{instructions} required to implement SIMDRAM operations, \revonur{thereby} reducing SIMDRAM's impact on the ISA. While a different implementation could use more \revonur{new} \omi{CPU} \omiii{instructions} to express finer-grained operations (e.g., an \texttt{AAP}), we believe that using a minimal set of \omiii{new \omiv{CPU} instructions} simplifies adoption and software design. Second, the \uprog{} \revonur{abstraction} enables a \revonur{smaller} application binary size since the \color{black}only information that needs to be placed in the application's binary is the address of the \uprog{} \nasirevi{in main memory}. Third, \revonur{the \uprog{}} provides an abstraction to relieve the end user \revonur{from} low-level programming 
with MAJ/NOT operations \nasirevi{that} is equivalent to programming with Boolean logic. \omi{We discuss how a user program invokes \mech \uprog{}s in \cmr{\Cref{sec:bbop}}.}

\subsection{\nasii{Step 3: \sgii{Operation} Execution}}
\label{sec:mc}
\label{sec:framework:step3}

\sgii{Once the framework stores the generated \uprog{} for a \mech operation in DRAM, \omi{the} SIMDRAM \omi{hardware} can now receive program requests to execute the operation.  To this end, we discuss the \mech \emph{control unit}, which handles the execution of the \uprog{} at runtime.}
\nasii{The control unit is designed} as an extension of the memory controller, \sgii{and is transparent to the programmer}. 
\sgii{A program issues a request to perform a \mech operation using a \emph{bbop} instruction \omi{(introduced by Ambit~\cite{seshadri2017ambit})}, which is one of \omi{the} \omiii{CPU} ISA extensions to allow programs to interact with the \mech framework (see \cmr{\Cref{sec:bbop}}). Each \mech operation corresponds to a different \emph{bbop} instruction. Upon receiving the request,}
the control unit loads the \uprog{} corresponding to the \sgii{requested} \emph{bbop} \sgii{from memory,} and \sgii{performs the \uop{}s in the \uprog{}}. 
\omi{Since all input data elements of a \mech operation may not fit in one DRAM row, the control unit repeats the \uprog{} $i$ times, where $i$ is the total number of data elements divided by the number of elements in a single DRAM row.}

\begin{figure}[ht]
    \centering
    \includegraphics[width=\linewidth]{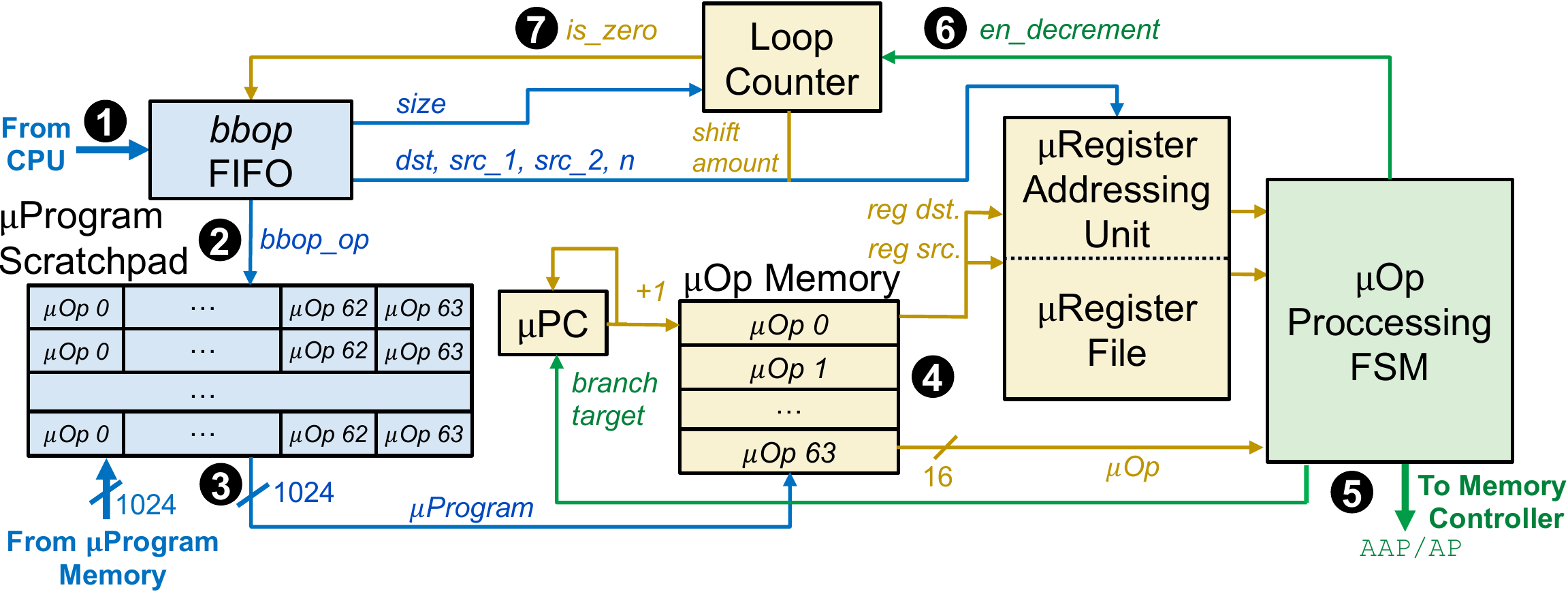}%
    \caption{\geraldo{\mech control unit.}}
    \label{fig_control}
\end{figure}

\cmr{Figure}~\ref{fig_control} shows a block diagram of the \mech control unit\geraldorevii{, which} consists of \nasrev{nine} main components:
(1)~a \emph{bbop FIFO} that receives the \emph{bbop}s from the program,
(2)~a \emph{\uprog{} Memory} allocated in DRAM (not shown in the figure), 
(3)~\geraldo{a \emph{\uprog{} Scratchpad} \sgii{that holds commonly-used \uprog{}s}},
(4)~a \emph{\uop{} Memory} that holds the \uop{}s of the currently running \geraldorevii{\uprog{}},
\sgii{(5)~a \emph{\ureg{} Addressing Unit} that generates the physical row addresses being used by the \ureg{}s that map to DRAM rows (based on the \ureg{}-to-row assignments for B0--B17 in \cmr{Figure}~\ref{fig_opcodes}),
(6)~a \emph{\ureg{} File} that holds the non-row-mapped \ureg{}s (B18--B31 in \cmr{Figure}~\ref{fig_opcodes}),
(7)~a \emph{Loop Counter} that tracks the number of remaining \omi{data elements that the \uprog{} needs to be performed on,}}
(8)~\omi{a \uop{} Processing FSM}  that controls the execution flow and issues \geraldorevii{\aap} \sgii{command sequences},
\sgii{and (9)~a \omi{\uprogc{}~(\upc{})}}. 
\geraldo{\mech reserves a region of DRAM for the \uprog{} Memory to store \uprog{}s corresponding to \emph{all} \mech operations. \geraldorevii{At runtime,} the control unit 
\cmr{stores the most commonly used \uprog{}s in the \uprog{} Scratchpad, to reduce} 
the overhead of fetching \uprog{}s from DRAM.
}

\sgii{At runtime, when a CPU running a user program reaches a \emph{bbop} instruction, it forwards the \emph{bbop}}
to the \mech control unit (\circled{1} \geraldorevii{in \cmr{Figure}~\ref{fig_control}}). 
The \sgii{control unit enqueues the \emph{bbop}} in the \emph{bbop} FIFO.
\sgii{The control unit goes through a four-stage procedure to execute the queued \emph{bbop}s one at a time.}

\sgii{In the first stage,} 
the control unit fetches and decodes the \emph{bbop} at the \sgii{head} of the FIFO (\circled{2}).
Decoding a \emph{bbop} involves
(1)~\sgii{setting the index of the \uprog{} Scratchpad to the \emph{bbop} opcode};
(2)~writing the number of \sgii{loop iterations required to perform the operation on all elements (i.e., the number of data elements divided by the number of elements in a single DRAM row) into the Loop Counter;} and
(3)~writing the base DRAM addresses of the source and destination arrays involved in the computation, and \sgii{the \omiii{size} of each data element, to the \ureg{} Addressing Unit}.

\sgii{In the second stage,}
the control unit \sgii{copies} the 
\uprog{} \sgii{currently indexed in the \uprog{} Scratchpad} to the \uop{} Memory (\circled{3}).
\geraldorevii{At this point}, the control unit is ready to start executing the \uprog{}, one \uop{} \sgii{at a} time. 

\sgii{In the third stage,}
\sgii{the current \uop{} is fetched} from the \uop{} Memory, \sgii{which is} indexed by the \omi{\upc{}}. The \omi{\uop{} Processing FSM} decodes the \uop{},
\sgii{and determines which \ureg{}s are needed (\ding{185}). For \ureg{}s B0--B17, the \omviii{\ureg{} Addressing Unit} generates the DRAM addresses that correspond to the requested registers (see \cmr{Figure}~\ref{fig_opcodes}) and sends the addresses to the \omi{\uop{} Processing FSM}. For \ureg{}s B18--B31, the \omviii{\ureg{} File} provides the register values to the \omi{\uop{} Processing FSM}.}

\sgii{In the fourth stage,}
\sgii{the \omi{\uop{} Processing FSM} executes the \uop{}. If the \uop{} is a command sequence, the corresponding commands are sent to the memory controller's request queue (\circled{5}) and the \omi{\upc{}} is incremented.  If the \uop{} is a \texttt{done} control operation, this indicates that all of the command sequence \uop{}s have been performed for the current iteration.
The \omi{\uop{} Processing FSM} then decrements the Loop Counter (\circled{6}).
If the \omi{decremented} Loop Counter is greater than zero, the \omi{\uop{} Processing FSM}
shifts the base source and destination addresses stored in the \ureg{} Addressing Unit to move onto the next set of data elements,\footnote{The \omi{source and destination} base addresses are \omi{incremented} by $n$ rows, where $n$ is the data element \omiii{size}. \omi{This is because each DRAM row contains one bit of a set of elements, so \mech uses $n$ consecutive rows to hold all $n$~bits of the set of elements.}} and resets the \omi{\upc{}} to the first \uop{} in the \uop{} Memory.}
\sgii{If the \omi{decremented} Loop Counter equals zero, this indicates that the control unit has completed executing the current \emph{bbop}.  The} control unit \sgii{then} fetches the next \emph{bbop} from the \emph{bbop} FIFO (\circled{7}), 
\sgii{and repeats all four stages for the next \emph{bbop}}.

\subsection{Supported Operations}
\label{sec:supported:ops}
\revGeraldo{We use our 
\nasii{framework} to} efficiently support a wide 
\juang{range} of operations of different types. \omi{In this work, we evaluate (in \cmr{\Cref{sec_evaluation}}) a set of 16 \mech operations of five different types for \sgii{$n$-bit data elements}:} (1)~\sgii{$N$}-input logic operations (OR-/AND-/XOR-reduction \sgii{across $N$ inputs}); (2)~relational operations (equality/inequality \omi{check}, greater\omi{-/less-than check, greater-than-or-equal-to check}, \omi{and maximum/minimum element in a set}); (3)~arithmetic operations (addition, subtraction, multiplication, division\omi{, and absolute value}); (4)~predication (if-then-else); and (5)~other complex operations (bitcount, and Re\geraldo{LU}). \omi{We support four different \omiv{element sizes} \omiii{that} correspond to \omiii{data type sizes} in popular programming languages (8-bit, 16-bit, 32-bit, 64-bit). }

\section{System Integration of SIMDRAM}
\label{implementation}

\nasrev{
\omi{We} discuss \sgii{several} challenges of integrating \mech in a real system, and how we address them: 
(1)~data \sgii{layout} and how \mech manages storing the data required for in-DRAM computation in a vertical layout (\Cref{sec:transposing}); 
(2)~ISA extensions \omiii{for} and programm\omiii{ing interface of SIMDRAM} (\Cref{sec:bbop}); 
(3)~how \mech handles 
page faults, address translation, \omi{coherence,} and interrupts (\Cref{sec:pagefaults});
(4)~how \mech manages computation \omiii{on} large amounts of data (\Cref{sec:limited});
\omi{(5)~security implications of \mech
(\Cref{sec:security})};
and (6)~current limitations of the \mech framework (\Cref{sec:limitations}).}

\subsection{Data Layout}
\label{sec:transposing}

\nasrev{We envision \mech as \emph{supplementing} (not \emph{replacing}) the traditional processing elements. As a result, a program in a \mech-enabled system can have a combination of CPU instructions and \mech instructions, with possible data sharing between the two. However, while \mech operates on vertically-laid-out data \omi{(\cmr{\Cref{sec:integ-overview}})}, the other system components (including the CPU) expect the data to be laid out in the traditional horizontal format, making it challenging to share data between \mech and CPU instructions. To address this challenge, memory management in \mech needs to (1)~support both horizontal and vertical data layouts in DRAM \emph{simultaneously}; and (2)~\omi{transform} vertically-laid-out data used by \mech to 
\omi{a} horizontal layout for CPU use, and vice versa. 
\sgii{We cannot rely on software (e.g., compiler or application support) to handle the data layout transformation, as this would go through the on-chip memory controller, and would introduce significant data movement, \omi{and thus latency}, between the DRAM and CPU during the transformation.
To avoid data movement during transformation,}
\mech uses a specialized hardware unit placed between the last-level cache (LLC) and the memory controller, called the \emph{\omi{data} transposition unit}, to transform 
data from horizontal data layout to vertical data layout, and vice versa.}
\omi{The transposition unit ensures that for every \mech object, its corresponding data is in a horizontal layout whenever the data is in the cache, and in a vertical layout whenever the data is in DRAM.}

\begin{figure}[ht]
    \centering
    \includegraphics[width=0.85\linewidth]{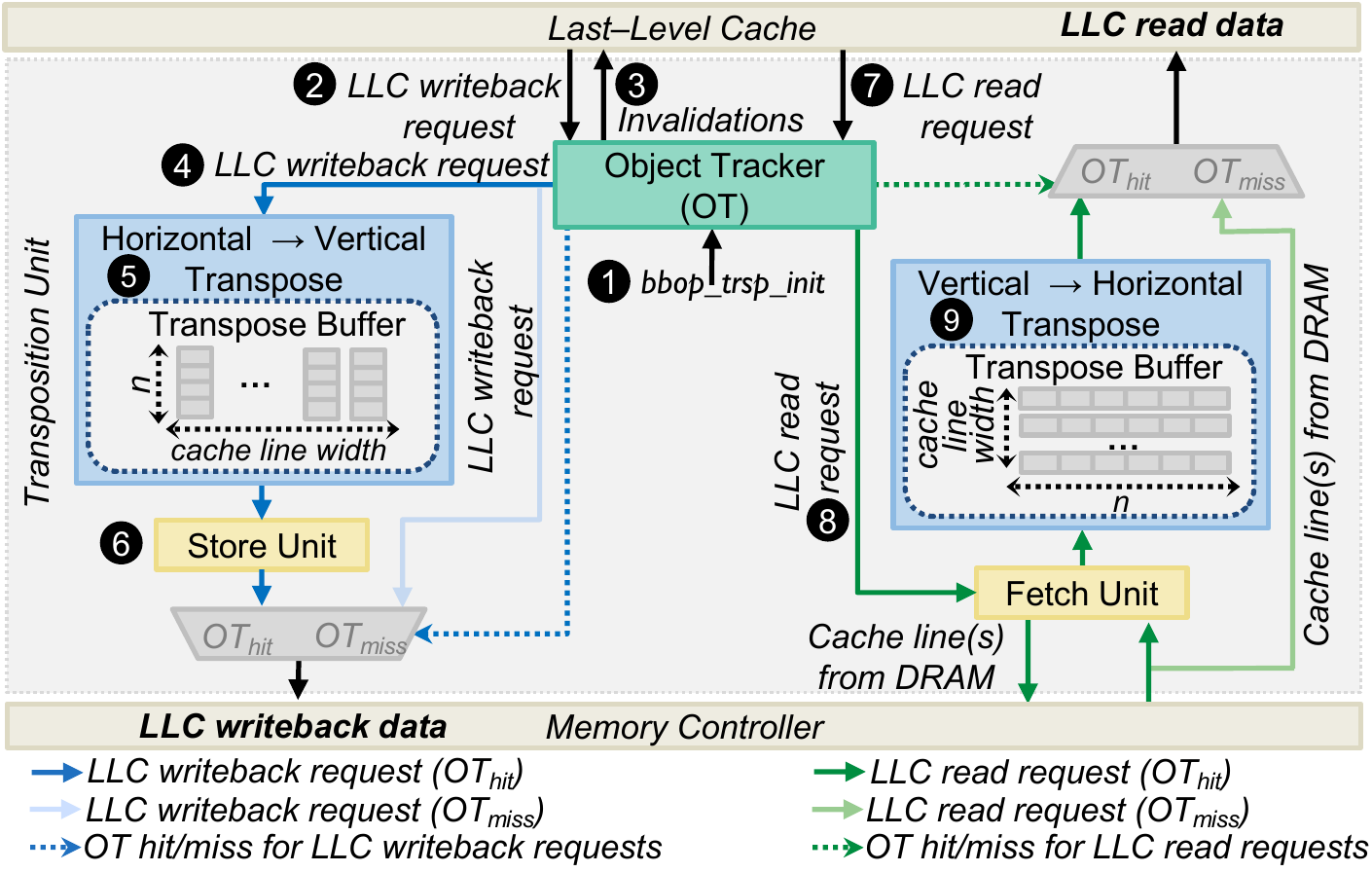}
    \caption{Major components of the \omi{data} transposition unit. 
    }
    \label{fig_transposing}
\end{figure}

\sgii{\cmr{Figure}~\ref{fig_transposing} shows the key components of the transposition unit.} \nasrev{\sgii{The transposition unit keeps track of the memory objects that are used by \mech operations in a small cache in the transposition unit, called the \emph{Object Tracker}. To add an entry to the Object Tracker} when allocating a memory object used by \mech, the programmer adds an \omi{initialization} instruction called \texttt{bbop\_trsp\_init} (\cmr{\Cref{sec:bbop}}) \sgii{\emph{immediately}} after the \texttt{malloc} that allocates the memory object \omiv{(\ding{182} in \cmr{Figure}~\ref{fig_transposing})}.
\sgii{Assuming a system that employs lazy allocation, the}
\texttt{bbop\_trsp\_init} instruction informs the operating system (OS) that the memory object is a \emph{SIMDRAM object}. 
This allows the OS to perform virtual-to-physical memory mapping optimizations \sgii{for the object before the allocation starts}
(e.g., mapping the arguments of an operation to the same row/column in the physical memory). 
\sgii{When the \mech object's physical memory is allocated, the OS inserts} the base physical address, \sgii{the total size of the allocated data, and the \omiii{size of each element} in the object \omi{(provided by \texttt{bbop\_trsp\_init})} into the Object Tracker}.} 
\omi{As the initially-allocated data is placed in the CPU cache, the data starts in a horizontal layout until it is evicted from the cache.}

\omi{\mech stores \mech objects in DRAM using a vertical layout, since this is the layout used for in-DRAM computation (\cmr{\Cref{sec:integ-overview}}). 
Since a vertically-laid-out $n$-bit element spans $n$ different cache lines in DRAM (with each cache line in a different DRAM row), \mech partitions \mech objects into \emph{\mech object slices}\omiii{, each of which is} $n$ cache lines in size. 
Thus, a \mech object slice in DRAM contains the vertically-laid-out bits of as many elements as bits in a cache line (e.g., 512 in a \SI{64}{\byte} cache line). 
Cache line $i$ ($0 \le i < n$) of an object slice contains bit~$i$ of \emph{all} elements stored in the slice. 
Whenever any one data element within a slice is requested by the CPU, the entire \mech object slice is brought into the LLC. 
Similarly, whenever a cache line from a \mech object is written back from the LLC to DRAM (i.e., it is evicted or flushed), \emph{all} $n-1$ remaining cache lines of the same \mech object slice are written back as well.\omiii{\footnote{\label{footnote:dbi}\omiii{The Dirty-Block Index~\cite{seshadri2014dirty} could be adapted for this purpose.}}} 
The use of object slices ensures correctness and simplifies the transposition unit.}

\sgii{\omi{Whenever the LLC writes back a cache line} to DRAM \omiii{(\circled{2} in \cmr{Figure}~\ref{fig_transposing})}, 
the transposition unit checks the Object Tracker to see whether the cache line belongs to a \mech object. If the LLC request misses in the Object Tracker, the cache line does not belong to any \mech object, and the writeback request is forwarded to the memory controller \omi{as in a conventional system}.
If the LLC request hits in the Object Tracker, the cache line belongs to a \mech object, and \omi{thus} must be transposed \cmr{from the horizontal layout to the vertical layout}. An Object Tracker hit triggers two actions.}

\omi{First, the Object Tracker issues invalidation requests to \emph{all} $n-1$ remaining cache lines of the same \mech object slice \omiii{(\circled{3} in \cmr{Figure}~\ref{fig_transposing})}.\textsuperscript{\ref{footnote:dbi}}
\omiii{We extend the LLC to support a special invalidation request type, which sends both dirty \emph{and} unmodified cache lines to the transposition unit (unlike a regular invalidation request, which simply invalidates unmodified cache lines).}
\omiii{The Object Tracker issues these invalidation requests for the remaining cache lines, ensuring that \emph{all}} cache lines of the object slice arrive at the transposition unit to perform the horizontal-to-vertical transpos\omiv{ition} correctly.}

\sgii{Second, the writeback request is forwarded \omiv{(\circled{4} in \cmr{Figure}~\ref{fig_transposing})} to a horizontal-to-vertical transpose buffer, which performs the bit-by-bit transposition. We design the transpose buffer \omiii{(\circled{5})} such that it can transpose all bits of a horizontally-laid-out cache line in a single cycle. As the other cache lines belonging from the slice are evicted (as a result of the Object Tracker's invalidation requests) and arrive at the transposition unit, they too are forwarded to the transpose buffer, and their bits are transposed.  Each horizontally-laid-out cache line maps to a specific set of bit columns in the vertically-laid-out cache line, which is determined using the physical address of the horizontally-laid-out cache line. Once all $n$ cache lines in the \mech object slice have been transposed, the Store Unit generates DRAM write requests for each vertically-laid-out cache line, and sends the requests to the memory controller (\circled{6}).}

\sgii{When a program wants to read data that belongs to a \mech object, 
\omi{and} the data is not in the CPU caches, the LLC issues a read request to DRAM \omiv{(\circled{7} in \cmr{Figure}~\ref{fig_transposing})}. If the address of the read request does not hit in the Object Tracker, the request is forwarded to the memory controller, 
\omi{as in a conventional system}. 
If the address of the read request hits in the Object Tracker, the read request is part of a \mech object, and the Object Tracker sends a signal \omi{(\circled{8})} to the Fetch Unit. The Fetch Unit generates the read requests for \emph{all} of the vertically-laid-out cache lines that belong to the same \mech object slice as the requested data, and sends these requests to the memory controller. When the request responses for 
\omi{the object slice's cache lines} arrive, the Fetch Unit sends the cache lines to a vertical-to-horizontal transpose buffer \omiii{(\circled{9})}, which can transpose all bits of one vertically-laid-out cache line into the horizontally-laid-out cache lines \omiii{in one cycle.} 
The horizontally-laid-out cache lines are then 
\omi{inserted into} the LLC. The \omi{$n-1$} cache lines that were \emph{not} part of the original memory request, \omi{but belong to the same object slice,} are inserted \omi{into} the LLC in a manner similar to \omiii{conventional} prefetch requests~\omiv{\cite{srinath2007feedback}}.}

\subsection{ISA Extensions and Programming Interface} 
\label{sec:bbop}

\nasrev{The lack of an efficient and expressive programmer/system interface can negatively impact the performance and usability of the \mech framework. 
This would \revonur{put} data transposition \revonur{on the critical path of} \mech computation, which would \revonur{cause large} performance overheads. To address such issues and to enable the programmer/system to efficiently communicate with \mech, we extend the ISA with specialized \mech instructions. The main goal of the \mech ISA extensions is to let the \mech control unit know (1)~what \mech operations need to be performed and when, and (2)~what the \mech memory objects are and when to transpose them.} 

\nasrev{Table~\ref{table_isa_format} shows the \omiii{CPU} ISA \omiii{extensions} that the \mech framework exposes to the programmer.
There are three types of instructions: 
(1)~\sgii{\mech object} initialization instructions, 
(2)~instructions to perform different \mech operations, and 
(3)~predication instructions.
We discuss \texttt{bbop\_trsp\_init}, our only \mech object initialization instruction, in \cmr{\Cref{sec:transposing}}.
The \omiii{CPU} ISA \omiii{extensions} for performing SIMDRAM \sgii{operations} can be further divided into two categories:
(1)~operations with one input operand (e.g., bitcount, ReLU), 
and (2)~operations with two input operands (e.g., addition, division, equal, maximum). \mech uses an array-based computation \omiii{model,} and \texttt{src} (i.e., \texttt{src} in 1-input operations and \texttt{src\_1, src\_2} in 2-input operations) and \texttt{dst} in these instructions represent source and destination \emph{arrays}.
\texttt{bbop\_op} represents the opcode of the \mech operation, while \texttt{size} and \texttt{n} represent the number of elements in the source and destination arrays, and the number of bits in each array element, respectively. To enable predication, \mech uses the \texttt{bbop\_if\_else} instruction in which, in addition to two source and one destination arrays, \texttt{select} represents the 
\omi{predicate} array \omi{(i.e., the predicate\omiii{, or mask,} bits)}.} 

\begin{table}[ht]
\tempcommand{0.8}
\caption{SIMDRAM ISA \sgii{extensions}.}
\label{table_isa_format}
\resizebox{\linewidth}{!}{
    \begin{tabular}{@{}cl@{}}
    \toprule
    \textbf{Type}                & \multicolumn{1}{c}{\textbf{ISA Format}}    \\ \midrule
    Initialization & \texttt{bbop\_trsp\_init address, \omi{size}, n}                         \\
    1-Input Operation           & \texttt{bbop\_op dst, src, \omi{size}, n}                                \\
    2-Input Operation           & \texttt{bbop\_op dst, src\_1, src\_2, size, n}               \\
    Predication                  & \texttt{bbop\_if\_else dst, src\_1, src\_2, select, size, n} \\ \bottomrule
    \end{tabular}
}
\end{table}

\nasrev{Listing~\ref{lst:simdraprog} shows how \omiii{\mech's CPU} ISA \omiii{extensions} can be used to perform in-DRAM computation, 
\omi{with an example code that performs element-wise addition or subtraction of two arrays (\texttt{A} and \texttt{B}) depending on the comparison of each element of \texttt{A} to the corresponding element of a third array (\texttt{pred}).} 
Listing~\ref{sublst:codea} shows the \omiv{original} C code for the computation, while Listing~\ref{sublst:codeb} shows the equivalent code using \mech operations. 
The lines that perform the same operations are highlighted using the same colors in both 
\omi{C code and SIMDRAM code}. 
The if-then-else condition in C code is performed in \mech using a predication instruction (i.e., \texttt{bbop\_if\_else} 
\omi{on} line \omiv{16} in Listing~\ref{sublst:codeb}). 
\mech treats the if-then-else condition as a multiplexer. Accordingly, \texttt{bbop\_if\_else} takes two source arrays and a 
\omi{predicate} array as inputs, where the 
\omi{predicate} is used to choose which source array should be selected as the output \omi{at the corresponding index}. 
To this end, we first 
\omi{allocate} 
\omi{two arrays to hold the addition and subtraction results} (i.e., arrays \texttt{D} and \texttt{E} 
\omi{on} line 10 in Listing~\ref{sublst:codeb}), and then populate them using \texttt{bbop\_add} and \texttt{bbop\_sub} (lines \omiv{13} and \omiv{14} in Listing~\ref{sublst:codeb}), respectively. 
We then 
\omi{allocate the predicate array} (i.e., array \texttt{F} on line \omiv{11} in Listing~\ref{sublst:codeb}) and populate \omi{it using \texttt{bbop\_greater}} (line \omiv{15} in Listing~\ref{sublst:codeb}). 
The addition, subtraction, and 
\omi{predicate} arrays form the three inputs \omiii{(arrays \texttt{D}, \texttt{E}, \texttt{F})} to the \texttt{bbop\_if\_else} instruction \omiii{(line \omiv{16} \omiii{ in Listing~\ref{sublst:codeb}})}, which stores the outcome \omi{of the predicated execution} to the destination array (i.e., array \texttt{C} in Listing~\ref{sublst:codeb}).}

\begin{figure}[h]
    \setcaptiontype{lstlisting}
    \begin{minipage}{.45\textwidth}
	    \begin{lstlisting}[style=myC]
int size = 65536;
int elm_size = sizeof(uint8_t);
uint8_t *A, *B, *C = (uint8_t*)malloc(size*elm_size);
uint8_t *pred = (uint8_t*)malloc(size*elm_size);
...
for(int i = 0; i < size; ++i) {
%\HilightGreen%    bool cond = A[i] > pred[i];
%\HilightPink%    if (cond)
%\HilightBlue%        C[i] = A[i] + B[i];
%\HilightPink%    else
%\HilightYellow%        C[i] = A[i] - B[i];
}
\end{lstlisting}
        \subcaption{C code for vector add/sub with \omi{predicated} execution}
        \label{sublst:codea}
    \end{minipage}
   ~
    \begin{minipage}{.45\textwidth}
	    \begin{lstlisting}[ style=myC]
int size = 65536;
int elm_size = sizeof(uint8_t);
uint8_t *A, *B, *C = (uint8_t*)malloc(size*elm_size);

bbop_trsp_init(A,size,elm_size);
bbop_trsp_init(B,size,elm_size);
bbop_trsp_init(C,size,elm_size);
uint8_t *pred = (uint8_t*)malloc(size*elm_size);
// D, E, F store intermediate data
uint8_t *D, *E = (uint8_t*)malloc(size*elm_size);
bool *F = (bool*)malloc(size*sizeof(bool));
...
%\HilightBlue%bbop_add(D, A, B, size, elm_size);
%\HilightYellow%bbop_sub(E, A, B, size, elm_size);
%\HilightGreen%bbop_greater(F, A, pred, size, elm_size);
%\HilightPink%bbop_if_else(C, D, E, F, size, elm_size);
\end{lstlisting}
       \subcaption{Equivalent code using SIMDRAM operations}
       \label{sublst:codeb}
   \end{minipage}
    \caption{\geraldorevi{Example code using SIMDRAM instructions. \gfrev{or operations?}}}
    \label{lst:simdraprog}


\end{figure}


\textfromsl{In this work, we assume that the programmer manually \revsgii{rewrites} the code to use SIMDRAM operations. We follow this approach when evaluating  real-world applications in \cmr{\Cref{sec_real_world_kernels}}. 
\geraldorevi{We envision two programming models for SIMDRAM. In the first programming model, SIMDRAM operations \revonurii{are} encapsulated within userspace library routines to ease programmability.} With this approach, the programmer can optimize the SIMDRAM-based code to \revonurii{make} the most out of the underlying in-DRAM computing mechanism. 
\geraldorevi{In the second programming model, SIMDRAM operations \revonurii{are} transparently inserted within the application's binary using compiler assistance.} Since SIMDRAM is a SIMD-like compute engine, we expect that the compiler can generate SIMDRAM code without \revonurii{programmer} intervention in at least two ways. First, it can leverage auto-vectorization routines already present in modern compilers~\revdelrefr{\cite{AutoVect51:online, Autovect33:online}}\revdelrefa{\hl{[39, 76]}} to generate SIMDRAM code, by setting the width of the SIMD lanes equivalent to a DRAM row. For example, in LLVM~\revdelrefr{\cite{lattner2004llvm}}\revdelrefa{\hl{[60]}}, the width of the SIMD units can be defined using the "\texttt{-force-vector-width}" flag~\revdelrefr{\cite{AutoVect51:online}}\revdelrefa{\hl{[76]}}. A SIMDRAM-based compiler back-end can convert the LLVM intermediate representation instructions into \emph{bbop} instructions.
Second, the compiler can compose groups of existing SIMD instructions generated by the compiler (e.g., AVX2 instructions~\revdelrefr{\cite{firasta2008intel}}\revdelrefa{\hl{[29]}}) into blocks that match the size of a DRAM row, and then convert such instructions into a single SIMDRAM operation. \omi{Prior work~\revdelrefr{\cite{ahmed2019compiler} uses a similar approach for 3D-stacked PIM}}\revdelrefa{\hl{[4]}}. We leave the design of a compiler for SIMDRAM for future work.}


SIMDRAM instructions can be implemented by extending the ISA of the host CPU. This is \color{black}possible since there is enough unused opcode space to support the extra opcodes that SIMDRAM requires. To illustrate, prior works~\revdelrefr{\cite{lopes2013isa,lopes2015shrink}}\revdelrefa{\hl{[75,76]}} show that there are 389 unused operation codes considering only the AVX and SSE \revonur{extensions} for the x86 ISA. Extending the instruction set 
is a common approach 
\omi{to interface a CPU with} PIM architectures~\revdelrefr{\cite{ahn2015scalable, seshadri2017ambit}}\revdelrefa{\hl{[4,93]}}. 





\subsection{Handling Page Faults, Address Translation, \omi{Coherence,} and Interrupts}
\label{sec:pagefaults}

\nasirevi{
SIMDRAM handles 
\omi{four} key system mechanisms \omi{as follows}:

\geraldorevi{\begin{itemize}[itemsep=0pt, topsep=0pt, leftmargin=*]
    \item  \textit{Page Faults:} \revonur{W}e assume that the pages that are touched during in-DRAM computation are already present and pinned in DRAM. In case the \revonur{required} data is not present in DRAM, we rely on the conventional page fault handling mechanism to bring the required pages into DRAM. 
    
    \item \textit{Address Translation:} Virtual memory and address translation \sr{
    \omi{are} challenging} for \nasirevi{many} PIM architectures~\revdelrefr{\cite{ghose2018enabling, PEI, vm30}}\revdelrefa{\hl{[5,37,86]}}. SIMDRAM is relieved of such challenge as it operates directly on physical addresses. When the CPU issues a SIMDRAM instruction, the instruction's \sr{virtual} memory addresses are translated \sr{into their corresponding physical addresses using} the same translation lookaside buffer (TLB) lookup mechanisms used by regular load/store operations.
    
    \item \omi{\textit{Coherence:} Input arrays to \mech may be generated or modified by the CPU, and the data updates may reside only in the cache (e.g., because the updates have not yet been written back to DRAM). To ensure that \mech does not operate on stale data, programmers are responsible for flushing cache lines~\cite{guide2016intel, manual2010arm} modified by the CPU. \cmr{\omiv{SIMDRAM can leverage coherence optimizations tailored to PIM to improve overall performance~\cite{lazypim,boroumand2019conda}}.}}
    
    \item  \textit{Interrupts:} Two cases where an interrupt could affect the execution of a SIMDRAM operation are (1)~on an application context switch, and (2)~on a page fault. In case of a context switch, the control unit's context needs to be saved and \nasirevi{then} restored \nasirevi{later} when the application resume\revonur{s} execution. \revonur{W}e do not expect to encounter a page fault during the execution of a SIMDRAM operation since\revonur{,} as previously mentioned, pages touched by SIMDRAM operations are \revonur{\cmr{expected to be loaded into and} pinned in DRAM.}
\end{itemize}}} 

\subsection{Handling Limited Subarray Size}
\label{sec:limited}

\textfromsl{\geraldorevi{SIMDRAM  
operates on data placed within the same subarray. However, a single subarray stores \revonur{only} 
\omi{several} megabytes of data. For example, a subarray with 1024 rows and a row size of \SI{8}{\kilo\byte} can only store \SI{8}{\mega\byte} of data. Therefore, SIMDRAM needs to use a mechanism that can efficiently move data within DRAM (e.g., across DRAM banks and subarrays). SIMDRAM can exploit (1)~RowClone Pipelined Serial Mode (PSM)~\revdelrefr{\cite{seshadri2013rowclone}}\revdelrefa{\hl{[92]}} to copy data between two banks by using the internal DRAM bus, or (2)~Low-Cost Inter-Linked Subarrays (LISA)~\revdelrefr{\cite{chang2016low}}\revdelrefa{\hl{[18]}} to copy rows between two subarrays within the same bank. We evaluate the performance overheads of using both mechanisms in~\Cref{sec:eval:datamovement}}.} 
\omi{Other mechanisms \omiii{for} fast in-DRAM data movement~\cite{nom2020,wang2020figaro} can also enhance \mech's capability.}

\subsection{Security Implications}
\label{sec:security}

SIMDRAM and other similar in-DRAM computation mechanisms that use dedicated DRAM rows to perform computation may 
\omi{increase} vulnerabilit\omiii{y} \omi{to} RowHammer attacks~\cite{kim2014flipping,revisitrh,frigo2020trrespass,mutlu2017rowhammer,mutlu2019rowhammer}\omii{.}  
\omii{W}e believe, and the literature suggests, that there 
\omi{should be} robust and scalable solutions to RowHammer, orthogonally to our work (e.g., \geraldorevii{BlockHammer~\cite{yaglikci2020blockhammerarxiv},} PARA~\cite{para}, TWiCe~\cite{lee2019twice}, \omi{Graphene~\cite{park2020graphene}}). 
Exploring RowHammer prevention and mitigation mechanisms 
\omi{in conjunction with} SIMDRAM (or other \cmr{PIM} 
approaches) requires special attention and research, which we leave for future work.

\subsection{SIMDRAM Limitations}
\label{sec:limitations}

\sgii{We note three key limitations of the current version of \revonur{the} SIMDRAM framework:}
\geraldorevi{
\begin{itemize}[itemsep=0pt, topsep=0pt, leftmargin=*]
    \item  \textit{Floating-Point Operations:} SIMDRAM supports \revonur{only} integer and fixed-\revonur{point} operations. Enabling floating-point operations in-DRAM while maintaining low area overheads is a challenge. For example, for floating-point addition, the IEEE 754 FP32 format~\cite{ieee754} requires \omiii{shifting} the mantissa by the difference of the exponents of elements. Since each bitline stores a data element in SIMDRAM, \revonur{shifting} \revonur{the value stored in one bitline} without compromising the values stored in other bitlines \revonur{at low cost is currently infeasible}.  
    
    \item \textit{Operations That Require Shuffling Data Across Bitlines:} Different from prior works (e.g., \revonur{DRISA}~\revdelrefr{\cite{li2017drisa}}\revdelrefa{\hl{[72]}}), SIMDRAM does \emph{not} add any extra circuitry to perform \sr{bit-shift} operations. Instead, SIMDRAM stores data in a vertical layout and \omi{can} perform 
    \omi{explicit} \sr{bit-shift} 
    operations \omi{(if needed)} by orchestrating row copies. 
    Even though this approach enables \mech to implement a large range of operations, it is not possible to perform shuffling and reduction operations \emph{across bitlines} without \sr{the inclusion of dedicated \omi{bit-shifting}} circuitry. This is due to the lack of physical connections across bitlines\sr{, which can} be solved by \sr{building} a \sr{bit-shift} engine \revonur{near} the sense amplifiers. 
    
    \item  \textit{Synchronization Between Concurrent In-DRAM Operations:} SIMDRAM can be easily modified to enable concurrent \revonur{execution of \emph{distinct operations} across different subarrays in DRAM}. However, this would require the implementation of software or hardware synchronization primitives to orchestrate the computation of a single task across different subarrays. 
    \omi{Ideas that are similar to SynCron~\cite{giannoula2021SynCron} can be beneficial.}
    \end{itemize}    }

\section{Methodology}
\label{methodology}

We implement SIMDRAM using the gem5 simulator~\cite{gem5} and compare it to a \omi{real} multicore CPU (Intel Skylake~\cite{intelskylake}), a \omii{real} high-end GPU (\omi{NVIDIA Titan V}~\cite{TitanV}), and a state-of-the-art \omi{processing-using-DRAM} mechanism (Ambit~\cite{seshadri2017ambit}). In all our evaluations, the CPU code is optimized to leverage AVX-512 instructions~\cite{firasta2008intel}. \geraldorevi{Table~\ref{table_parameters} shows the system parameters we use in our evaluations.} To measure CPU performance, we implement a set of timers in \texttt{sys/time.h}~\cite{systime}\omii{. T}o measure CPU energy consumption, we use Intel RAPL~\cite{hahnel2012measuring}. To measure GPU performance, we implement a set of timers using the \texttt{cudaEvents} API~\cite{cheng2014professional}. We capture GPU kernel execution time \sr{that excludes} data initialization/transfer time. To measure GPU energy consumption, we use the \texttt{nvml} API~\cite{NVIDIAMa14}. We report the average of five runs for each CPU/GPU data point, each with a \omii{warmup} phase to avoid cold cache effects. \omii{W}e implement Ambit \omii{on} gem5 \omii{and} validate our implementation rigorously with the results \omi{reported in \cite{seshadri2017ambit}}. \omii{We \omiii{use} \omiii{the same} vertical data layout in our Ambit \omiii{and SIMDRAM implementations}, \omiii{which} enables us to (1) evaluate all 16 SIMDRAM operations in Ambit using their equivalent AND/OR/NOT-based implementation\omiii{s}, and (2) highlight the benefits of Step 1 in the \mech framework (i.e., using an optimized MAJ/NOT-based implementation of the operations).} 
\omvuii{Our synthetic throughput analysis \omviii{(\Cref{sec_performance})} uses 64M-element input
arrays.}


\begin{table}[ht]
   \caption{\omiii{Evaluated s}ystem configuration\omiii{s}.}
   \centering
   \footnotesize
   \tempcommand{1.3}
   \renewcommand{\arraystretch}{0.7}
   \resizebox{\columnwidth}{!}{
   \begin{tabular}{@{} c l @{}}
   \toprule
   \multirow{5}{*}{\shortstack{\textbf{Intel}\\ \textbf{Skylake CPU\omii{~\cite{intelskylake}}}}} & x86~\cite{guide2016intel}, 16~cores, 8-wide, out-of-order, \SI{4}{\giga\hertz};  \\
                                                                           & \emph{L1 Data + Inst. Private Cache:} \SI{32}{\kilo\byte}, 8-way, \SI{64}{\byte} line; \\
                                                                           & \emph{L2 Private Cache:} \SI{256}{\kilo\byte}, 4-way, \SI{64}{\byte} line; \\
                                                                           & \emph{L3 Shared Cache:} \SI{8}{\mega\byte}, 16-way, \SI{64}{\byte} line; \\
                                                                           & \emph{Main Memory:} \SI{32}{\giga\byte} DDR4-2400, 4~channels, 4~ranks \\
   \midrule
   \multirow{3}{*}{\shortstack{\textbf{\omi{NVIDIA}}\\ \textbf{\omi{Titan V} GPU\omii{~\cite{TitanV}}}}} & 6 graphics processing clusters, 5120 CUDA Cores;\\ 
                                                                            & 80 streaming multiprocessors, \SI{1.2}{\giga\hertz} base clock; \\
                                                                            & \emph{L2 Cache:} \SI{4.5}{\mega\byte} L2 Cache; \emph{Main Memory:} \SI{12}{\giga\byte} HBM~\omii{\cite{HBM,lee2016simultaneous}} \\
   \midrule
   \multirow{5}{*}{\shortstack{\omi{\textbf{Ambit~\cite{seshadri2017ambit}}}\\ \textbf{\omi{and SIMDRAM}}}} &  gem5 system emulation;  x86~\cite{guide2016intel}, 1-core, out-of-order, \SI{4}{\giga\hertz};\\
                                                                             & \emph{L1 Data + Inst. Cache:} \SI{32}{\kilo\byte}, 8-way, \SI{64}{\byte} line;\\
                                                                             & \emph{L2 Cache:} \SI{256}{\kilo\byte}, 4-way, \SI{64}{\byte} line; \\
                                                                             & \emph{Memory Controller:}  \SI{8}{\kilo\byte} row size, FR-FCFS~\cite{mutlu2007stall,zuravleff1997controller} scheduling\\
                                                                             & \emph{Main Memory:}  DDR4-2400, 1~channel, 1~rank, 16~banks \\
   \bottomrule
   \end{tabular}
   }
   \label{table_parameters}
\end{table}

We evaluate three different configurations of SIMDRAM \omii{where} 1 \geraldorevii{(\emph{SIMDRAM:1})}, 4 \geraldorevii{(\emph{SIMDRAM:4})}, and 16 \geraldorevii{(\emph{SIMDRAM:16})} banks out of all the banks in one channel (16 banks in our evaluations) have SIMDRAM computation capability. In the SIMDRAM 1-bank configuration, our mechanism exploits 65536 (\omi{i.e., size of an} \SI{8}{\kilo\byte} row buffer) SIMD lanes. Conventional DRAM architectures exploit bank-level parallelism (BLP) to maximize DRAM throughput~\omii{\cite{mutlu2008parallelism,kim2010thread,salp,ramulator,lee2009improving}}. The memory controller can issue commands to different banks (one-per-cycle) on the same channel such that banks can operate in parallel. In SIMDRAM, banks in the same channel can operate in parallel, just like conventional banks. Therefore, to enable the required parallelism, SIMDRAM requires no more modifications. Accordingly, the number of available SIMD lanes\omii{, i.e., SIMDRAM's computation capability,} increases by exploiting BLP in SIMDRAM configurations (i.e., the number of available SIMD lanes in the \omii{16}-bank configuration is \omii{16} $\times$ 65536).\omii{\footnote{\omii{SIMDRAM computation capability can be further increased by enabling and exploiting subarray-level parallelism in each bank~\omiii{\cite{salp,chang2016low,chang2014improving,kang2014co}}.}}} 

\section{Evaluation}
\label{evaluation}
\label{sec_evaluation}



\omii{We demonstrate the advantages of the \mech framework by evaluating: (1)~SIMDRAM's throughput and energy consumption \omii{for a wide range of operations}; (2)~SIMDRAM's performance benefits \omiii{on} real-world applications; (3)~SIMDRAM's performance and energy benefits \omiii{over} a closely\omiii{-}related processing-using-\omiii{cache} architecture~\cite{dualitycache}; and (4)~the reliability of SIMDRAM operations. Finally, we evaluate three key overheads in SIMDRAM\omii{:} in-DRAM data movement, data transposition, and area cost.}


\subsection{Throughput Analysis}
\label{sec_performance}

\begin{figure}[ht]
   \centering
   \includegraphics[width=1.0\linewidth]{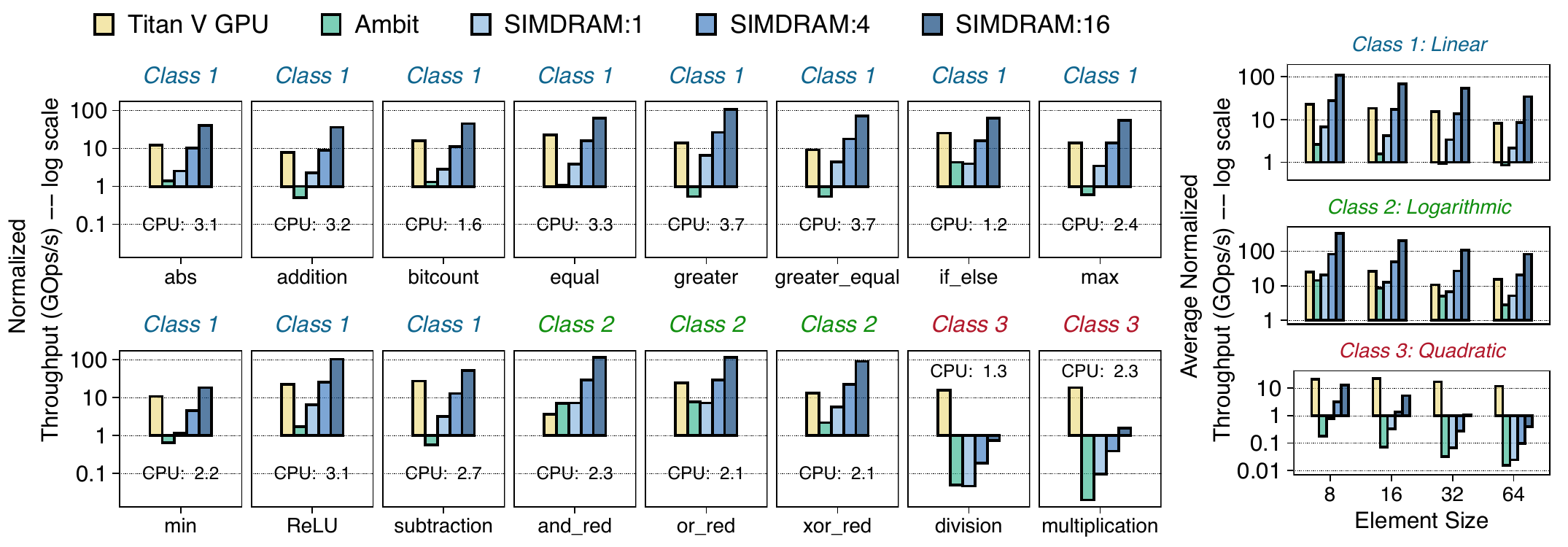}
   \caption{\omvi{Normalized throughput of 16  operations. \omvuii{SIMDRAM:\emph{X} uses \emph{X} DRAM banks for computation.}}}
 
   \label{fig_throughput}
\end{figure}

 \cmr{Figure}~\ref{fig_throughput} (\omi{left}) shows the \omii{normalized} throughput of all \omii{16} \mech operations \omi{(\cmr{\Cref{sec:supported:ops}}) compared to \omii{those on CPU, GPU, and Ambit} (normalized to the multicore CPU throughput), for 
 a\omii{n element size} of 32~bits. \omv{We provide the} \omiii{absolute throughput of the baseline CPU (in GOps/s) in each graph.} 
 We classify each operation based on how the latency of the operation scales with respect to \omii{element size} $n$\omii{.}\footnote{\omdef{Appendix~\ref{apdx:op-class} \omiii{discusses the scalability} of each operation\omii{.}}}
 Class~1\omii{, 2, and 3 operations} scale linearly\omii{,} logarithmically, and quadratically with $n$\omii{, respectively}.} 
\omi{\cmr{Figure}~\ref{fig_throughput} (right) shows how the average throughput across all operations of the same class scales relative to \omii{element size}.
 We evaluate element sizes \revonur{of} 8, 16, 32, 64 bits. We normalize the figure to the average throughput on a CPU}. 

\omi{W}e \omi{make} \omi{four} observations \omi{from \cmr{Figure}~\ref{fig_throughput}}. \omi{First, we observe that SIMDRAM outperforms the three state-of-the-art baseline systems i.e., CPU/GPU/Ambit. \omii{Compared to CPU/GPU, SIMDRAM\omiii{'s throughput is} 
5.5$\times$/0.4$\times$, 
22.0$\times$/1.5$\times$, and 
88.0$\times$/5.8$\times$ \omiii{that of \omiv{the} CPU/GPU,} 
\omii{averaged across all 16 SIMDRAM operations} for 1, 4, and 16~banks, respectively. To ensure fairness, we compare Ambit, which uses a single DRAM bank in our evaluations, only against \emph{SIMDRAM:1}.\footnote{\omiii{Ambit's throughput scales proportionally to bank count, just like SIMDRAM's.}} Our evaluations show that \emph{SIMDRAM:1} outperforms Ambit by 2.0$\times$, averaged across all 16 SIMDRAM operations.}} \omi{Second, SIMDRAM outperforms the GPU baseline when we use more than four DRAM banks for \emph{all} the linear and logarithmic operations. \emph{SIMDRAM:16} provides 5.7$\times$ (9.3$\times$) \omiii{the throughput of} the GPU across all linear (logarithmic) operations, on average. \emph{SIMDRAM:16}\omiii{'s throughput is} 83$\times$ (189$\times$) and 45.2$\times$ (19.9$\times$) \omiv{that of CPU and Ambit, respectively}, averaged across all linear (logarithmic) operations.} \omi{Third}, we observe that both  \omi{the multicore CPU} baseline and GPU outperform \geraldorevii{\emph{SIMDRAM:1}, \emph{SIMDRAM:4}, and \emph{SIMDRAM:16}} \omi{\emph{only} for the} division and multiplication operations. This is due to the quadratic nature of our bit-serial implementation of \omi{these two} operations. \omi{Fourth}, as expected, we observe a drop in the throughput for \omii{\emph{all}} operations \omii{with} increasing element size, since the latency \omii{of} each operation increases \omi{with} element size. \omi{W}e conclude that SIMDRAM \omii{significantly} \omi{outperforms all three state-of-the-art baselines} for a wide range of operations. 




\subsection{Energy Analysis}
\label{sec_energy}

\begin{figure}[h]
   \centering
   \includegraphics[width=1.0\linewidth]{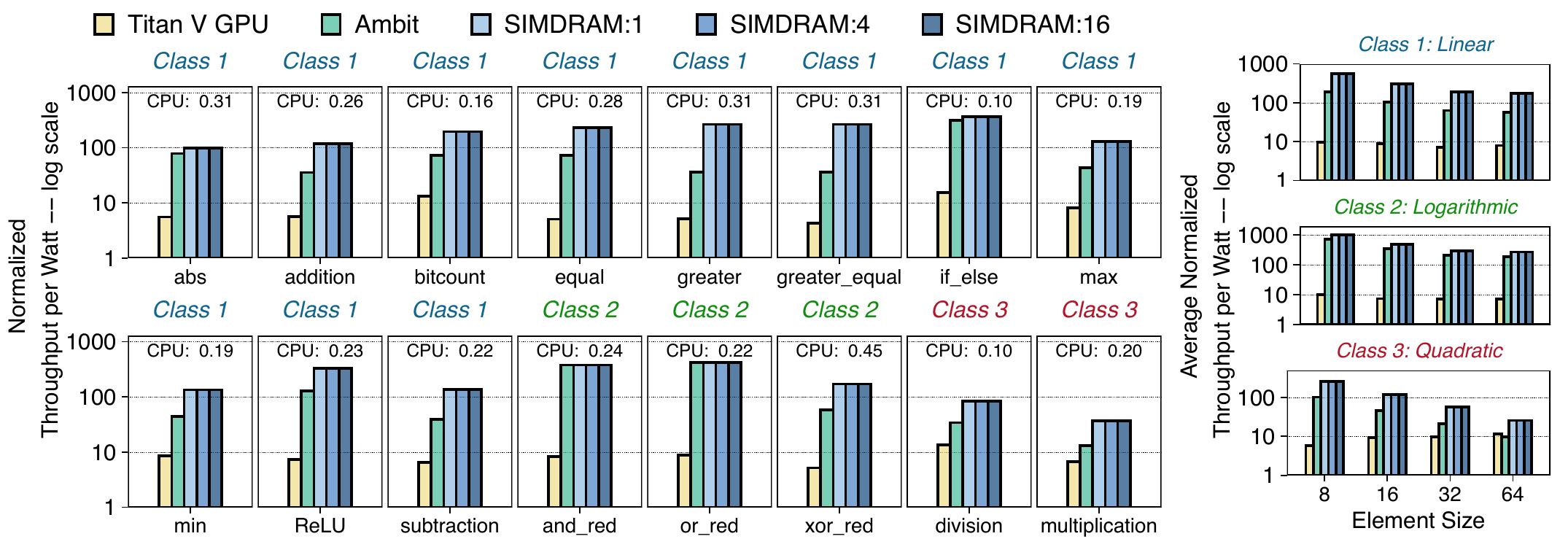}%
   \caption{\omvi{Normalized  energy efficiency of 16 operations.}
   }
   \label{fig_energy}
\end{figure}

We use \omi{CACTI}~\cite{cacti} to evaluate SIMDRAM's \omi{energy} consumption. Prior work~\cite{seshadri2017ambit} shows that each additional simultaneous row activation increases energy consumption by 22\%. We \omii{use} this observation in evaluating the energy consumption of SIMDRAM, which requires TRAs. \cmr{Figure}~\ref{fig_energy} compares the energy efficiency (\omii{\omii{T}hroughput} per Watt) of SIMDRAM against the GPU and Ambit baselines\omi{, normalized to the CPU baseline}. \omiv{We provide the absolute Throughput per Watt of the \omv{baseline} CPU in each graph}. We make \omii{four} observations. First, SIMDRAM significantly increases energy efficiency for \emph{all} operations \sgii{over \omii{\emph{all} three} baselines}. 
\omi{SIMDRAM’s energy efficiency is 257$\times$, 31$\times$, and 2.6$\times$ that of CPU, GPU, and Ambit, respectively, averaged across all 16 operations.} The energy savings in SIMDRAM directly result from (1)~avoiding the costly off-chip round-trips to load\omiii{/store} data from\omiii{/to} memory, (2)~exploiting the abundant memory bandwidth within the memory device, reducing execution time\omii{, and (3)~reducing the number of TRAs required to compute a given operation by leveraging an optimized majority-based implementation of the operation}. Second, similar to \omi{our results on} throughput \omi{(\cmr{\Cref{sec_performance}})}, the energy efficiency \omi{of} SIMDRAM reduces \omi{as} element size \omi{increases}. \omi{However, the energy efficiency of the CPU \omiii{or} GPU does not}. This is because (1)~for \omii{all SIMDRAM operations}, the number of TRAs increases with element size; and (2) \omii{CPU and GPU can} fully utiliz\omii{e} their \omii{wider} arithmetic units with larger (i.e., 32\omi{-} and 64\omi{-}bit) element sizes. 
Third, even though \omii{SIMDRAM} multiplication and division operations \omi{scale} poorly \omii{with} element size, the SIMDRAM implementations of these operations are significantly more energy-efficient compared to the CPU and GPU baselines,  
\geraldorevi{making SIMDRAM \revonur{a} competitive \revonur{candidate} \omiii{even} for multiplication and division operations.} \omii{Fourth, since \emph{both} SIMDRAM's throughput and power consumption increase proportionally to the number of banks, the \omii{T}hroughput per Watt for SIMDRAM 1-, 4-, and 16-bank configurations is the same.} \omi{We conclude that SIMDRAM is more energy\omii{-}efficient than all three state-of-the-art baselines for a wide range of operations. }


\subsection{Effect on Real-World Kernels}
\label{sec_real_world_kernels}

We evaluate SIMDRAM with a set of kernels that represent the behavior of \omi{selected} important real-world applications from different domains. The evaluated kernels \omi{come from} databases (TPC-H query 1~\cite{tpch}, BitWeaving~\cite{li2013bitweaving}), convolutional neural networks (LeNET-5~\cite{lecun2015lenet}, VGG-13~\cite{simonyan2014very}, VGG-16~\cite{simonyan2014very}), \omi{classification algorithms} (k-nearest neighbors~\cite{lee1991handwritten}), and  image processing (brightness\omii{~\cite{gonzales2002digital}}). These kernels rely on many of the basic operations we evaluate in \cmr{\Cref{sec_performance}}. \omdef{We provide a brief description of each kernel and the SIMDRAM operations that they utilize in Appendix~\ref{appendix}}. 

\cmr{Figure}~\ref{fig_speedup_real_world} shows the \omi{performance of \mech and our baseline configurations} \omiii{for} each kernel, normalized to \omii{that} of the multicore CPU. 
We make \omi{four} observations. 
\omi{First, \emph{SIMDRAM:16} greatly outperforms the CPU \omii{and} GPU baselines, \omiv{providing 21$\times$ \omii{and} 2.1$\times$ 
the performance of the CPU and GPU, respectively, on average across all \omii{seven} kernels.
\omii{\omiv{SIMDRAM has a maximum performance of 65$\times$ \omii{and} 5.4$\times$ that of the CPU and GPU, respectively (for the BitWeaving kernel in both cases).}}}
\omii{Similarly, \emph{SIMDRAM:1} \omiv{provides 2.5$\times$ the performance of Ambit (which also uses a single bank for in-DRAM computation)}, on average across all seven kernels, with a maximum \omiv{of 4.8$\times$ the performance of Ambit} for the TPC-H kernel.}} 
\omi{Second, even with a single DRAM bank, SIMDRAM \emph{always} outperforms the CPU baseline, \omiv{providing 2.9$\times$ the performance of the CPU on average} across all kernels.} \omi{Third, 
\emph{SIMDRAM:4} provides 2$\times$ and 1.1$\times$ \omiv{the performance of} the GPU \omii{baseline} for \omiv{the BitWeaving and brightness kernels, respectively}.}  Fourth, \omii{despite GPU's higher multiplication throughput compared to \mech (\cmr{\Cref{sec_performance}}), \emph{SIMDRAM\omii{:16}} outperforms the GPU baseline \omiii{even} for kernels 
that heavily rely on multiplication \omdef{(Appendix~\ref{appendix})} (e.g., by 1.03$\times$ and 2.5$\times$ for kNN and TPC-H kernels, respectively). This speedup is a direct result of exploiting the high \omii{in-DRAM} bandwidth in \mech to avoid the memory bottleneck in GPU caused by the large amounts of intermediate data generated in such kernels.} \revonur{We conclude that} SIMDRAM is \omi{an effective and efficient} substrate to accelerate many common\revonuri{ly}-used real-world applications.

\begin{figure}[h]
   \centering
   \includegraphics[width=\linewidth]{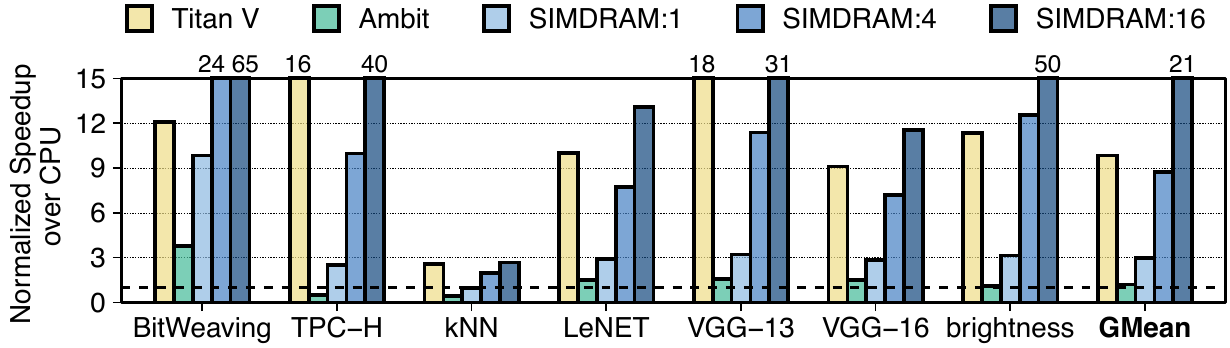}%
   \caption{\omvi{\omii{Normalized s}peedup of real-world kernels.} 
   }
   \label{fig_speedup_real_world}
\end{figure}

\subsection{\omi{Comparison to DualityCache}}
\label{sec:eva:dualitycache}

\nasirevi{We compare SIMDRAM to 
\omi{DualityCache~\cite{dualitycache}, a closely-related 
\omiii{processing-using-cache} architecture.}
DualityCache is an in-cache computing framework that performs computation using discrete logic elements (e.g., logic gates, latches, muxes) that are added to the SRAM peripheral circuitry. \revonur{In-cache computing approaches \revonuri{(such as DualityCache)} need  data to be brought into the cache first\revonuri{, which} requires extra data movement (and even more \omii{if} the working set of the application does not fit in the cache) compared to in-memory computing approaches (like SIMDRAM).}}

\begin{figure}[h]
   \centering
   \includegraphics[width=\linewidth]{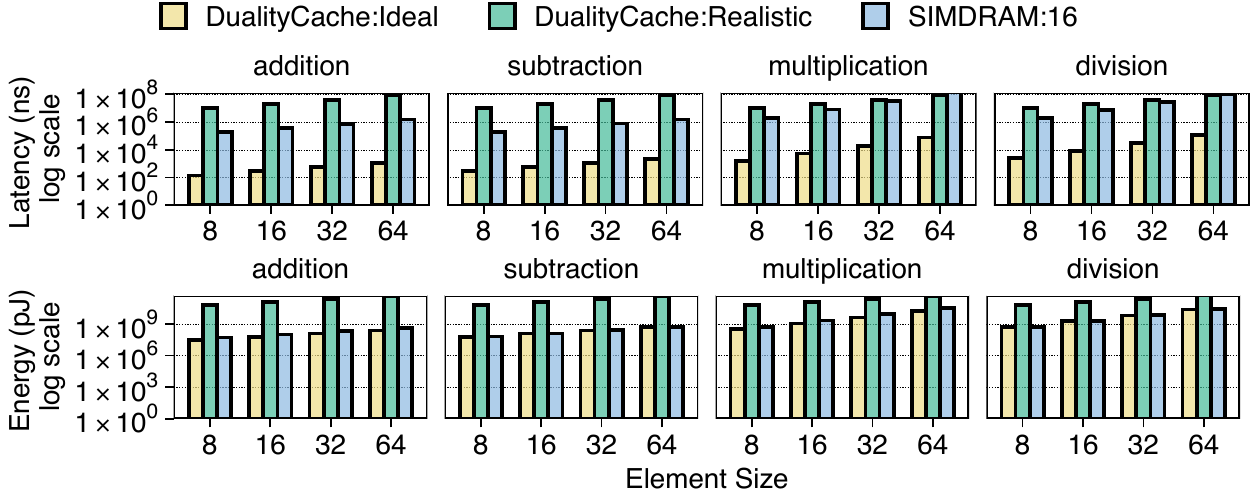}%
   \caption{\omvi{Latency and energy to execute 64M operations.}
   }
   \label{fig_dualitycache}
\end{figure}

\cmr{Figure}~\ref{fig_dualitycache} \omii{(top)} compares the \omii{latency} of SIMDRAM against DualityCache~\cite{dualitycache} for the subset of operations that \emph{both} SIMDRAM and DualityCache support \revonur{(i.e., addition, subtraction, multiplication, and division)}. In this experiment, we study three different configurations. \revonuri{First,} \emph{\geraldorevii{DualityCache:\omiii{Ideal}}} has \revonuri{\emph{all}} data \revonuri{required} for DualityCache residing in the cache. Therefore, results for \emph{\geraldorevii{DualityCache:\omiii{Ideal}}} do \omiii{\emph{not}} include the overhead of moving data from DRAM to the cache\revonur{, making it an unrealistic configuration that needs the data to already reside and fit in the cache}. \revonuri{Second,} \emph{\geraldorevii{DualityCache:\omiii{Realistic}}} includes the overhead of data movement from DRAM to the cache. Both DualityCache configurations compute \omiv{on} an input array of \SI{45}{\mega\byte}. \omii{\revonur{Third}, \omii{\emph{SIMDRAM:16}.}}
\omiv{For all three configurations, we use the same cache size (\SI{35}{\mega\byte}) as the original DualityCache work~\cite{dualitycache} to provide a fair comparison}. 
As shown in the figure, \revonur{SIMDRAM \omii{greatly outperforms} \revonuri{DualityCache} when data movement is \revonurii{realistically} taken into account.} \geraldorevii{\emph{SIMDRAM:16}} \omii{outperforms} \emph{\geraldorevii{DualityCache:\omiii{Realistic}}} for all four operations \revonuri{(by 52.9$\times$, 52.\omii{4}$\times$, 1.\omii{8}$\times$, and 2.\omii{1}$\times$ for addition, subtraction, multiplication, and division} respectively, on average across all element sizes). \nasirevi{\revonur{SIMDRAM\omii{'s performance improvement} comes at a much lower area overhead compared to DualityCache. DualityCache \revsgii{(including its peripherals, transpose memory unit, controller, miss status holding registers, and crossbar network)} \omiv{has an area overhead of 3.5\% in a high-end CPU, whereas SIMDRAM has an area overhead of only 0.2\%~\omv{(\cmr{\Cref{sec_area_overhead}})}.}
As a result, SIMDRAM can \revonuri{actually fit} \revsgii{a} significantly higher number of SIMD lanes in a given area compared to DualityCache. 
\revsgii{Therefore, SIMDRAM's \omii{performance improvement \omiv{per unit area} would be much larger \omiii{than \omv{that} we observe in \cmr{Figure}~\ref{fig_dualitycache}.}}}}
\revonuri{We conclude that SIMDRAM achieve\omii{s} higher performance at lower area \omix{cost over} DualityCache\omii{, when we consider DRAM-to-cache data movement}.}}

\omi{\cmr{Figure}~\ref{fig_dualitycache} \omii{(bottom)} shows} the energy consumption of \emph{DualityCache:\omiii{Realistic}}, \emph{DualityCache:\omiii{Ideal}}, and \emph{SIMDRAM:16} when performing \omi{64M addition, subtraction, multiplication, and division operations}. We make two observations. First, compared to \emph{\omii{Duality}Cache:\omiii{Ideal}}, \emph{SIMDRAM:16} increases average energy consumption by 60\%. This \omii{is} because while the energy per bit to \omii{perform} computation in DRAM (\SI{13.3}{\nano\joule}/bit~\cite{micropower,vogelsang2010understanding}) is smaller than the energy per bit to \omii{perform} computation in the cache (\SI{60.1}{\nano\joule}/bit~\cite{eckert2018neural}), the DualityCache implementation of \omiii{each} operation requires fewer iterations than \omiv{its} equivalent SIMDRAM implementation. Second, \emph{SIMDRAM:16} reduces average energy by 600$\times$ over \emph{\omii{Duality}Cache:\omiii{Realistic}} \omii{because} \emph{DualityCache\omii{:\omiii{Realistic}}} needs to load \emph{all} input data from DRAM, \omiv{incurring} high energy overhead (a DRAM access consumes 650$\times$ \omiii{the energy-per-bit of} a DualityCache operation~\omii{\cite{eckert2018neural, dualitycache}}). \omii{In contrast,} SIMDRAM operates on data that is already present in DRAM, eliminating any data movement \omii{overhead}. \omii{We conclude that SIMDRAM \omiii{is} much more efficient than DualityCache, when cache-to-DRAM data movement is \omiv{realistically} considered.}






\subsection{Reliability}
\label{sec_reliability}

We use SPICE simulations to test the reliability of SIMDRAM for different technology nodes and \revonur{varying} \revonuri{amounts} of process variation. At the core of SIMDRAM, \omiii{there are} two back-to-back triple-row activations (TRAs). Table~\revdelrefr{\ref{table:variation}}\revdelrefa{\hl{3}} shows the \revonur{characteristics} of TRA and two back-to-back TRAs (TRAb2b) for the \omii{45, 32, and } \SI{22}{\nano\metre} technology nodes. We compare these with the reliability of quintuple-row activations (QRA\revonur{s}), used by prior works~\revdelrefr{\cite{ali2019memory, angizi2019graphide}}\revdelrefa{\hl{[7,9]}} to implement bit-serial addition. We use the reference \SI{55}{\nano\metre} DRAM model from Rambus~\revdelrefr{\cite{rambus_model}}\revdelrefa{\hl{[89]}} and scale it based on the ITRS roadmap~\revdelrefr{\cite{itrs_model,vogelsang2010understanding}}\revdelrefa{\hl{[48,104]}} to model smaller technology nodes 
following the PTM transistor models~\revdelrefr{\cite{ptm_model}}\revdelrefa{[83]}.  
\geraldorevi{The goal of our analysis is to understand the reliability trends for TRA and QRA operations \omii{with} technology \omii{scaling}.} For each technology \revonur{node} and \revonurii{process} variation \revonuri{amount}, we run Monte-Carlo simulations for $10^4$ iterations. 

\begin{table}[h]
\footnotesize
\tempcommand{0.8}
\renewcommand{\arraystretch}{1.1}
\centering
\caption{\cmr{Process variation's effect on TRA/QRA failure rates.}}
\begin{tabular}{ |c||c||c|c|c|c|c| } 
\hline
& \textbf{Variation (\%)} & \textbf{$\pm$ 0} & \textbf{$\pm$ 5} & \textbf{$\pm$ 10} & \textbf{$\pm$ 20}\\
\hhline{|=#=#=|=|=|=|}
\multirow{3}{*}{\textbf{45~nm}} & TRA Failure (\%)& 0 & 0 & 0.02 & 3.01 \\ 
& TRAb2b Failure (\%)& 0 & 0 & 0.04 & 5.93\\ 
& QRA Failure (\%) & 0 & 0 & 0.35 & 6.54\\ 
\hline
\multirow{3}{*}{\textbf{32~nm}} & TRA Failure (\%) & 0 & 0 & 0.35 & 3.90 \\ 
& TRAb2b Failure (\%) & 0 & 0 & 0.69 & 7.64\\ 
& QRA Failure (\%) & 0 & 0.42 & 6.33 & 11.52\\
\hline
\multirow{3}{*}{\revDMicro{\textbf{22~nm}}} & \revDMicro{TRA Failure (\%)} & \revDMicro{0} & \revDMicro{0} & \revDMicro{0.42} & \revDMicro{4.50} \\ 
& \revDMicro{TRAb2b Failure (\%)} & \revDMicro{0} & \revDMicro{0} & \revDMicro{0.84} & \revDMicro{8.83}\\ 
& \revDMicro{QRA Failure (\%)} & \omii{error} & \omii{error} & \omii{error} & \omii{error}\\
\hline
\end{tabular}
\label{table:variation}
\end{table}

We make \geraldorevi{four} observations. First, for all \geraldorevii{process} variation ranges
, TRA and TRAb2b perform more reliably than QRA. Specifically, TRA and TRAb2b perform without errors \omii{for} 5\% variation. Second, while moving from \SI{45}{\nano\metre} to \SI{32}{\nano\metre}, we observe that the error rate of QRA \omi{increases} faster than \omiii{than that of} TRA, making \omi{QRA} less reliable as the technology \omii{node} size \omii{reduces}. Third, for TRA and TRAb2b in \SI{22}{\nano\metre}, we observe a similar trend of increased error rate while still having zero error rate for 5\% process variation. In our simulations, QRA does not perform correctly 
\omii{in} the projected \SI{22}{\nano\metre} DRAM. For example, \texttt{MAJ(11100)} \omiii{always} leads to the incorrect outcome of \texttt{\omii{`}0'}. This is because charge sharing between five capacitors in QRA does not lead to enough voltage on the bitline for the sense amplifier to pull up the bitline to the value \texttt{\omii{`}1'}. We believe that proposals based on QRA require changes to the circuit elements (e.g., transistors in the sense amplifier) to enable \omii{correct} operation in \omiii{the} \SI{22}{\nano\metre} technology \omii{node}. 
\geraldorevi{Fourth, a TRA can fail depending on the amount of manufacturing process variation. We observe that a TRA starts to fail \revonur{when} process variation \revonur{is} larger than 10\%, for all \omii{technology nodes}. Since SIMDRAM operations are executed within a DRAM module, it is quite challenging to leverage existing \revonur{in-}DRAM \revonur{or in-memory-controller} error
correction mechanisms~\omiii{\cite{patel19, patel20, yixin2014, meza2015revisiting}}. The same problem exists for \omii{other} \omi{processing-using-DRAM} mechanisms~\omiv{\cite{seshadri2017ambit, seshadri2013rowclone, li2017drisa, deng2018dracc, xin2020elp2im, ali2019memory,angizi2019graphide, li2018scope,seshadri2019dram, seshadri2017simple,seshadri2015fast, seshadri2016processing,subramaniyan2017parallel,dlugosch2014efficient}}.} 
\omii{We conclude that the TRA operations SIMDRAM relies on are much more scalable and variation-tolerant than QRA operations some prior works rely on.} \omi{We leave a study of
reliability solutions for future work.}

\subsection{Data Movement Overhead}
\label{sec:eval:datamovement}

\begin{figure}[h]
   \centering
   \includegraphics[width=\linewidth]{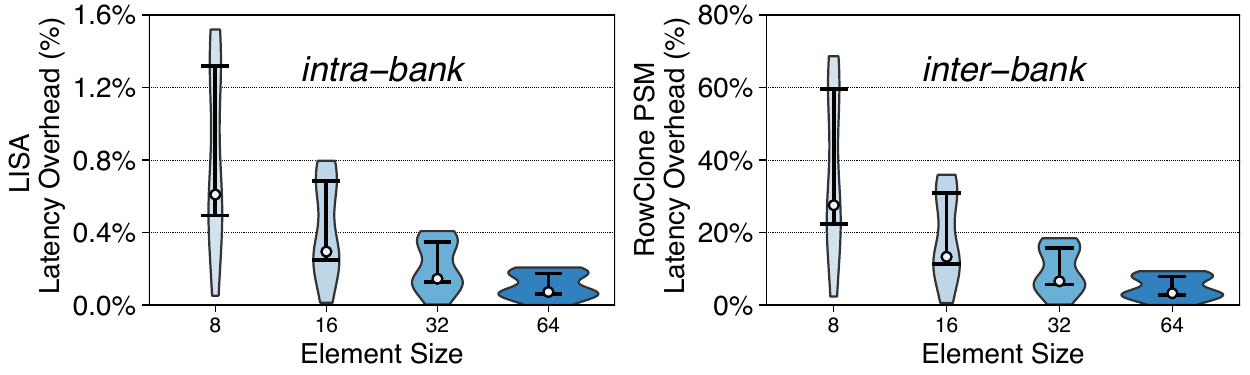}%
   \caption{\omii{Latency overhead distribution of \omiii{worst-case} intra-bank (left) and inter-bank (right) data movement} \cmr{for \emph{SIMDRAM:1}. Error bars depict the 25th and 75th percentiles.}} 
   \label{fig_data_movement}
\end{figure}

\omi{There may be cases where the output of a \mech operation that is used as an input to a subsequent operation does not reside in the same subarray as other inputs.}  
\omii{For example, consider the computation $C = OP(A, B)$. If the output of the SIMDRAM operation $OP$ is an input to a subsequent SIMDRAM operation, \revonur{$C$} needs to move to the same subarray as the other inputs of the subsequent operation, before the operation can start.} 
\geraldorevi{\cmr{Figure}~\ref{fig_data_movement} shows the \omii{distribution of the worst-case latency} overhead of moving the output of each of our \omii{16 SIMDRAM operations with 8-, 16-, 32-, and 64-bit element sizes \omii{in \emph{SIMDRAM:1}}} \omii{to a different subarray within the same bank, i.e., \emph{intra-bank} (using LISA~\revdelrefr{\cite{chang2016low}}\revdelrefa{\hl{[18]}}) or a different bank, i.e., \emph{inter-bank} (using RowClone PSM~\revdelrefr{\cite{seshadri2013rowclone}}\revdelrefa{\hl{[92]}})}}. 
We make two observations. First, \omii{intra-bank data movement \omii{(\cmr{Figure}~\ref{fig_data_movement}, left)} results in \omii{only} 0.39\% latency overhead, averaged across all 16 \omii{SIMDRAM operations and four different element sizes}} (max. 1.52\% for 8-bit reduction, min. 0.001\% for 64-bit multiplication). 
Second, 
\omii{inter-bank data movement} \omii{(\cmr{Figure}~\ref{fig_data_movement}, right) results in 17.5\% \revonur{\omii{latency}} overhead, averaged across all \omii{16 SIMDRAM operations and four different element sizes}} (max. 68.7\% for 8-bit reduction, min. 0.03\% for 64-bit multiplication). 
\omiv{We observe that the latency overhead of moving data, as a fraction of the total \omv{computation} latency \emph{decreases} \omiii{with} element size, \omii{because} the \omv{computation} latency of \omii{each} SIMDRAM operation increases \omii{with element size}.} 
\revonur{We conclude that while efficient data movement is a challenge in processing-in-memory architectures that rely on moving and aligning operands, the performance overhead of data movement in SIMDRAM stays within \revonuri{an} acceptable range} \omii{even under worst-case assumptions.}


\subsection{Data Transposi\omii{tion} Overhead}
\label{sec:eval:transposing}

\nasirevi{
\revonuriii{\omii{T}ransposition} of the data in one subarray can overlap with in-DRAM computation in another subarray. As a result, if the data required for in-DRAM computation spans over multiple subarrays, only the transposition of the data in the first subarray is on the critical path of \omii{SIMDRAM} execution. The data in each remaining subarray is then transposed simultaneously with the in-DRAM computation in the previous subarray.} 

\begin{figure}[h]
   \centering
   \includegraphics[width=\linewidth]{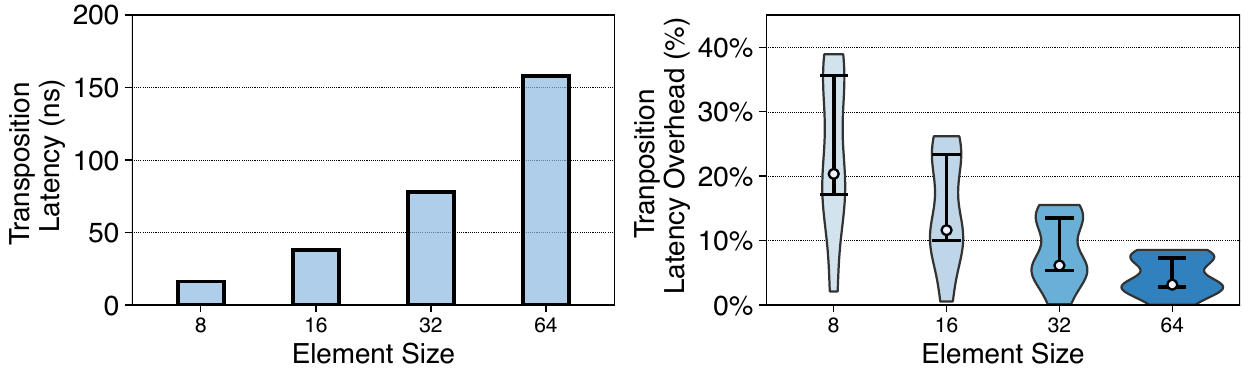}%
   \caption{\omiii{Worst-case latency} \omii{(left)} and \omiii{worst-case} \omii{latency overhead distribution (\omii{right}) of \omiii{data} transposition in 16 SIMDRAM operations for \emph{SIMDRAM:1}. \omii{Error bars depict the 25th and 75th percentiles, and a \omiv{bubble} depicts the 50th percentile.}}}
  \label{fig_transposing_latency_transpose}
 \end{figure}

\nasirevi{To better understand the overhead of transposing data, we evaluate the \omiii{worst-case} \geraldorevii{latency of data transposition\omiii{, which is when SIMDRAM's data initially resides in the cache in a horizontal layout.  Before the computation of the SIMDRAM operation can start, this data needs to be transposed to a vertical layout and transferred to DRAM, incurring additional latency.} 
\cmr{Figure}~\ref{fig_transposing_latency_transpose} shows \omiii{this worst-case data} \geraldorevii{transposition latency and the \omii{distribution of latency overhead of} data transposi\omii{tion in \emph{SIMDRAM:1}} across all \omii{16} SIMDRAM operations,} as a function of \omi{element \omii{size}}.  We make \geraldorevii{three} observations. \omii{First}, \omii{in \emph{SIMDRAM:1} (\emph{SIMDRAM:16}),} \omii{data} transposition incurs \geraldorevii{7.1}\% \omii{(44.6\%)} \omii{latency overhead} across all SIMDRAM operations (min.\ \geraldorevii{0.03\% \omii{(0.55\%)} for 64-bit multiplication, max.\ 38.9\% \omii{(91.1\%)} for 8-bit \revonuriii{AND-reduction} and {OR-reduction}}). \omii{\omiii{As shown in \cmr{\Cref{sec_performance}},} for all the evaluated element sizes, \emph{SIMDRAM:1} (\emph{SIMDRAM:16}) outperforms the CPU \omii{and GPU} baselines by 5.5$\times$ and \omii{0.4$\times$} \omii{(88.0$\times$ and 5.8$\times$) on average across all 16 SIMDRAM operations, respectively.} Even \revonuriii{when we include the data} transposition overhead, \emph{SIMDRAM:1} (\emph{SIMDRAM:16}) still outperforms both the CPU and GPU baselines by  
\geraldorevii{4.0$\times$ and \omii{0.24$\times$} \omii{(20.0$\times$ and 1.4$\times$)}} on average across all 16 SIMDRAM operations. }
Our analysis \revonuriii{for kernels that represent the behavior of real-world applications} (\cmr{\Cref{sec_real_world_kernels}}) \omiii{\emph{already includes}} the  data transposition \omii{overhead}. 
\omii{Second}, the data transposi\omii{tion} latency \omii{significantly} increases  \omii{with element} size (\omiii{by} 9.7$\times$ from 8-bit \omii{elements} to 64-bit \omii{elements}). \geraldorevii{\omii{T}he number of cache lines that need to be transposed increases linearly \omii{with element size}, which, in turn, increases the total transposi\omii{tion} latency. Third, even though the \omii{transposition} latency increases with element size, the \omii{transposition} overhead \omi{as a fraction of the total latency} \emph{decreases} \omiii{with} element size, \omii{because} the latency of \omii{each} SIMDRAM operation also increases \omii{with element size}. \omii{Since 
the transposition of data in each subarray is overlapped with the computation in another subarray,} 
the increase in \omii{transposition} latency is amortized over an even higher increase in the SIMDRAM operation latency.} We conclude that SIMDRAM \revonuriii{can} efficiently perform in-DRAM computation even when \omiii{worst-case} data transposition overhead is taken into account.}

}

\subsection{Area Overhead}
\label{sec_area_overhead}

We use \revonur{CACTI}~\cite{cacti} to evaluate the area overhead of the primary components in the SIMDRAM design using a \SI{22}{\nano\meter} technology node. SIMDRAM does not \omi{introduce any modifications} to DRAM circuitry \revonur{other} than those proposed by Ambit\geraldorevii{,} \geraldorevi{which \revonur{has} an area overhead of $<$1\% \revonur{in a commodity} DRAM chip~\revdelrefr{\cite{seshadri2017ambit}}\revdelrefa{\hl{[93]}}}. \revonur{Therefore, SIMDRAM's area overhead over Ambit is only}
two structures in the memory controller: the control and transposition units.


\textbf{Control Unit Area Overhead.} \geraldorevii{The main component\omi{s} in \omi{the} SIMDRAM control unit are the (1)~\emph{bbop} FIFO, (2)~\uprog{} Scratchpad, (3)~\uop{} Memory.} We size the \emph{bbop} FIFO and \uprog{} Scratchpad to \SI{2}{\kilo\byte} each. The size of the \emph{bbop} FIFO is enough to hold up to 1024 \emph{bbop} \omii{instructions}, which we observe is more than enough for our real-world applications. \omi{T}he size of the \uprog{} Scratchpad is large enough to store the \uprog{}s for all 16 SIMDRAM \omii{operations} that we evaluate in this work (16~\uprog{}s $\times$ \SI{128}{\byte} max per \uprog{}). We use a \SI{128}{\byte} scratchpad for the \uop{} Memory.\textsuperscript{\ref{footnote:upgram}} 
We estimate that the \omi{SIMDRAM} control unit area is \SI{0.04}{\milli\meter\squared}. 

 \geraldorevii{\textbf{Transposition Unit Area Overhead.} The primary components in the transposition unit are (1)~the Object Tracker and (2)~two transposition buffers. 
 We use an \SI{8}{\kilo\byte} fully-associative cache with a 64-bit cache line size for the Object Tracker. This is enough to store 1024~entries in the Object Tracker, where each entry holds the base physical address of a SIMDRAM object (19~bits), the total size of the allocated data (32~bits), and the \omii{size} of each element in the object (6~bits). Each transposition buffer is \SI{4}{\kilo\byte}, to transpose up to a 64-bit SIMDRAM object (64-bit $\times$ \SI{64}{\byte}).}  
 \omi{We estimate the transposition unit area is \SI{0.06}{\milli\meter\squared}.} 
\omi{Considering the area of the control and transposition units,
SIMDRAM \omii{has} an area overhead of \omii{only} 0.2\% compared to the die area of an Intel Xeon E5-2697 v3 CPU~\cite{dualitycache}.} 
\omi{W}e conclude that SIMDRAM has low area cost. 

\section{Related Work}


To our knowledge, SIMDRAM is the first end-to-end framework that supports in-DRAM computation flexibly and transparently \omi{to the user}. We highlight SIMDRAM’s key contributions by contrasting it with state-of-the-art \omii{processing-in-memory} designs.

\noindent\textbf{Processing-\omiii{near}-Memory \omiii{(PnM)} \sgii{\omii{w}ithin} 3D-Stacked Memories.} Many recent works (e.g.,~\cite{ahn2015scalable,
nai2017graphpim,
boroumand2018google, 
lazypim, 
top-pim, 
gao2016hrl, 
santos2017operand, 
NIM, 
gu2020ipim, 
lenjani2020fulcrum, 
gokhale1995processing, 
fernandez2020natsa, 
cali2020genasm, 
boroumand2019conda, 
giannoula2021SynCron,
santos2018processing, 
PEI,
hsieh2016accelerating,
hsieh2016transparent,
kim2018grim,
drumond2017mondrian,
gao2017tetris,
Kim2016,
farmahini2015nda,
loh2013processing,
devaux2019true,
vm30,
Polynesia,
Mensa,
deoliveira2021optane, Guo2014, napel, lazypimarx, mohammedprimer}) explore adding logic directly to the logic layer \sgii{of} 3D-stacked memories (e.g., \sgii{High-Bandwidth Memory}~\revdelrefr{\cite{HBM, lee2016simultaneous}}\revdelrefa{\hl{[62\revonurii{, 63}]}}, \sgii{Hybrid Memory Cube}~\revdelrefr{\cite{HMC2}}\revdelrefa{\hl{[49]}}). \omii{T}he implementation of SIMDRAM is considerably simpler, and relies on minimal modifications to \omii{commodity} DRAM \omii{chips}.

\noindent \textbf{Processing-\omiii{using}-Memory (\omiii{PuM}).} Prior works propose mechanisms wherein the memory arrays themselves perform various operations
~\cite{ataberk2021, pluto, wang2020figaro, nom2020, chang2016low, kim2019d, changhpca2018, paternpum, Chi2016, Shafiee2016, seshadri2017ambit, seshadri2019dram, li2017drisa, seshadri2013rowclone, seshadri2016processing, deng2018dracc, xin2020elp2im, song2018graphr, song2017pipelayer, gao2019computedram, eckert2018neural, aga2017compute, dualitycache, ali2019memory, angizi2019graphide, li2018scope, seshadri2017simple, seshadri2015fast, dlugosch2014efficient, subramaniyan2017parallel, angizi2018imce, pinatubo2016, he2020sparse, angizi2019redram, imani2019floatpim, angizi2018dima, vivekbook1, vivekphd}. SIMDRAM supports a much \sgii{wider} range of operations (compared to \revdelrefr{\cite{xin2020elp2im,li2017drisa,deng2018dracc,seshadri2017ambit, seshadri2013rowclone, ali2019memory,angizi2019graphide, li2018scope}}\revdelrefa{\hl{[8, 10, 23, \color{blue}72\color{black}, 73, 97, 98, 111]}}), at lower computational cost (compared to \revdelrefr{\cite{xin2020elp2im,seshadri2017ambit}}\revdelrefa{\hl{[98, 111]}}), \sgii{at lower} area overhead (compared to \revdelrefr{\cite{li2017drisa}}\revdelrefa{\hl{[73]}}), and \sgii{with more} reliable execution (compared to \revdelrefr{\cite{ali2019memory,angizi2019graphide}}\revdelrefa{\hl{[8, 10]}}).



\noindent \textbf{Processing-\omiii{in}-Cache.} \omi{Recent works~\cite{aga2017compute, eckert2018neural, dualitycache} propose} in-SRAM accelerators that take advantage of the SRAM bitline structures to perform bit-serial computation \omi{in caches}.  SIMDRAM shares similarities with these approaches, but offers a significantly lower cost per bit by exploiting the high density and low \omii{cost of} DRAM technology. \omi{We show the large performance and energy advantages of SIMDRAM compared to \omii{DualityCache}~\cite{dualitycache} in \cmr{\Cref{sec:eva:dualitycache}}.}

\noindent \textbf{Frameworks for \omii{PIM}.
} 
\omi{F}ew prior works tackle the challenge of providing end-to-end support for \omii{PIM}. \omi{W}e describe these frameworks and their limitations for in-DRAM computing. \sgii{DualityCache~\cite{dualitycache} is} an end-to-end framework for in-cache computing. DualityCache utilizes \omiii{the} \sgii{CUDA}/OpenAcc programming languages~\omiii{\cite{cheng2014professional, OpenACCA1:online}} to generate code for an in-cache mechanism that execute\geraldorevi{s} a fixed set of operations in a \revsgii{single\omii{-}instruction multiple\omii{-}thread (SIMT)} manner. Like SIMDRAM, DualityCache stores data in a vertical layout through the \revonurii{bitlines} of the SRAM array. It \revonurii{treats} each \revonurii{bitline} as an independent execution thread and utilizes a crossbar network to allow inter-thread communication across \revonurii{bitlines}. Despite its benefits, employing DualityCache \revonurii{in DRAM} is not straightforward for two reasons. 
\omii{First, extending the DRAM subarray with the crossbar network utilized by DualityCache \revonurii{in SRAM} to allow inter-thread communication \omi{would} impose \omiii{a} prohibitive area overhead \omii{in DRAM} (9$\times$ the DRAM subarray area).} 
Second, as an in-cache \revonurii{computing} solution, DualityCache does \revonurii{\emph{not}} account for the limitations of in-DRAM computing, \omiii{i.e.,} \omiv{DRAM
operations that destroy input data, limited number of DRAM rows
that are capable of processing-using-DRAM, and the need to avoid
costly in-DRAM copies.} 
\omi{\omiv{We have already shown that SIMDRAM achieves higher performance at lower area overhead than DualityCache, \omii{when DRAM-to-cache data movement is \omiii{realistically} taken into account} (\cmr{\Cref{sec:eva:dualitycache}}).}}


Two prior works propose frameworks targeting ReRAM devices. \revsgii{Hyper-AP~\revdelrefr{\cite{zha2020hyper}}\revdelrefa{\hl{[112]}} is} a framework for \revsgii{associative processing} using ReRAM. Since Hyper-AP targets \sgii{associative processing}, the proposed framework is \emph{fundamentally} different from \omii{SIMDRAM}. \revsgii{IMP~\revdelrefr{\cite{fujiki2018memory}}\revdelrefa{\hl{[31]}} is} a framework for in-situ ReRAM operations. Like DualityCache, the IMP framework depends on particular structures of the ReRAM array (\revonurii{such} as \revonurii{analog-to-digital/digital-to-analog converters}) to \omi{perform} computation \sgii{and, thus,} \revonurii{is not} applicable to an in-DRAM substrate \revonurii{that performs \revsgii{bulk bitwise} operations}. Moreover, \geraldorevi{DualityCache, Hyper-AP, and IMP} \omi{each} have a rigid ISA that enables \revonurii{only} a \revonurii{limited set} of in-memory operations \revonurii{(DualityCache supports 16 in-memory operations\revsgii{, while} both Hyper-AP and IMP support 12)}. In contrast, SIMDRAM is the first framework \geraldorevi{for \omiii{PuM}} that is flexible, \sgii{providing} a methodology that allows new operations to be \omi{integrated and} computed in memory as needed. \revonurii{In summary}, SIMDRAM fills the gap for a \revonurii{flexible} end-to-end framework \revonurii{that targets} \omii{processing-using-DRAM}. 

\section{Summary \onurtt{and Contributions}}
\label{sec:conclusion_simd}

\omii{We introduce \mech, a massively\omii{-}parallel general-purpose processing-using-DRAM framework that (1)~enables the efficient implementation of \omii{a wide variety of operations in DRAM,} in SIMD fashion, and (2)~provides a flexible mechanism to support the implementation of arbitrary user-defined operations. \mech introduces a \omii{new} three-\omiii{step} framework to enable efficient MAJ/NOT-based in-DRAM implementation for complex operations of different categories (e.g., arithmetic, relational, predication), and is applicable to a wide range of real-world applications.  We design the hardware and
ISA support for SIMDRAM framework to (1)~address key system
integration challenges, and (2)~allow programmers to employ new
SIMDRAM operations without hardware changes.} We 
\omii{experimentally demonstrate} that \mech provides significant \omiii{performance and energy benefits} over \omi{state-of-the-art CPU, GPU, and \omiii{PuM} systems. We hope that future work builds on our framework to \omiii{further} \omii{ease} \omiii{the} adoption \omiii{and improve the performance and efficiency} of processing-using-DRAM architectures and applications.}\\

\noindent \onurtt{In this chapter, we make the following key contributions:\\
\begin{itemize}[noitemsep,topsep=0pt,parsep=0pt,partopsep=0pt,labelindent=0pt,itemindent=0pt,leftmargin=*]
\item We propose the first framework, called SIMDRAM, to enable efficient computation of a flexible \om{set} and wide range of operations in a massively parallel SIMD substrate built via processing-using-DRAM. 
\item We demonstrate that \mech is a three-step framework to develop efficient and reliable MAJ/NOT-based implementations of a wide range of operations.  We design this framework, and add hardware\omiv{, programming,} and ISA support, to (1)~address key system integration challenges and (2)~allow programmers to \omiv{define and} employ new \mech operations without hardware changes.
\item We provide a detailed reference implementation of \mech, including required changes to applications, ISA, and hardware.
\item We evaluate the reliability of \mech under different degrees of process variation and observe that it guarantees correct operation as the DRAM technology scales to smaller node sizes.
\end{itemize}}
\chapter{The Virtual Block Interface}
\label{sec:data-aware}


Virtual memory~\cite{denning1970, atlas, multics, wsdenning, fotheringham1961, kilburn1962} was originally designed for systems whose memory hierarchy fit a simple two-\onurii{level} model of small-but-fast main memory that can be directly accessed via CPU instructions and large-but-slow external storage accessed with the help of the \onur{operating system (OS)}. In such a configuration, the OS can easily abstract away the underlying memory architecture details and present applications with a unified view of memory. 

However, continuing to efficiently support the conventional virtual memory framework requires significant effort due to (1)~high memory demand and diverse memory requirements of modern applications,  \onurthird{(2)~emerging} memory technologies (e.g., DRAM--NVM hybrid memories), and \onurthird{(3)~diverse} system architectures. 
The \onurthird{OS} must now efficiently meet the wide range of application memory requirements that leverage the advantages offered by emerging memory architectures \onur{and new system designs} while simultaneously hiding \onur{the complexity of the underlying memory and system architecture} from the applications. Unfortunately, this is a difficult problem to tackle in a generalized manner. We describe three examples of challenges that arise when adapting conventional virtual memory frameworks to today's \nas{diverse system configurations}.

\textbf{Virtualized Environments.} 
\nas{In a virtual machine,} 
the guest OS performs virtual memory management on the emulated ``physical memory'' while the host OS performs a second round of memory management to map the emulated physical memory to the actual physical memory. This extra level of indirection results in three problems:
(1)~two-dimensional page walks~\cite{vm11, vm25, vm35, vm37, merrifield2016, pham2015},  where the number of memory accesses required to serve a TLB miss increases dramatically
(e.g., up to 24~accesses in \xeightsix with 4-level page tables);
(2)~performance loss in case of miscoordination between the guest and host OS mapping and allocation mechanisms (e.g., \onur{when} the guest supports superpages, but the host does not); and
(3)~inefficiency in virtualizing increasingly complex physical memory architectures \onur{(e.g., hybrid memory systems)} \nas{for the guest OS}. 
These problems worsen with \nas{more page table levels}~\cite{fivelevel}, and in systems that \nas{support} nested virtualization (i.e., a virtual machine running \onurii{inside} another)~\cite{google-nested, azure-nested}.

\textbf{Address Translation.}
In existing virtual memory frameworks, \nas{the} OS manages virtual-to-physical address mapping. However, the hardware must be able to traverse these mappings to handle memory access operations (e.g., TLB lookups). This arrangement requires using \emph{rigid} address-translation structures that are shared between and understood by both the hardware and the OS. 
Prior works show that many applications can benefit from flexible page tables, which cater to the application's actual memory footprint and access patterns\sg{~\cite{vm2,vm19,engler1995, kaashoek1997}}. Unfortunately, enabling such flexibility in conventional virtual memory frameworks requires more complex 
\onurthird{address translation structures \emph{every time} a new address translation} approach is proposed. \onur{For example, a recent work~\cite{vm2} proposes using direct segments to accelerate big-memory \onurthird{applications. 
However,} in order to \onurvi{support direct segments}, the virtual memory contract needs to change to enable the OS to specify which \onurthird{regions of memory are} directly mapped to physical memory.} 
Despite the potential performance benefits, 
\onurthird{this approach is} not easily scalable \nas{to today's increasingly diverse system architectures}.

\textbf{Memory Heterogeneity.}
\nas{Prior works propose many performance-enhancing techniques that require (1)~dynamically \emph{mapping} data to different physical memory regions according to application requirements (e.g., mapping frequently-accessed data to fast memory), 
and (2)~\emph{migrating} data when those requirements change (e.g., \cite{charm, diva-dram, tldram, dynsub, chang.sigmetrics2016, chang2016low, kim2018solar,clrdram,yoon2012,refree, raoux2008, li2017utility,dhiman2009pdram, ramos11,het2,zhang2009exploring,chop})}. Efficiently implementing such functionality faces two challenges. \onurthird{First, a customized data mapping requires 
\onurthird{the OS to be aware of microarchitectural properties of the underlying memory}.} Second, even if this can be achieved, the OS has low visibility \onurii{into} \onurv{rich fine-grain\onur{ed} runtime} memory behavior information \onur{(e.g., access \onurthird{pattern, memory} bandwidth availability)}, especially at the main memory level. 
\nas{While hardware has access to such fine-grained information, \onur{informing} the OS \emph{frequently enough} such that it can react to changes in the memory behavior of an application in a \emph{timely} manner is challenging~\cite{sim14, meswani15, ramos11, tumanov13,banshee}.}

A wide body of research \onurv{(e.g., \cite{vm1,vm2,vm3,karakostas2015,vm5,vm6,pichai2014,vm8,mask,vm9,vm10,vm11,vm12,vm13,vm14,pham2014,vm16,pham2015,vm18,vm19,vm20,vm21,vm22,vm23,vm24,vm25,vm26,vm27,vm28,vm29,vm30,wood1986, teller1990, teller1988, black1989, vm32,vm33, ritchie1985, vm34,vm35,vm36,vm37,vm38,vm39,vm40,vm41,vm42,meswani15,het9,het10,sim14,het12,mitosis-asplos20,elastic-cuckoo-asplos20,meza2013, mondrian, page_overlays, tlbpref, multics, denning1970, wsdenning, atlas, Kavita1994, seshadri2015gather, mondrianthesis, sharedvm, Abhishek2010, Saulsbury2000, Artemiy2019, yale2020, latr, Xingbo2017, Licheng2013, Harsh2020, Zhulin2020, Nadav2017, Artemiy2021, Bogdan2010})} proposes mechanisms to alleviate the overheads of conventional memory allocation and address translation by exploiting specific trends observed in modern systems (e.g., the behavior of emerging applications).
Despite notable improvements, these solutions have two major shortcomings. \onurvi{First, these solutions mainly exploit specific system or workload characteristics and, thus, are applicable to a limited set of problems or applications. Second, \onurthird{each solution requires specialized and not necessarily compatible changes to both the OS and hardware. Therefore, implementing all of these proposals at the same time in a system is a daunting prospect.}}

\textbf{Our goal} in this work is to \onurthird{design} \emph{\onurthird{a general-purpose} alternative virtual memory framework 
\onurthird{that} \onur{naturally supports and better extracts performance from a wide variety of new system configurations}, \onurthird{while still providing} the key features of conventional virtual memory frameworks.} \onurthird{To this end,} we propose the \sysfull (\sys), an alternative approach to memory virtualization that is inspired by the logical block abstraction used by solid-state drives to hide the underlying device details from the rest of the system. In a similar way, we envision the memory controller as the primary provider of an abstract interface that hides the details of the underlying physical memory architecture, \onurthird{including the physical addresses of the memory locations}. 

\onurthird{VBI is based on three guiding principles. First, \emph{programs should be allowed to choose the \emph{size} of their virtual address space}, to \onurii{mitigate translation} overheads associated with very large virtual address spaces. Second, \emph{address translation should be decoupled from memory protection}, since they are logically separate and need not be managed at the same granularity by the same structures.
Third, \emph{\sgii{software should be allowed} to communicate semantic information about application data to the hardware}, so that the hardware can \sgii{more intelligently manage} the underlying hardware resources.}

\onurthird{\sys introduces a \emph{globally-visible} address space called the \emph{\sys Address Space}, that consists of a large set of \emph{virtual blocks (VBs)} of different sizes. 
\nasi{For any semantically meaningful
unit of information (e.g., a data structure, a shared library)}, the program can choose a VB of appropriate size, and tag the VB with properties that describe the contents of the VB.
\textbf{The key idea} of \sys is to delegate physical memory allocation and address translation to a hardware-based \mtlfull (MTL) at the memory controller. This idea is enabled by the fact that the globally-visible \sys address space provides \sys with system-wide unique \emph{\sys addresses} that can be \emph{directly} used by on-chip caches without requiring address translation. In \sys, the OS no longer needs to manage address translation and memory allocation for the physical memory devices. Instead, the OS (1)~retains full control over access protection by controlling which programs have access to which virtual blocks, and (2)~uses VB properties to communicate the data's memory requirements (e.g., latency sensitivity) and characteristics (e.g., access pattern) to the memory controller.}

\begin{figure}[h]
    \centering
    \begin{subfigure}[b]{0.455\columnwidth}
      \centering
      \includegraphics[width=\textwidth, trim=0 35 266 0, clip]{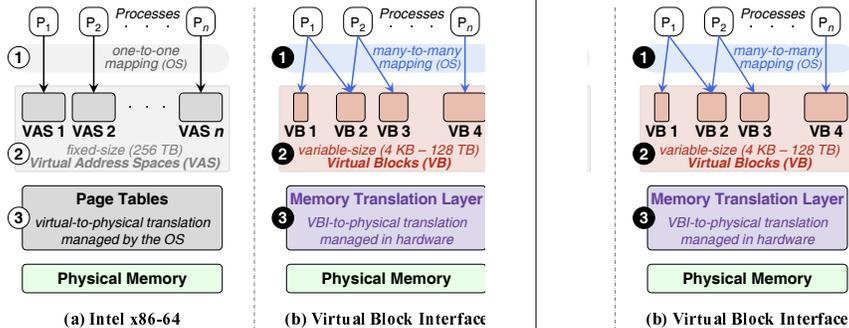}
      \caption{Intel x86-64}
    \end{subfigure}%
    \hfill\vline\hfill%
    \begin{subfigure}[b]{0.455\columnwidth}
      \centering
      \includegraphics[width=\textwidth, trim=266 35 0 0, clip]{figures/vbi-intro.pdf}
      \caption{Virtual Block Interface}
    \end{subfigure}%
    \caption{Virtual memory management in \xeightsix and in \sys.}
    \label{fig:vbi-intro}
\end{figure}

Figure~\ref{fig:vbi-intro} illustrates the differences between virtual memory management in \onurii{state-of-the-art production Intel \xeightsix systems and in} \sys. In \xeightsix (Figure~\ref{fig:vbi-intro}a), the OS manages a single private virtual address space (VAS) for each process (\inhollowcircle{1}), providing each process with a fixed-size 256 TB VAS irrespective of the actual memory requirements of the process (\inhollowcircle{2}). The OS uses a set of page tables\onurii{, one per process}, to define how each VAS maps to physical memory (\inhollowcircle{3}). In contrast, \sys (Figure~\ref{fig:vbi-intro}b) \onurvi{makes} \emph{all} virtual blocks (VBs) \onurvi{visible} to \emph{all} processes, and the OS controls which \onurvi{processes can access which VBs} (\incircle{1}). Therefore, a process' total virtual address space is defined by which VBs are attached to it, i.e., by the process' actual memory needs (\incircle{2}). In \sys, the \mtl has full control over mapping of data from each VB to physical memory, invisibly to the system software (\incircle{3}).

\sys seamlessly and efficiently supports important optimizations that improve overall system performance, including: (1)~enabling benefits akin to using virtually-indexed virtually-tagged (VIVT) caches (e.g., reduced address translation overhead), (2)~eliminating two-dimensional page table walks in virtual machine environments, (3)~delaying physical memory allocation until the first dirty last-level cache line eviction, and (4)~flexibly supporting different virtual-to-physical address translation structures for different memory regions. \pointer{\Cref{sec:optimizations}} describes these optimizations in detail.

We evaluate VBI for two important and emerging use-cases. First, we demonstrate that 
\sys significantly reduces the address translation overhead both for \emph{natively-running programs} and for programs running inside a virtual machine (\emph{VM programs}). Quantitative evaluations using workloads from SPEC CPU 2006~\cite{spec2006}, SPEC CPU 2017~\cite{spec2017}, TailBench~\cite{tailbench}, and Graph 500~\cite{graph500} show that 
\onurii{a simplified version of \sys that maps VBs using 4~KB granularity only} improves the performance of native programs by 2.18$\times$ and VM programs by 3.8$\times$. Even when \onurii{enabling support for large pages} for \emph{all data}, which significantly lowers translation overheads, \sys improves performance by 77\% for native programs and 89\% for VM programs. Second, we demonstrate that \sys significantly improves the performance of heterogeneous memory architectures by evaluating two 
\onurvi{heterogeneous} memory systems (PCM--DRAM~\cite{ramos11} and Tiered-Latency-DRAM~\cite{tldram}). We show that \sys, by intelligently mapping frequently-accessed data to the low-latency region of memory, improves overall performance of these two systems by 33\% and 21\% respectively, compared to systems that employ a heterogeneity-unaware data mapping scheme. \pointer{\Cref{sec:methodology2}} describes our methodology, results, and insights from these evaluations.

\section{Design Principles}
\label{sec:design-principles}



To \onurii{minimize} performance and complexity overheads \onurii{of memory virtualization}, our virtual memory framework is grounded on three key design principles.


\paragraph{Appropriately-Sized Virtual Address Spaces.}
The virtual memory framework should \emph{allow each application to have control over the size of its virtual address space}. The majority of applications far underutilize the large virtual address space offered by modern architectures (e.g., 256~TB in Intel \xeightsix). Even \onurii{demanding} applications such as databases~\cite{db1,db2,db3,db4,db5,db6} and caching servers~\cite{memcache,memcached1} are cognizant of the amount of available physical memory and of the size of virtual memory they need. Unfortunately, a larger virtual address space results in 
%
\onurii{larger or }deeper page tables (i.e., page tables with more levels). \onurii{A larger page table increases TLB contention, while a deeper page table requires a greater number of page table accesses to retrieve the physical address for each TLB miss.  In both cases, the address translation overhead increases.} 
Therefore, \onurthird{allowing applications to choose an appropriately-sized virtual address space based on their actual needs,} avoids the higher translation overheads associated with a larger address space.

\paragraph{Decoupling Address Translation from Access Protection.}
The virtual memory framework should \emph{decouple address translation from access protection checks}, as the two have inherently different characteristics. While address translation is typically performed at page granularity, protection information is typically the same for an entire data structure, which can span multiple pages. Moreover, protection information is purely a function of the virtual address, and does not require address translation. However, existing systems store both translation and protection information for each virtual page as part of the page table. Decoupling address translation from protection checking can enable opportunities to remove address translation from the critical path of an access protection check, deferring the translation until physical memory \mhp{\emph{must}} be accessed, thereby lowering the performance overheads of virtual memory.

\paragraph{Better Partitioning of Duties Between Software and Hardware.}
The virtual memory framework should \emph{allow software to easily communicate semantic information about application data to hardware and allow hardware to manage the \onurthird{physical} memory resources}. Different pieces of program data have different performance characteristics (latency, bandwidth, and parallelism), and have other inherent properties (e.g., compressibility, persistence) at the software level. As highlighted by recent work~\cite{xmem,vijaykumar2018}, while software is aware of this semantic information, the hardware is privy to \mhp{fine-grained} dynamic runtime information \mhp{(e.g., memory access} behavior, phase changes, \mhp{memory} bandwidth availability) that can enable vastly more intelligent management of the underlying hardware resources (e.g., better data mapping, migration, and scheduling decisions). Therefore, conveying semantic information to the hardware (\onurii{i.e., memory controller)} that manages the \mhp{physical} memory resources can enable a host of new optimization opportunities. 

\section{Virtual Block Interface\mhp{:} Overview}
\label{sec:overview}

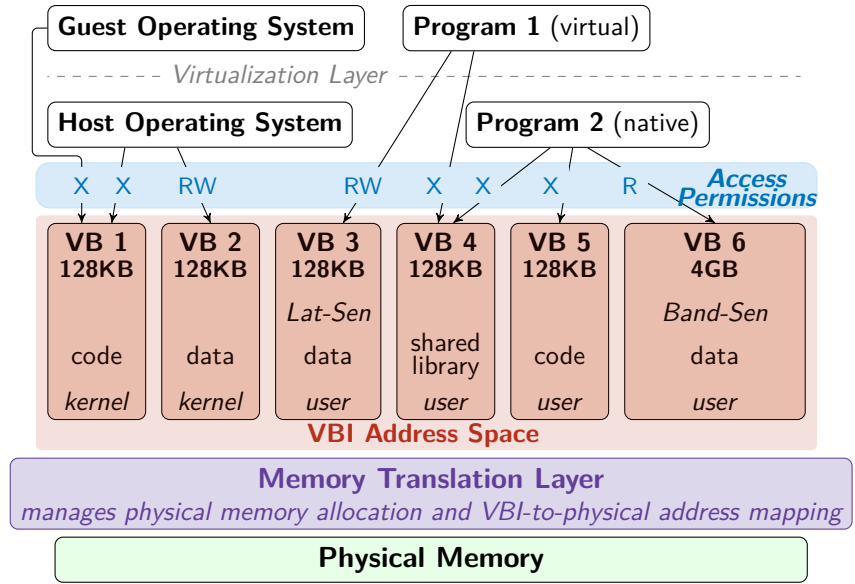
\begin{figure}[h]
    \centering
    \scalebox{1}{\begin{tikzpicture}[>=stealth',font=\small\sffamily]

  \tikzset{block/.style={draw,rounded corners=3pt,minimum height=0.6cm,align=center}};
  \tikzset{acllabel/.style={fill=CornflowerBlue!20,text=RoyalBlue,font=\footnotesize\sffamily}};
  \tikzset{acl/.style={->,rounded corners=2pt}};
  \tikzset{mmap/.style={->,rounded corners=2pt}};
  \tikzset{vb/.style={draw,rounded corners=3pt,minimum height=1.2cm,minimum width=1.3cm,align=center,fill=BrickRed!15}};
  \tikzset{client/.style={draw,rounded corners=3pt,minimum height=1cm,align=center,fill=black!20}};
  
  \node (mml) [block, minimum width=10cm,minimum height=0.8cm,RoyalPurple, fill=RoyalPurple!15] {\normalsize\textbf{Memory Translation Layer}\\\emph{manages physical memory allocation and VBI-to-physical address mapping\vspace{-2pt}}};
  
  \node (pmem1) [block, minimum width=10cm,minimum height=0.6cm, fill=green!10, yshift=-4mm] at (mml.south) {\normalsize\textbf{Physical Memory}};

  \node (vb1) at (mml.north west) [anchor=south west,yshift=5mm,xshift=5mm,vb] {\textbf{VB 1}\\[-0.5ex]\textbf{\footnotesize{128KB}}\\[1ex]\\[1ex]code\\[-1.5ex]\\\emph{kernel}};
  \node (vb2) at (vb1.east) [anchor=west,xshift=2mm,vb] {\textbf{VB 2}\\[-0.5ex]\textbf{\footnotesize{128KB}}\\[1ex]\\[1ex]data\\[-1.5ex]\\\emph{kernel}};
  \node (vb4) at (vb2.east) [anchor=west,xshift=2mm,minimum width=1.75cm,vb] {\textbf{VB 3}\\[-0.5ex]\textbf{\footnotesize{128KB}}\\[1ex]\emph{Lat-Sen}\\[1ex]data\\[-1.5ex]\\\emph{user}};
  \node (vb5) at (vb4.east) [anchor=west,xshift=2mm,vb] {\textbf{VB 4}\\[-0.5ex]\textbf{\footnotesize{128KB}}\\[0.9ex]\\shared\\[-0.8ex]library\\[0.4ex]\emph{user}};
  \node (vb6) at (vb5.east) [anchor=west,xshift=2mm,vb] {\textbf{VB 5}\\[-0.5ex]\textbf{\footnotesize{128KB}}\\[1ex]\\[1ex]code\\[-1.5ex]\\\emph{user}};
  \node (vb7) at (vb6.east) [anchor=west,xshift=2mm,vb,minimum width=2.4cm] {\textbf{VB 6}\\[-0.5ex]\textbf{\footnotesize{4GB}}\\[1ex]\emph{Band-Sen}\\[1ex]data\\[-1.5ex]\\\emph{user}};
  
       \node at (vb5.south) [yshift=-2mm, xshift=-3mm,text=BrickRed]{\textbf{VBI Address Space}};
  
  \node (hostos) at (vb1.north west) [anchor=south west,minimum width=4cm,yshift=10mm,block] {\textbf{Host Operating System}};
  \node (p1) at (hostos.east) [anchor=west,xshift=15mm,block] {\textbf{Program 2} (native)};

  \node (vl) at (hostos.north west) [anchor=south west,yshift=2mm,minimum width=10cm,align=center] {};
  \node (vllabel) at (vl.west) [anchor=west,xshift=15mm,black!50] {\emph{Virtualization Layer}};
  \draw [dashed,black!50] (vllabel.west) -- (vl.west);
  \draw [dashed,black!50] (vllabel.east) -- (vl.east);

  \node (guestos) at (vl.north west) [anchor=south west,yshift=2mm,minimum width=4cm,block] {\textbf{Guest Operating System}};
  \node (p2) at (guestos.east) [anchor=west,xshift=5mm,block] {\textbf{Program 1} (virtual)};
  
  \draw [rounded corners=5pt,color=CornflowerBlue!30,fill=CornflowerBlue!20] ([yshift=2mm,xshift=-1.5mm]vb1.north west) rectangle ([yshift=8mm,xshift=1.5mm]vb7.north east);

     \fill [BrickRed,fill opacity=0.15,rounded corners=2pt] ([yshift=1mm,xshift=-1.5mm]vb1.north west) rectangle ([yshift=-4mm,xshift=1.5mm]vb7.south east);

  \draw [acl] ([xshift=-10mm]hostos.south) -- ([xshift=2mm]vb1.north);
  \draw [acl] ([xshift=-3mm]hostos.south) -- ([xshift=0mm]vb2.north);
  \draw [acl] (guestos.west) -- ++(-2mm,0) -- ++(0,-1.7cm) -| ([xshift=-2mm]vb1.north);

  \draw [acl] (p1.south) -- (vb7.north);
  \draw [acl] ([xshift=-2mm]p1.south) -- (vb6.north);
  \draw [acl] ([xshift=-5mm]p1.south) -- ([xshift=1mm]vb5.north);

  \draw [acl] ([xshift=-10mm]p2.south) -- ([xshift=2mm]vb4.north);
  \draw [acl] ([xshift=-7mm]p2.south) -- ([xshift=-1mm]vb5.north);



  \node (l1) [acllabel] at ([yshift=5mm,xshift=-2mm]vb1.north) {X};
  \node (l2) [acllabel,anchor=west] at ([xshift=0.7mm]l1.east) {X};
  \node (l2) [acllabel,anchor=west] at ([xshift=9mm]l1.east) {RW};
  \node (l3) [acllabel,anchor=west] at ([xshift=31mm]l1.east) {RW};
  \node (l4) [acllabel,anchor=west] at ([xshift=42mm]l1.east) {X};
  \node (l4) [acllabel,anchor=west] at ([xshift=48.5mm]l1.east) {X};
  \node (l4) [acllabel,anchor=west] at ([xshift=57.5mm]l1.east) {X};
  \node (l4) [acllabel,anchor=west] at ([xshift=68mm]l1.east) {R};
  
    \node (label) at (l1.east) [yshift=1mm,xshift=86mm,text=RoyalBlue,align=center]{\textbf{\emph{Access}}};
    \node at (label.south) [text=RoyalBlue,align=center]{\textbf{\emph{Permissions}}};

\end{tikzpicture}

}
    \caption{\onur{Overview of \sys. \emph{Lat-Sen} and \emph{Band-Sen} represent latency-sensitive and bandwidth-sensitive,
respectively.}}
    \label{fig:design-overview}
\end{figure}

\nas{Figure~\ref{fig:design-overview} shows an overview of \sys.} There are three major aspects of the \sys design: (1)~the \mhp{\sys address space}, (2)~\sys access permissions, and (3)~the \mtlfull. We \mhp{first} describe these aspects in detail (\pointer{\Cref{sec:add-space}}--\pointer{\Cref{sec:mtl-overview}}). Next, \nasi{we explain the implementation of key OS functionalities in \sys}~\mhp{(\pointer{\Cref{sec:os-functions}}). Finally, we discuss some of the \mhp{key} optimizations that \sys enables \mhp{(\pointer{\Cref{sec:optimizations}})}}.

\subsection{\sys Address Space}
\label{sec:add-space}

Unlike most existing architectures wherein each process has its own virtual address space, virtual memory in \sys is a single, \onur{globally-visible} address space called the \emph{\sys Address Space}. As shown in Figure~\ref{fig:design-overview}, the \sys Address Space consists of a finite set of \emph{Virtual Blocks} (VBs). Each VB is a contiguous region of VBI address space that does not overlap with any other VB. \nasi{Each VB contains a semantically meaningful
unit of information (e.g., a data structure, a shared library)} and is associated with (1)~a system-wide unique ID, (2)~a specific size (chosen from a set of pre-determined size classes), and (3)~a set of properties that specify the semantics of the content of the VB and its desired characteristics. For example, in the figure, \onur{\texttt{VB~1} indicates the VB with ID 1}; its size is 128~KB, and it contains code that is accessible only to the kernel. On the other hand, \onur{\texttt{VB~6} is the VB} with ID 6; its size is 4~GB, and it contains data that is bandwidth-sensitive. 
In contrast to conventional systems, where the mapping from the process' \nasi{virtual-to-physical address space} is stored in a per-process page table~\cite{intelx86manual}, \sys maintains the VBI-to-physical address mapping information of each VB in a \emph{separate} translation structure. This approach enables \sys to flexibly tune the type of translation structure for each VB to the characteristics of the VB~(as described in \Cref{sec:flexible-translations}). \sys stores the above information and a pointer to the translation structure of each VB
in a set of \emph{VB Info Tables} (VITs; described in \pointer{\Cref{sec:vit}}).

\subsection{\sys Access \onurii{Permissions}}
\label{sec:access-semantics}

As the \sys Address Space is global, all VBs in the system are \emph{visible} to all processes. However, a program can \emph{access} data within a VB \emph{only if} it is attached to the VB with appropriate permissions. In Figure~\ref{fig:design-overview}, \texttt{Program 2} can only execute from \texttt{VB 4} or \texttt{VB 5}, only read from \texttt{VB 6}, and cannot access \texttt{VB 3} at all; \texttt{Program 1} and \texttt{Program 2} both share \texttt{VB 4}. For each process, \sys maintains information about the set of VBs attached to the process in an \onurthird{OS-managed} \onurvi{per-process} table called \mhp{the} \emph{Client--VB Table} (CVT) (described in \pointer{\Cref{sec:client}}). \sys provides the OS with a set of instructions with which the OS can control which processes have what type of access permissions to which VBs. On each memory access, the processor checks the CVT to ensure that the program has the necessary permission to perform the access. With this approach, \sys \onur{\emph{decouples protection checks from address translation}}, which allows it to defer the address translation to the memory controller where the physical address is required to access \mhp{main} memory.

\subsection{\mtlfull}
\label{sec:mtl-overview}

In \sys, to access a piece of data, a program must specify the ID of the VB that contains the data and the offset of the data within the VB. Since the \onurvi{ID of the VB} is unique system-wide, \onurvi{the} combination of \onurvi{the ID} and offset points to the address of a specific byte of data in the \sys address space. We call this address the \emph{\sys address}. As the \sys address space is globally visible, similar to the physical address in existing architectures, the \sys address points to a unique piece of data in the system. As a result, \sys uses the VBI address \emph{directly} (i.e., without requiring address translation) to locate data within the on-chip caches without worrying about the complexity of homonyms and synonyms~\cite{cekleov1997a, cekleov1997b, jacob1998}\onurthird{, which cannot exist in VBI (see \pointer{\Cref{sec:optimizations}})}. Address translation is required only when an access misses in all levels of on-chip caches.

\onurthird{To perform address translation,} \sys uses the \mtlfull (\mtl). The \mtl, \onur{implemented in the memory controller with an interface to the system software,} manages both allocation of physical memory to VBs  
and VBI-to-physical address translation \onurvi{(relieving the OS of these duties)}. 
\onurii{Memory-controller-based memory management} enables a number of performance optimizations \onur{(e.g., avoiding 2D page walks in virtual machines, flexible address translation structures)}, which we describe in \pointer{\Cref{sec:optimizations}}.

\subsection{Implementing Key OS Functionalities}
\label{sec:os-functions}

\onur{\sys allows the system to efficiently implement existing OS functionalities. In this section, we describe five key functionalities \onurthird{and how VBI enables them}}.

\paragraph{Physical Memory Capacity Management.} In \sys, the \mtl allocates physical memory for VBs as and when required. 
To handle situations when the \mtl runs out of physical memory, \sys provides two system calls that allow the \mtl to move data from physical memory to the backing store and vice versa. The \mtl maintains information about swapped-out data as part of the VB's translation structures.

\paragraph{Data Protection.} The goal of data protection is to prevent a malicious program from accessing kernel data or private data of other programs. In \sys, the OS ensures such protection by appropriately setting the permissions with which each process can access different VBs. Before each \sgii{memory access}, the CPU checks if the executing thread has appropriate access permissions to the corresponding VB (\pointer{\Cref{sec:load}}).

\paragraph{Inter-Process Data Sharing (True Sharing).} When two processes share data (e.g., via pipes), both processes have a coherent view of the shared memory, i.e., modifications made by one process should be visible to \onur{the other} process. In \sys, the OS supports such \emph{true} sharing by granting both processes permission to access the VB containing the shared data.

\paragraph{Data Deduplication (Copy-on-Write Sharing).} In most modern systems, the OS reduces redundancy in physical memory by mapping virtual pages \onurv{containing} the \emph{same} data to the same physical page. On a write to one of the virtual pages, the OS copies the data 
\onurv{to a new physical page,} and remaps the written virtual page to the new physical page before performing the write.  
In \sys, the \mtl performs data deduplication when a VB is cloned by sharing both translation structures and data pages between the two VBs (\pointer{\Cref{sec:pop-interaction}}), and using the copy-on-write mechanism to ensure consistency.  

\paragraph{Memory-Mapped Files.} 
\nasi{To support memory-mapped files, existing systems map a region of the virtual address space to a file in storage, and loads/stores to that region are} \onur{used}  \onur{to access/update the file content.} 
\sys naturally supports memory-mapped files as the OS simply associates the file to a VB of appropriate size. An offset within the VB maps to the same offset within the file. The \mtl uses the same system calls \onur{used to manage physical memory capacity (described under \emph{Physical Memory Capacity Management} above)} to move data between the VB in memory and the file in storage.

\subsection{ Optimizations Supported by \sys}
\label{sec:optimizations}

In this section, we describe four \onurthird{key} optimizations that the \sys design enables.

\paragraph{Virtually-Indexed Virtually-Tagged Caches.}
Using \onurii{fully-virtual (i.e., VIVT) caches} enables the system to delay address translation and \onurii{reduce accesses to translation structures} such as the \onurii{TLBs}. However, most modern architectures do not support VIVT caches due to \onur{two main reasons. First, }handling homonyms (i.e., where the same virtual address maps to multiple physical addresses) and synonyms (i.e., where multiple virtual addresses map to the same physical address) \onur{introduces complexity to the system}~\cite{cekleov1997a, cekleov1997b, jacob1998}. \onur{Second, although address translation is not required to access VIVT caches, the access permission check required prior to the cache access still necessitates accessing the TLB and can induce a page table walk on a TLB miss. This is due to the fact that the protection bits are stored as part of the page table entry for each page in current systems.} VBI avoids both of these problems. 

First, VBI addresses are unique system-wide, eliminating the possibility of homonyms. Furthermore, since VBs do not overlap, each VBI address appears in \emph{at most one} VB, \onurii{avoiding the possibility of} synonyms. In case of true sharing (\pointer{\Cref{sec:os-functions}}), different processes are attached to the same VB. Therefore, the VBI address that each process uses to access the shared region refers to the \emph{same} VB. In case of copy-on-write sharing, where the MTL may map two VBI addresses to the same physical memory for deduplication, the MTL creates a new copy of the data before any write to either address. \onurii{Thus, neither form of sharing can lead to synonyms.} As a result, by using VBI addresses directly to access on-chip caches, \sys achieves benefits akin to VIVT caches without the complexity of dealing with synonyms and homonyms. Additionally, since the VBI address acts as a system-wide single point of reference for the data that it refers to, all coherence-related requests can use VBI addresses without introducing any ambiguity.

Second, \sys decouples protection checks from address translation, by storing protection and address translation information in \emph{separate} sets of tables and delegating access permission management to the OS\onurii{, avoiding the need to access translation structures for protection purposes (as done in existing systems).}

\paragraph{Avoiding 2D Page Walks in Virtual Machines.}
In \sys, once a process inside a VM attaches itself to a VB (with the help of the host and guest OSes), any memory access \onur{from the VM} directly 
\onurvi{uses} a VBI address. As described in \pointer{\Cref{sec:mtl-overview}}, this address is directly used to address the on-chip caches. In case of an LLC miss, the \mtl translates the VBI address to physical address. As a result, unlike existing systems, \onurthird{address translation for a VM under VBI is no different from that for a host}, enabling significant performance improvements. We expect these benefits to further increase in systems supporting nested virtualization~\cite{google-nested,azure-nested}. \pointer{\Cref{sec:virtual-machines}} discusses the implementation of \onur{\sys in virtualized environments}.

\paragraph{Delayed Physical Memory Allocation.}
As \sys uses VBI addresses to access all on-chip caches, it is no longer necessary for a cache line to be backed by physical memory \emph{before} it can be accessed. This enables the opportunity to delay physical memory allocation for a VB (or a region of a VB) until a dirty cache line from the VB is evicted from the last-level cache. Delayed allocation has three benefits. First, the allocation process is removed from the critical path of execution, as cache line evictions are not on the critical path. Second, for VBs that never leave the cache during the lifetime of the VB (\onur{likely more} common with growing cache sizes in modern hardware), \sys avoids physical memory allocation altogether. Third, \nas{when using delayed physical memory allocation, }for an access to a region with no physical memory allocated yet, \sys simply returns a zero cache line, thereby avoiding \emph{both} address translation and a main memory access, which improves performance. \pointer{\Cref{sec:delayed-allocation}} describes the implementation of \onur{delayed physical memory allocation} in \sys.

\paragraph{Flexible \onur{Address} Translation Structures.}
A recent work~\cite{vm19} shows that different data structures benefit from different types of \onur{address} translation structures depending on their data layout and access patterns. However, since in \onurv{conventional virtual memory, the} hardware needs to read the OS-managed page tables to perform page table walks, the structure of the page table \onur{needs to be understood by both the hardware and OS, thereby limiting the flexibility of the page table structure.} 
In contrast, in \sys, the \mtl is the \emph{only} component that manages and accesses translation structures. Therefore, the constraint of sharing address translation structures with the  OS is relaxed, providing \sys with more flexibility in employing different types of translation structures \mhp{in the MTL}. Accordingly, \sys \onurvi{maintains a separate translation structure for each VB, and} can tune 
\onurvi{it} to suit the properties of the VB \onur{(e.g., multi-level tables for large VBs or those with many sparsely-allocated regions, and single-level tables for small VBs or those with many \nas{large} contiguously-allocated regions)}. This optimization reduces the number of memory accesses necessary to serve a TLB miss.

\section{\sys: Detailed Design}
\label{sec:detailed}

In this section, we present the detailed design and a reference implementation of the \sysfull. \onurthird{We describe (1)}~the components architecturally exposed by \sys to the rest of the system (\pointer{\Cref{sec:arch-interface}})\onurthird{, (2)}~the life-cycle of \onurii{allocated memory} (\pointer{\Cref{sec:ds-life-cycle}})\onurthird{, (3)}~the interactions between the processor, OS, and the process in \sys (\pointer{\Cref{sec:pop-interaction}})\onurthird{, and (4)}~the operation of the \mtlfull in detail (\pointer{\Cref{sec:mtl}}). 

\subsection{Architectural Components}
\label{sec:arch-interface}

\sys exposes two architectural components to the rest of the system that form the contract between hardware and software: (1)~\onurii{virtual blocks}, and (2)~\emph{memory clients}.

\subsubsection{\onurii{Virtual  Blocks (VBs)}}
\label{sec:vb-detail}

\begin{figure}[h]
  \centering
  \begin{tikzpicture}[inner sep=1pt,outer sep=0pt,>=stealth']

\draw [rounded corners=2pt] (0,0) rectangle (8,0.7);
\draw (4.4,0) -- ++(0,0.7);
\draw[dotted] (1.4,0) -- ++(0,0.7);

\node at (2.2, 0.5) {\small\texttt{VBUID}};
\node at (0.7, 0.2) {\footnotesize\color{Gray}\texttt{\emph{SizeID}}};
\node at (2.8, 0.2) {\footnotesize\color{Gray}\texttt{\emph{VBID}}};
\node at (6.2, 0.35) {\small\texttt{offset}};

\draw[->] (1.75, 0.5) -- ++ (-1.7,0);
\draw[->] (2.65, 0.5) -- ++ (1.7,0);

\end{tikzpicture}
  \caption{Components of a VBI address.}
  \label{fig:vbi-address}
\end{figure}
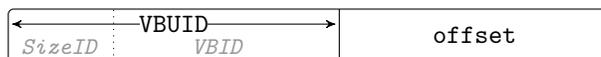

The \mhp{\sys address space} in \sys is characterized by three parameters: (1)~the size of the address space, which is determined by the bit width of the processor's address bus (64 in our implementation); (2)~the number of VB size classes (8 in our implementation); and (3)~the list of size classes (4~KB, 128~KB, 4~MB, 128~MB, 4~GB, 128~GB, 4~TB, and 128~TB). \nas{Each size class in \sys is associated with an ID (\texttt{SizeID}), and each VB is assigned an ID \emph{within its size class} (\texttt{VBID}).
Every VB is identified system-wide by its \emph{VBI unique ID} (\texttt{VBUID}), which is the concatenation of \texttt{SizeID} and \texttt{VBID}.
As shown in Figure~\ref{fig:vbi-address}, \sys constructs a \mhp{\emph{\sys address}} 
 using two components:
(1)~\texttt{VBUID}, and (2)~the offset of the addressed data within the VB.
In our implementation, \texttt{SizeID} uses three bits to represent each of our eight possible size classes.
The remaining address bits are split between 
\texttt{VBID} and the
offset. The precise number of bits required for the offset is determined by the size of the VB, and the remaining bits are used for \texttt{VBID}.}
For example, the 4~KB size class in our implementation uses 12~bits for the offset, leaving 49~bits for \texttt{VBID}, i.e., 2$^{49}$ VBs of size 4~KB. In contrast, the 128~TB size class uses 47 bits for the offset, leaving 14~bits for \texttt{VBID}, i.e., 2$^{14}$ VBs of size 128~TB.

As \pointer{\Cref{sec:overview}} describes, \sys associates each VB with a set of flags that characterize the contents of the VB (e.g, \texttt{code}, \texttt{read-only}, \texttt{kernel}, \texttt{compressible}, \texttt{persistent}). In addition to these flags,  software may also provide hints to describe the memory behavior of the data that the VB contains (e.g., latency sensitivity, bandwidth sensitivity, \onurii{compressibility, error tolerance}). Prior work extensively studies a set of useful properties\onurv{~\cite{xmem,vijaykumar2018,yixin2014,yu2017}}. 
\mhp{Software} specifies these properties via a bitvector \mhp{that is defined as part of the ISA specification}. \sys maintains the flags and the \mhp{software-provided} hints as a \emph{property bitvector}.

For each VB \onurvi{in the system}, \sys stores (1)~\mhp{an \emph{enable} bit to describe whether the VB is currently assigned to any process}, (2)~the property bitvector, (3)~the number of processes attached to the VB (i.e., a reference count), \nas{(4)~the type of VBI-to-physical address translation structure being used for the VB, and} (5)~a pointer to the VB's address translation structure. All \onurv{of this information is} stored as \onurv{an entry in} the VB Info Tables (\pointer{\Cref{sec:vit}}).

\subsubsection{Memory Clients}
\label{sec:client}

Similar to \sg{address space identifiers~\cite{ahearn1973}} in existing architectures, \sys introduces the notion of \emph{memory client} to communicate the concept of a process in \sys. \nasi{A memory client refers to any entity that needs to allocate and use memory, such as the OS itself, and any process running on the system (natively or inside a virtual machine).} In order to track the
permissions with which a client can access different VBs,
each client in VBI is assigned a unique ID to identify
the client system-wide. During execution, \sys tags each core with the client ID of the process currently running on it.

As \pointer{\Cref{sec:overview}} discusses, the set of VBs that a client can access and their associated permissions are stored in a per-client table called the \emph{Client--VB Table} (CVT). Each entry in the CVT contains 
(1)~a valid bit, 
(2)~\texttt{VBUID} of the VB, and
(3)~a three-bit field representing the read-write-execute permissions (RWX) with which the client can access that VB. For each memory access, the processor checks the CVT to ensure that the client has appropriate access to the VB. The OS implicitly manages the CVTs using the following two new instructions:

\begin{figure}[h!]\small
  \centering
  \begin{tikzpicture}
    \node (attach) [draw,rounded corners=2pt,inner sep=5pt] {\tt \textcolor{blue}{\attachvb} CID, VBUID, RWX};
    \node (detach) at (attach.east) [draw,rounded corners=2pt,inner sep=5pt,anchor=west,xshift=8mm] {\tt \textcolor{blue}{\detachvb} CID, VBUID};
  \end{tikzpicture}
\end{figure}

The \attachvb instruction adds an entry for VB \texttt{VBUID} in the CVT of client \texttt{CID} with the specified \texttt{RWX} permissions (either by replacing an invalid entry in the CVT, or being inserted at the end of the CVT). This instruction returns the index of the CVT entry to the OS \onurii{and increments the reference count of the VB (stored in the VIT entry of the VB; see \pointer{\Cref{sec:vit}})}. The \detachvb instruction resets the valid bit of the entry corresponding to VB \texttt{VBUID} in the CVT of client \texttt{CID} and decrements the reference count of the VB.

The processor maintains the location and size of the CVT for each client in a reserved region of physical memory. As clients are visible to both the hardware and the software, the number of clients is an architectural parameter determined at design time and exposed to the OS. In our implementation, we use 16-bit client IDs (supporting 2$^{16}$ clients).

\subsection{Life Cycle of \nas{Allocated Memory}}
\label{sec:ds-life-cycle}

\begin{figure*}[t]
  \centering
  \includegraphics[width=\textwidth]{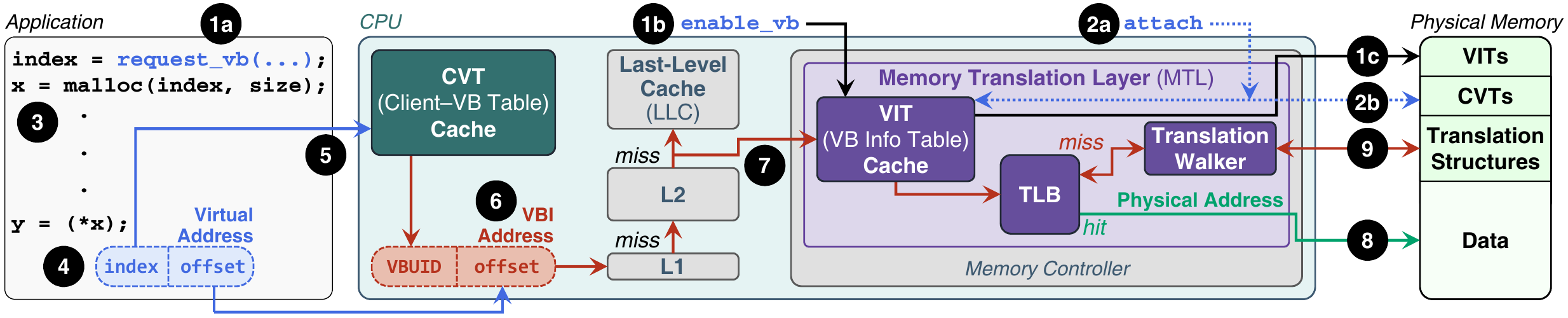}
  \caption{Reference microarchitectural implementation of the Virtual Block Interface.}
  \label{fig:vbi-uarch}
\end{figure*}

In this section, we describe the phases in the life cycle of \nas{dynamically-allocated memory}: memory allocation, address specification, data access, and deallocation. Figure~\ref{fig:vbi-uarch} shows this flow in detail, including the hardware components that aid \sys in efficiently executing memory operations. In \pointer{\Cref{sec:pop-interaction}}, we discuss how VBI \nas{manages 
code}, shared libraries, \onurii{static data}, and the life cycle of an entire process.

\nas{When} a program needs to allocate memory for a new data structure, it first requests a new VB from the OS. For this purpose, we introduce a new system call, \requestvas. The program invokes \requestvas with two parameters: (1)~the \emph{expected} size of the data structure, and (2)~a bitvector of the desired properties for the data structure \sg{(\incircle{1a} in Figure~\ref{fig:vbi-uarch})}.

In response, the OS first scans the VB Info Table to identify the smallest free VB that can accommodate the data structure. 
\sg{The OS then uses the \enablevb instruction (\incircle{1b}) to inform the MTL that the VB is now \nas{enabled}}.
The \enablevb instruction takes the \texttt{VBUID} of the VB to be enabled along with the properties bitvector as arguments. Upon executing this instruction, the MTL updates the entry for the VB in the VB Info Table to reflect that it is now enabled with the appropriate properties (\incircle{1c}).

\begin{figure}[h!]\small
  \centering
  \begin{tikzpicture}
    \node (enable) [draw,rounded corners=2pt,inner sep=5pt] {\tt \textcolor{blue}{\enablevb} VBUID, props};
  \end{tikzpicture}
\end{figure}

\subsubsection{Dynamic Memory Allocation}
\label{sec:dynamic_alloc}

After enabling the VB, the OS \sg{uses the \attachvb instruction (\incircle{2a}) to add the VB to the CVT of the calling process \onurii{and increment the \sgii{VB's} reference count in its VIT entry}
(\incircle{2b}; \pointer{\Cref{sec:client}}).} The OS then returns the index of the \mhp{newly-added} CVT entry as the return value of the \requestvas system call (stored as \texttt{index} in the application code \mhp{example} of Figure~\ref{fig:vbi-uarch}). This \texttt{index} serves as a pointer to the VB. As we discuss in \mhp{\pointer{\Cref{subsubsec:addr_spec}}}, the program \sgii{uses} this index to specify virtual addresses to the processor.

After the VB is attached to the process, the process can access any location within the VB with the appropriate permissions. It can also dynamically manage memory inside the VB using modified versions of \texttt{malloc} and \texttt{free} that take the CVT entry \texttt{index} as an additional argument \sg{(\incircle{3})}. During execution, it is possible that the process runs out of memory within a VB (\nas{e.g., }due to an incorrect estimate of the expected size of the data structure). In such a case, \sys allows automatic promotion of the \nas{allocated data} to a VB of a larger size class. \pointer{\Cref{sec:vb-promotion}} discusses VB promotion in detail.

\subsubsection{Address Specification}
\label{subsubsec:addr_spec}

In order to access data inside a VB, the process generates a two-part virtual address in the format of $\{${\tt \cvt index, offset}$\}$. The \texttt{CVT index} specifies the CVT entry that points to the corresponding VB, and the \texttt{offset} is the location of the data inside the VB. Accessing the data indirectly through the \texttt{CVT index} as opposed to direcly using the VBI address allows \sys to 
\nas{not require relocatable code and maintain the validity of the pointers (i.e., virtual addresses) within a VB when migrating/copying the content of a VB to another VB. \mhp{With CVT indirection, VBI can} seamlessly \mhp{migrate/copy VBs by just} updating the \texttt{VBUID} of the corresponding CVT entry with the \texttt{VBUID} of the new VB.}



\subsubsection{Operation of a Memory Load}
\label{sec:load}

Figure~\ref{fig:vbi-uarch} shows the execution of the memory load instruction triggered by the code \texttt{y~=~(*x)}, where the pointer \texttt{x} contains 
the virtual address consisting of (1)~the index of the corresponding VB in the \sg{process'} CVT, and (2)~the offset within the VB \sg{(\incircle{4} in Figure~\ref{fig:vbi-uarch})}.
\nas{When performing a load operation, the CPU first checks whether \texttt{index} is within the range of the client's CVT.  Next, the CPU needs to fetch the corresponding CVT entry in order to perform the permissions check. The CPU uses a \onurvi{per-process} small direct-mapped CVT cache to speed up accesses to the client's recently-accessed CVT entries (\pointer{\Cref{sec:cvt}}). Therefore, the CPU \sg{looks up} the corresponding CVT cache entry using \texttt{index} as the key \sg{(\incircle{5})}}, \sg{and checks} if (1)~the client has permission to read from the VB, and (2)~\texttt{offset} is smaller than the size of the VB. If either of these checks fail, the CPU raises an exception. If the access is allowed, the CPU constructs the VBI address \onurthird{by concatenating the 
\texttt{VBUID}} stored in the CVT entry with \texttt{offset} \sg{(\incircle{6})}. The processor directly uses the generated VBI address to access the on-chip caches. If the data is present in any of the on-chip caches, it is returned to the CPU, thereby completing the load operation.

\nas{\sys performs address translation in parallel with the cache lookup in order to minimize the address translation overhead on the critical path of the access. Accordingly, when an access misses in \sg{the} L2 cache, \nasi{the processor requests the \mtl to perform the VBI-to-physical address translation}. To this end, \mtl fetches the pointer to the VB's translation structure from the VBI Info Table (VIT) entry associated with the VB. \sys uses a VIT cache to speed up accesses to recently-accessed VIT entries (\incircle{7}). In order to facilitate the VBI-to-physical address translation, \mtl employs a translation lookaside buffer (TLB). On a TLB hit, the memory controller accesses the cache line using the physical address in the corresponding TLB entry \sg{(\incircle{8})}. \onurii{On a TLB miss, the \mtl performs the address translation by traversing the VB's translation structure \sg{(\incircle{9})}, and inserts the mapping information in\onurv{to} the TLB once the physical address is obtained. Next, the memory controller fetches} the corresponding cache line from main memory and returns it to the processor. The processor inserts the cache line into the on-chip caches using the VBI address, and returns the cache line to the CPU to complete the load. \pointer{\Cref{sec:mtl}} describes the operation of the \mtl in detail.}



\subsubsection{Memory Deallocation}
\label{sec:mem-dealloc}

\onurvi{The program can deallocate the memory allocated inside a VB using \texttt{free} (\Cref{sec:dynamic_alloc}).} \onurii{When a process terminates, the OS traverses the CVT of the process and} detaches all of the VBs attached to the process using the \detachvb instruction. For each VB whose reference count \onurii{(stored as part of VIT entry of the VB; see \pointer{\Cref{sec:vit}})} drops to zero, the OS informs \sys that the VB is no longer in use via the \disablevb instruction.

\begin{figure}[h!]\small
  \centering
  \begin{tikzpicture}
    \node (enable) [draw,rounded corners=2pt,inner sep=5pt] {\tt \textcolor{blue}{\disablevb} VBUID};
  \end{tikzpicture}
\end{figure}

In response to the \disablevb instruction, the MTL destroys all state associated with VB \texttt{VBUID}. To avoid stale data in the cache, all of the VB's cache lines are invalidated \emph{before} \mhp{the \texttt{VBUID} is reused for another memory allocation}. 
\mpt{Because} there are a large number of VBs in each size class, \mpt{it is likely that} the disabled \texttt{VBUID} does not need to be reused immediately, and the cache cleanup can be performed lazily in the background.


\subsection{CVT Cache}
\label{sec:cvt}

For every memory operation, the CPU must check if the operation is permitted by accessing the information in the corresponding CVT entry. To exploit locality in the CVT, \sys uses a per-core \emph{CVT cache} to store recently-accessed entries in the client's CVT. The CVT cache is similar to the TLB in existing processors. However, unlike a TLB that caches virtual-to-physical address mappings of page-sized memory regions, the CVT cache maintains information at the VB granularity, and only for VBs that can be accessed by the program. While programs may typically access hundreds or thousands of pages, our evaluations show that most programs only need a few \emph{tens} of VBs to subsume all their data. With the exception of \emph{GemsFDTD} (which allocates 195 VBs),\footnote{\emph{GemsFDTD} performs computations in the time domain on 3D grids. It involves multiple execution timesteps, each of which allocates new 3D grids to store the computation output. Multiple allocations are also needed during the post-processing Fourier transformation performed in \emph{GemsFDTD}.} all applications use fewer than 48 VBs. Therefore, the processor can achieve a near-100\% hit rate even with a 64-entry \emph{direct-mapped} CVT cache, which is faster and more efficient than the large set-associative TLBs employed by modern processors. 
\subsection{Processor, OS, and Process Interactions}
\label{sec:pop-interaction}

\sys handles basic process lifetime operations similar to  current systems. This section describes in detail how these operations work with \sys. 

\paragraph{System Booting.}
When the system is booted, the processor initializes the data structures relevant to \sys (e.g., pointers to VIT tables) with the help of the \sgii{MTL} (discussed in \pointer{\Cref{sec:mtl}}). An initial ROM program runs as a privileged client, copies the bootloader code from bootable storage to a newly enabled VB, and jumps to the bootloader's entry point. This process initiates the usual sequence of chain loading until the OS is finally loaded into a VB. The OS reads the parameters of \sys, namely, the number of bits of virtual address, the number and sizes of the virtual block size classes, and the maximum number of memory clients supported by the system, to initialize \mhp{the OS-level} memory management subsystem.

\paragraph{Process Creation.} When a binary is executed, the OS creates a new process by associating it with one of the available client IDs. For each \mpt{section of} the binary (e.g., code, static data), the OS (1)~enables the smallest VB that can fit the contents of the \mpt{section} and associates the VB with the appropriate properties using the \enablevb instruction, (2)~attaches itself to the VB with write permissions using the \attachvb instruction, (3)~copies the contents from the application binary into the VB, and (4)~detaches itself from the VB using the \detachvb instruction. The OS then attaches the client to the newly enabled VBs and jumps to program's entry point.

\paragraph{Shared Libraries.}
The OS loads the executable code of each shared library into a separate VB. While \mpt{a} shared library can dynamically allocate data using the \requestvas system call, any \mpt{static per-process} data associated with the library should be loaded \mhp{into} a separate VB for each process that uses the library. In existing systems, access to \mpt{static} data is \mpt{typically} performed using PC-relative addressing. \sys~\mpt{provides an analogous memory} addressing mode that we call \emph{CVT-relative addressing}. In this addressing mode, the CVT index of a memory reference is specified relative to the CVT index of the VB containing the reference. Specifically, in shared libraries, all references to static data use +1 CVT-relative addressing, i.e., the CVT index of the data is one more than the CVT index of the code. After process creation, the OS iterates over the list of shared libraries requested by the process. For each shared library, the OS attaches the client to the VB containing the corresponding library code and ensures that the subsequent CVT entry is allocated to the VB containing the static data associated with the shared library. This solution avoids the \mpt{need to perform load-time relocation for each data reference in the executable code}, although VBI can use relocations in the same manner as current systems, \mpt{if required}.

\paragraph{Process Destruction.}
When a process terminates, the OS deallocates all VBs for the process using the mechanism described in \pointer{\Cref{sec:mem-dealloc}}, and then frees the client ID for reuse.

\paragraph{Process Forking.}
When a process forks, all of its memory state must be replicated for the newly created process. In \sys, forking entails creating copies of all the private VBs attached to a process. To reduce the overhead of this operation, \sys introduces the following instruction:
\begin{figure}[h!]\small
  \centering
  \begin{tikzpicture}
    \node [draw,rounded corners=2pt,inner sep=5pt] {\tt \textcolor{blue}{\clonevb} SVBUID, DVBUID};
  \end{tikzpicture}
\end{figure}

\clonevb  instructs \sys to make the destination VB \texttt{DVBUID} a clone of the source VB \texttt{SVBUID}. To efficiently implement \clonevb, the MTL marks all translation structures and physical pages of the VB as copy-on-write, and lazily copies the relevant regions if they receive a write operation.\footnote{\onurii{The actual physical copy can be accelerated using in-DRAM copy mechanisms such as RowClone~\cite{seshadri2013rowclone},  LISA~\cite{chang2016low}, \onurv{and NoM~\cite{nom2020}}.}}

When forking a process, the OS first copies all CVT entries of the parent to the CVT of the child so that the child VBs have the same CVT index
as the parent VBs. This maintains the validity of the pointers
in the child VBs after cloning. Next, for each CVT entry corresponding to a private VB (shared VBs are already enabled), the OS (1)~enables a new VB of the same size class and executes the \clonevb instruction, and (2)~updates the \texttt{VBUID} in the CVT entry to point to the newly enabled clone. The fork returns after all the \clonevb operations are completed.

\paragraph{VB Promotion.}
\label{sec:vb-promotion}
As described in \pointer{\Cref{sec:dynamic_alloc}}, when a program runs out of memory for a data structure within the assigned VB, the OS can automatically promote the data structure to a VB of higher size class. To perform such a promotion, the OS first suspends the program. It enables a new VB of the higher size class, and executes the \promotevb instruction.
\begin{figure}[h!]\small
  \centering
  \begin{tikzpicture}
    \node [draw,rounded corners=2pt,inner sep=5pt] {\tt \textcolor{blue}{\promotevb} SVBUID, LVBUID};
  \end{tikzpicture}
\end{figure}

In response to this instruction, \sys first flushes all dirty cache lines from the smaller VB with the unique ID of \texttt{SVBUID}. This operation can be sped up using structures like the Dirty Block Index~\cite{seshadri2014dirty}. \sys then copies all the translation information from the smaller VB appropriately to the larger VB with the unique ID of \texttt{LVBUID}. After this operation, in effect, the early portion of the larger VB is mapped to the same region in the physical memory as the smaller VB. The remaining portions of the larger VB are unallocated and can be used by the program to expand its data structures \onurii{and allocate more memory using \texttt{malloc}}. \sys updates the entry in the program's CVT \onurii{that points} to \texttt{SVBUID} to now point to \texttt{LVBUID}.

\subsection{\mtlfull}
\label{sec:mtl}

The \mtlfull (\mtl) centers around the VB Info Tables (VITs)\onurthird{, which store the metadata associated with each VB}. In this section, we discuss (1)~the design of the VITs, (2)~the two main responsibilities of the \mtl; memory allocation and address translation, and (3) the hardware complexity of the \mtl.


\subsubsection{\vit (\vitshort)}
\label{sec:vit}

\onurthird{As \pointer{\Cref{sec:vb-detail}} briefly describes, MTL uses a set of VB Info Tables (VITs) to maintain information about VBs. Specifically, 
for each VB \onurvi{in the system}, a \vit stores an entry that consists of (1)~an \emph{enable} bit, which indicates if the VB is currently assigned to a process; (2)~\texttt{props}, a bitvector that describes the VB properties; \nas{(3)~the number of processes attached to the VB (i.e., a reference count);} (4)~the type of VBI-to-physical address translation structure being used for the VB; and (5)~a pointer to the translation structure.} For ease of access, the MTL maintains a \emph{separate} VIT for each size class. The ID of a VB within its size class (VBID) is used as an index into the corresponding VIT. When a VB is enabled (using \enablevb), the MTL \onurthird{finds the corresponding VIT and entry using the \texttt{SizeID} and \texttt{VBID}, respectively (both extracted from \texttt{VBUID}). \mtl then sets the} \emph{enabled} bit of the entry 
and updates \texttt{props}. \nas{The reference counter of the VB is also set to 0, indicating that no process is attached to this VB.} The type and pointer of the translation structure \onurthird{of the VB} are updated \onurthird{in its VIT entry} at the time of physical memory allocation (as we discuss in \pointer{\Cref{sec:flexible-translations}}). \onurthird{Since a VIT contains \onurii{entries for the} VBs of only a single size class, the number of entries in each VIT equals the number of VBs that the associated size class supports (\pointer{\Cref{sec:vb-detail}}). However, }\sys limits the size of each \vit by storing entries only \mhp{up to the currently-enabled} VB with the largest \texttt{VBID} \onurii{in the size class associated with that \vit}. The OS ensures that the table does not become prohibitively large by reusing previously-disabled VBs for subsequent requests (\pointer{\Cref{sec:mem-dealloc}}).


\subsubsection{\onurthird{Base Memory Allocation and Address Translation}}
\label{sec:base}
\onurthird{Our \emph{base} memory allocation algorithm} allocates physical memory at 4~KB granularity. Similar to \xeightsix~\cite{intelx86manual}, \onurthird{Our \emph{base} address translation mechanism} stores VBI-to-physical address translation information in multi-level tables. However, unlike the 4-level page tables in \xeightsix, \sys uses tables with varying number of levels according to the size of the VB. \onurthird{For example, a 4~KB VB does not require a translation structure (i.e., can be direct-mapped) since 4~KB is the minimum granularity of meomry allocation. On the other hand, a 128~KB VB requires a one-level table for translating address to 4~KB regions}. As a result, smaller VBs require fewer memory accesses to serve a TLB miss. For each VB, the \vitshort stores a pointer to the address of the root of the multi-level table (or the base physical address of the directly mapped VBs). 


\subsubsection{\mhp{\mtl} Hardware Complexity}

We envision the \mtl as software running on a programmable low-power core within the memory controller. While conventional OSes are responsible for memory allocation, virtual-to-physical mapping, and memory protection, the \mtl does not need to deal with protection, so we expect the \mtl code to be simpler than typical OS memory management software. As a result, the complexity of the \mtl hardware is similar to that of prior proposals such as Pinnacle~\cite{pinnacle} (commercially available) and Page Overlays~\cite{page_overlays}, which perform memory allocation and remapping in the memory controller. While both Pinnacle and Page Overlays are hardware solutions, VBI provides flexibility by making the \mtl programmable, thereby allowing software updates for different memory management policies (e.g., address translation, mapping, migration, scheduling). Our goal in this work is to understand the potential of hardware-based memory allocation and address translation.

\section{Allocation and Translation Optimizations}
\label{sec:mtl_optimizations}

The \mtl employs three techniques to optimize the base memory allocation and address translation \onurii{described in \pointer{\Cref{sec:base}}}. \nasi{We explain these techniques in the following subsections.}

\subsection{Delayed Physical Memory Allocation}
\label{sec:delayed-allocation}

As described in \pointer{\Cref{sec:optimizations}}, \sys delays physical memory allocation for a VB (or a region of a VB) until a dirty cache line from that \onurvii{VB (or a region of the VB)} is evicted from the last-level cache (LLC). This optimization is enabled by the fact that \sys uses VBI address directly to access \mhp{\emph{all}} on-chip caches. Therefore, a cache line does \emph{not} need to be backed by a physical memory mapping in order to be accessed.

\onurii{In this approach,} when a VB is enabled, \sys does not immediately allocate physical memory to the VB. 
On an LLC miss \onurvii{to the VB}, \sys checks the status of the VB in its corresponding VIT entry. If there is no physical memory backing the data, \sys does one of two things. (1)~If the VB corresponds to a memory-mapped file or if the required data was \onurii{allocated before but }swapped out to a backing store, then \sys allocates physical memory for the region, interrupts the OS to copy the relevant data from storage into the allocated memory, and then returns the relevant cache line to the processor. (2)~If this is the first time the cache line is being accessed from memory, \sys simply returns a zeroed cache line without allocating physical memory to the VB.


\onurii{On a dirty cache line \onurthird{writeback from the LLC}, if physical memory is yet to be allocated for the region that the cache line maps to, \sys first allocates physical memory for the region, and then performs the writeback. \sys allocates only the region of the VB containing the evicted cache line. 
As \pointer{\Cref{sec:base}} describes, our base memory allocation mechanism allocates physical memory at a 4~KB granularity. Therefore, the region allocated for the evicted cache line is 4~KB.} \pointer{\Cref{sec:early-reservation}} describes an optimization that eagerly \emph{reserves} \onurii{a larger amount of} physical memory for a VB \onurii{during allocation}, to reduce \onurthird{the} overall translation overhead.

\subsection{Flexible \nasi{Address} Translation Structures}
\label{sec:flexible-translations}
For each VB, \sys chooses one of three types of address translation structures, depending on the needs of the VB and the physical memory availability. The first type \emph{directly} maps the VB to physical memory when enough contiguous memory is available. With this mapping, a single TLB entry
is sufficient to maintain the translation for the entire VB. The second type
uses a single-level table, where the VB is divided into equal-sized blocks of one of the supported size classes. Each entry in the table maintains the mapping for the corresponding block. This mapping exploits the fact that \mhp{a} majority of the data structures are densely allocated inside their respective VBs. With \mhp{a} single-level table, the mapping for any region of the VB can be retrieved with a single memory access. The third type,
suitable for sparsely-allocated VBs, is 
\onurii{our base address translation mechanism (described in \pointer{\Cref{sec:mtl}}), which uses multi-level page tables where the table depth is chosen based on the size of the VB.} 

\onurii{\sgii{In our evaluation, we implement a flexible mechanism that}
statically chooses a translation structure type based on the size of the VB. Each 4~KB VB is directly mapped. 128~KB and 4~MB VBs use a single-level table. VBs of a larger size class use a multi-level table with as many levels as necessary to map the VB using 4~KB pages.\footnote{\onurvi{For fair comparison with conventional virtual memory, our evaluations use a 4~KB granularity to map VBs to physical memory. However, \sys can flexibly map VBs at the granularity of any available size class.}} The \onurthird{\emph{early reservation}} optimization (described in \pointer{\Cref{sec:early-reservation}}) improves upon this static policy by dynamically choosing a translation structure type from the three types mentioned above based on the available contiguous physical memory. While we evaluate table-based translation structures in this work, \sys can be easily extended to support other structures (e.g., customized per-application translation structures as proposed in DVMT~\cite{vm19}).}



Similar to \xeightsix, \sys uses multiple types of TLBs to cache mappings of different granularity. The type of translation structure used for a VB is stored in the \vitshort and is cached in the on-chip \vitshort Cache. This information enables \sys to access the right type of TLB. For a fair comparison, our evaluations use the same TLB type and size for all baselines and variants of \sys.

\subsection{Early Reservation of Physical Memory}
\label{sec:early-reservation}

\sys can perform early reservation of the physical memory for a VB. \onurii{To this end, \sys reserves (but does not allocate) physical memory for the entire VB at the time of memory allocation, and treats the VB as \emph{directly mapped} by serving future memory allocation requests for that VB from that contiguous reserved region.}
This optimization is inspired by prior work on super-page management~\cite{reserve}, which reserves a larger contiguous region of memory than the requested size, and upgrades the allocated pages to larger super-pages when enough contiguous pages are allocated in that region.

For \sys's early reservation optimization, at the time of the \emph{first} physical memory allocation request for a VB, the \mtl checks if there is enough contiguous free space in physical memory to fit the entire VB. If so, it allocates the requested memory from that contiguous space, and marks the remaining free blocks in that contiguous space as reserved for that specific VB. In order to reduce internal fragmentation when free physical memory is running low, physical blocks reserved for a VB may \mhp{be} used by \mhp{another} VB when no unreserved blocks are available.  As a result, the \mtl uses a three-level priority when allocating physical blocks: (1)~free blocks reserved for the VB that is demanding allocation, (2)~unreserved free blocks, and (3)~free blocks reserved for other VBs. A VB is considered directly mapped as long as all its allocated memory is mapped to a single contiguous region of memory, thereby requiring just a single TLB entry for the entire VB.
If there is not enough contiguous physical memory available to fit the entire VB, the early reservation mechanism allocates the VB sparsely by reserving blocks of the largest size class that can be allocated contiguously. 

\onurthird{With the early reservation approach, memory allocation is performed at a different granularity than mapping, which enables \sys to benefit from larger mapping granularities and \mhp{thereby} minimize the address translation latency, while eliminating memory allocation for regions that may never be accessed. }
To support the early reservation mechanism, \sys uses the Buddy algorithm~\cite{buddy,knowlton65} to manage free and reserved \nasi{regions} of different 
\onurvii{size classes}.



\section{VBI in Other System Architectures}

\sys is designed to easily and efficiently function in various system designs. \mhp{We} describe the implementation of \sys in two important examples of modern system architectures: virtualized environments and multi-node systems.

\subsection{Supporting Virtual Machines}
\label{sec:virtual-machines}

\begin{figure}[h]
  \centering
  \begin{tikzpicture}[inner sep=1pt,outer sep=0pt,>=stealth', font=\sffamily]

\draw [rounded corners=2pt] (0,0) rectangle (8,0.5);
\draw (4.4,0) -- ++(0,0.5);
\draw (0.8,0) -- ++(0,0.5);
\draw (1.8,0) -- ++(0,0.5);

\node at (0.4,0.25) {\small\texttt{100}};
\node at (1.3,0.25) {\small{\texttt{VM ID}}};
\node at (3.1,0.25) {\small\texttt{VBID}};
\node at (6.2,0.25) {\small\texttt{offset}};

\node (s) at (0.4,-0.3) {\small{3b}};
\node (m) at (1.3,-0.3) {\small{5b}};
\node (v) at (3.1,-0.3) {\small{24 bits}};
\node (o) at (6.2,-0.3) {\small{32 bits}};

\draw [->] (s.west) -- (0,-0.3);
\draw [->] (s.east) -- (0.8,-0.3);
\draw [->] (m.west) -- (0.8,-0.3);
\draw [->] (m.east) -- (1.8,-0.3);
\draw [->] (v.west) -- (1.8,-0.3);
\draw [->] (v.east) -- (4.4,-0.3);
\draw [->] (o.west) -- (4.4,-0.3);
\draw [->] (o.east) -- (8,-0.3);

\end{tikzpicture}
  \caption{Partitioning the VBI address space among virtual machines, using the 4~GB size class (\texttt{100}) as an example.}
  \label{fig:vaddr-vm}
\end{figure}
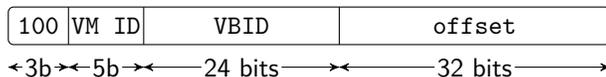

\sys implements address space isolation between virtual machines (VMs) by \emph{partitioning} the global VBI address space among multiple VMs and the host OS. To this end, \sys reserves a few bits in the VBI address for the \texttt{VM ID}. Figure~\ref{fig:vaddr-vm} shows how \sys implements this for a system supporting 31 virtual machines (ID 0 is reserved for the host). In the \onurii{VBI} address, the 5 bits following the size class bits are used to denote the \texttt{VM ID}. For every new virtual machine in the system, the host OS assigns a \texttt{VM ID} to be used by the guest OS while assigning virtual blocks to processes inside the virtual machine. \sys partitions client IDs using a similar approach. With address space division between VMs, a guest VM is unaware that it is virtualized, and it can allocate/deallocate/access VBs \onurii{\emph{without} having to} coordinate with the host OS. Sharing VBs across multiple VMs is possible, but requires explicit coordination with the host OS.

\subsection{Supporting Multi-Node Systems}
\label{sec:multi-node}

There are many ways to implement \sys in multi-node systems. Our initial approach provides each node with its own \mtl. \sys equally partitions VBs of each size class among the {\mtl}s, with the higher order bits of \texttt{VBID} indicating the \sgii{VB's} home \mtl. The home \mtl of a VB is the only \mtl that manages the VB's physical memory allocation and address translation. When allocating a VB to a process, the OS attempts to ensure that the \sgii{VB's} home \mtl is in the same node as the core executing the process. During phase changes, the OS can seamlessly migrate data from a VB hosted by one \mtl to a VB hosted by another \mtl. We leave the evaluation of this approach and exploration of other ways of integrating \sys with multi-node systems to future work.


\section{Evaluation}
\label{sec:methodology2}

We evaluate VBI for two concrete use cases. First, we evaluate how \sys reduces address translation overheads in native and virtualized environments (\pointer{\Cref{sec:4k-perf}} and \pointer{\Cref{sec:2m-perf}}\mhp{, respectively}). Second, we evaluate the benefits that \sys offers in harnessing the full potential of two \mhp{main memory} architectures that are tightly dependent on the 
\nas{data mapping}: (1)~a hybrid PCM--DRAM memory architecture; and (2)~TL-DRAM\mhp{\cite{tldram}}, a \mhp{heterogeneous-latency} DRAM \onurii{architecture} (\pointer{\Cref{sec:tld}}). 

\subsection{Methodology}
\label{sec:methodology}

For our evaluations, we use a heavily-customized version of Ramulator~\cite{ramulator} to faithfully model all components of the memory subsystem (including TLBs, page tables, the page table walker, and the page walk cache), as well as the functionality of memory management calls (e.g., \texttt{malloc}, \texttt{realloc}, \texttt{free}). We have released this modified version of Ramulator~\cite{vbi}. Table~\ref{tbl:simconfig} summarizes the main simulation parameters. Our workloads consist of benchmarks from SPECspeed 2017~\cite{spec2017}, SPEC CPU 2006~\cite{spec2006}, TailBench~\cite{tailbench}, and Graph 500~\cite{graph500}. 
We identify representative code regions for the SPEC benchmarks using SimPoint~\cite{sim-point}. For TailBench applications, we skip the first five billion instructions. For Graph 500, we mark the region of interest directly in the source code. We use an Intel Pintool~\cite{luk2005} to collect traces of the representative regions of each of our benchmarks. For our evaluations, we first warm up the system with 100~million instructions, and then run the benchmark for 1~billion instructions.

\begin{table}[h]\scriptsize
  \centering
  \setlength{\aboverulesep}{0pt}
    \setlength{\belowrulesep}{0pt}
  \def\arraystretch{0.9}
   \begin{tabular}{ll} 
        \toprule
        \textbf{CPU} & 4-wide issue, OOO, 128-entry ROB\\
        \cmidrule(rl){1-2}
        \textbf{L1 Cache} & 32~KB, 8-way associative, 4 cycles \\
        \cmidrule(rl){1-2}
        \textbf{L2 Cache} & 256~KB, 8-way associative, 8 cycles\\
        \cmidrule(rl){1-2}
        \textbf{L3 Cache} & 8~MB (2~MB per-core), 16-way associative, 31 cycles\\
        \cmidrule(rl){1-2}
        \multirow{2}{*}{\textbf{L1 DTLB}} & 4~KB pages: 64-entry, fully associative\\
        & 2~MB pages: 32-entry, fully associative\\
        \cmidrule(rl){1-2}
        \textbf{L2 DTLB} & 4~KB and 2~MB pages: 512-entry, 4-way associative\\
        \cmidrule(rl){1-2}
        \textbf{Page Walk Cache} & 32-entry, fully associative\\
        \cmidrule(rl){1-2}
        \multirow{2}{*}{\textbf{DRAM}} & DDR3-1600, 1 channel, 1 rank/channel\\
         & 8 banks/rank, open-page policy\\
        \cmidrule(rl){1-2}
        \multirow{1}{*}{\textbf{DRAM Timing~\cite{dramtiming}}} & tRCD=5cy, tRP=5cy, tRRDact=3cy, tRRDpre=3cy\\
        \cmidrule(rl){1-2}
        \textbf{PCM} & PCM-800, 1 channel, 1 rank/channel, 8 banks/rank\\
        \cmidrule(rl){1-2}
        \multirow{1}{*}{\textbf{PCM Timing~\cite{lee2009}}} & tRCD=22cy, tRP=60cy, tRRDact=2cy, tRRDpre=11cy\\
        \bottomrule
    \end{tabular}%


  \caption{Simulation configuration.}
  \label{tbl:simconfig}
\end{table}%

\begin{figure*}[h]
  \centering
  \includegraphics[scale=0.75]{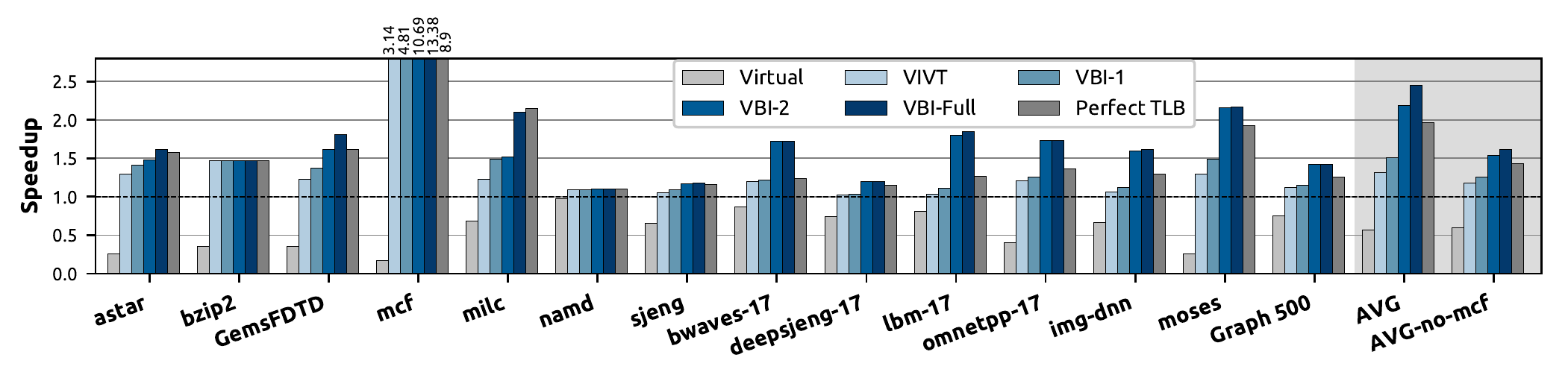}
  \caption{Performance of systems with 4KB pages (normalized to \textsf{Native}).}
  \label{fig:4k-perf}
\end{figure*}

\subsection{Use Case 1: Address Translation}
\label{sec:vm-perf}

We evaluate the performance of seven baseline systems to compare with \sys:
(1)~\textbf{\sffamily Native}: applications run natively on an \xeightsix system with only 4~KB pages;
(2)~\textbf{\sffamily Native-2M}: \textsf{Native} but with only 2~MB pages; 
(3)~\textbf{\sffamily Virtual}: applications run inside a virtual machine with only 4~KB pages;
(4)~\textbf{\sffamily Virtual-2M}: \textsf{Virtual} but with only 2~MB pages;\footnote{We augment this system with a 2D page walk cache, which is shown to improve the performance of guest workloads~\cite{vm11}.} 
(5)~\textbf{\sffamily Perfect TLB}: an unrealistic version of \textsf{Native} with no L1 TLB misses \onurii{(i.e., no address translation overhead)};
(6)~\textbf{\sffamily VIVT}: \textsf{Native} with VIVT on-chip caches; and
(7)~\onurii{\textbf{\sffamily Enigma-HW-2M}}: applications run natively in a system with Enigma~\cite{enigma}. \onurii{Enigma uses a system-wide unique intermediate address space to defer address translation until data must be retrieved from physical memory. A centralized translation cache (CTC) at the memory controller \sgii{performs intermediate-to-physical address translation}. However, \sgii{unlike \sys, Enigma asks the OS to perform the translation on a CTC miss, and} to explicitly manage address mapping. Therefore, Enigma's benefits do not seamlessly extend to programs running inside a virtual machine. We evaluate Enigma with a 16K-entry centralized translation cache (CTC) that we enhance with hardware-managed page walks and 2 MB pages.}


We evaluate the performance of three \sys systems:
(1)~\textbf{\sffamily \mbox{\sys-1}}: \onurii{inherently virtual caches (\pointer{\Cref{sec:optimizations}}) along with} \onurii{our} \onurii{\emph{flexible translation mechanism} \onurii{that} maps VBs using \sgii{a 4~KB granularity}~(\pointer{\Cref{sec:base}})} ,
(2)~\textbf{\sffamily \mbox{\sys-2}}: \textsf{\sys-1} with \emph{delayed physical memory allocation} \onurii{(allocates the 4~KB region of the VB that the dirty cache line evicted from the last-level cache belongs to)}. (\pointer{\Cref{sec:delayed-allocation}}), and
(3)~\textbf{\sffamily \mbox{\sys-Full}}: \textsf{\sys-2} with \emph{early reservation} (\pointer{\Cref{sec:early-reservation}}).
\textsf{\sys-1} and \textsf{\sys-2} manage memory at 4~KB granularity, while \textsf{\sys-Full} uses early reservation to support all of the size classes listed in \pointer{\Cref{sec:vb-detail}} for VB allocation, providing similar benefits to large page support \onurii{and direct mapping}. \onurv{We first present results comparing \textsf{\sys-1} and \textsf{\sys-2}} with \textsf{Native}, \textsf{Virtual}, \textsf{VIVT}, and \textsf{Perfect TLB} (\pointer{\Cref{sec:4k-perf}}). We then present results comparing \textsf{\sys-Full} with \textsf{Native-2M}, \onurii{\textsf{Enigma-HW-2M}}, and \textsf{Perfect TLB} (\pointer{\Cref{sec:2m-perf}}).

\subsubsection{Results with 4~KB Pages}
\label{sec:4k-perf}

Figure~\ref{fig:4k-perf} plots the performance of \textsf{Virtual}, \textsf{VIVT}, \textsf{\sys-1}, \textsf{\sys-2}, and \textsf{Perfect TLB} normalized to the performance of \textsf{Native}, for a single-core system. \onurii{We also show \textsf{\sys-Full} as a reference} \onurv{that shows the full potentials of \sys which \textsf{\sys-1} and \textsf{\sys-2} do not enable}. \emph{mcf} has an overwhelmingly high number of TLB misses. Consequently, mechanisms that reduce TLB misses greatly improve \emph{mcf}'s performance, to the point of skewing the \onurvii{average} significantly. Therefore, the figure also presents the \onurvii{average} speedup without \emph{mcf}. We draw five observations from the figure.

First, \textsf{\sys-1} outperforms \textsf{Native} by 50\%, averaged across all benchmarks (25\% without \emph{mcf}). This performance gain is a direct result of \onurii{(1) inherently virtual on-chip caches in \sys that reduce the number of address translation requests, and (2) }fewer levels of address translation for smaller VBs, \onurv{which reduces} the number of translation-related memory accesses (i.e., page walks). 

Second, \textsf{Perfect TLB} serves as an upper bound for the performance benefits of \textsf{\sys-1}. 
\onurii{However, by employing flexible translation structures,} \textsf{\sys-1} bridges the performance gap between \textsf{Native} and \textsf{Perfect TLB} by 52\%\mhp{, on average}.

Third, when accessing regions for which no physical memory is allocated yet, \textsf{\sys-2} avoids \emph{both} the memory \onurii{requests themselves and any} translation-related memory accesses \onurvii{for those requests}. \onurii{Therefore, \textsf{\sys-2} enables benefits over and beyond solely reducing the number of page walks, as it further improves the overall performance by reducing the number of memory requests accessing the main memory as well.} Consequently, for many memory-intensive applications, \textsf{\sys-2} outperforms \textsf{Perfect TLB}. Compared to \textsf{Perfect TLB}, \textsf{\sys-2} reduces the total number of DRAM accesses (including the translation-related memory accesses) by 62\%, averaged across applications that outperform \textsf{Perfect TLB}, and by 46\% across all applications. Overall, \textsf{\sys-2} outperforms \textsf{Native} by an average of 118\% (53\% without \emph{mcf}).

Fourth, by performing address translations \emph{only for and in parallel with} LLC accesses, \textsf{VIVT} outperforms \textsf{Native} by 31\% on average (17\% without \emph{mcf}). This performance gain is due to reducing the number of translation requests and therefore decreasing the number of TLB misses using VIVT caches. However, \textsf{\sys-1} and \textsf{\sys-2} gain an extra 19\% and 87\% performance on average, respectively, over \textsf{VIVT}. These improvements highlight \sys's ability to improve performance beyond \mhp{only} employing VIVT caches.

Finally, our results indicate \onurii{that} due to considerably higher translation overhead, \textsf{Virtual} significantly slows down applications compared to \textsf{Native} (44\% on average). As described in \pointer{\Cref{sec:optimizations}}, once an application running inside a virtual machine 
\onurvi{is attached to its VBs}, \sys incurs no additional translation overhead compared to running natively. As a result, in virtualized environments that use only 4K pages, \textsf{\sys-1} and \textsf{\sys-2} achieve an average performance of 2.6$\times$ and 3.8$\times$, respectively, compared to \textsf{Virtual}.

\onurthird{We conclude that even when mapping and allocating VBs using 4~KB granularity only,} both \textsf{\sys-1} and \textsf{\sys-2} provide \mhp{large} benefits over a wide range of baseline systems, due to their effective optimizations to reduce address translation and memory allocation overheads. 
\onurvi{\textsf{\sys-Full} further improves performance by mapping VBs using larger granularities (as we elaborate in \Cref{sec:2m-perf}).}


\subsubsection{Results with Large Pages}
\label{sec:2m-perf}

\begin{figure}[h]
  \centering
  \includegraphics[width=\linewidth]{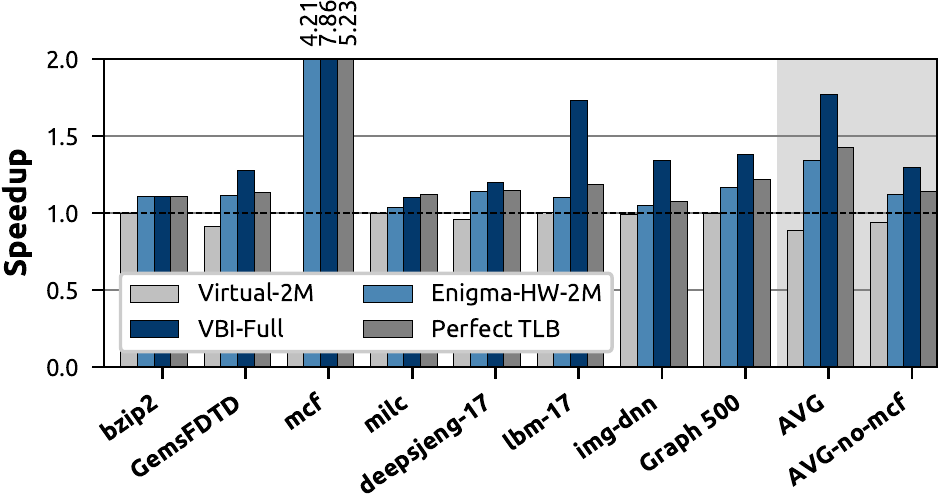}
  \caption{Performance with large pages (\mhp{norm.} to \textsf{Native-2M}).}
  \label{fig:2m-perf}
\end{figure}

Figure~\ref{fig:2m-perf} plots the performance of \textsf{Virtual-2M}, \onurii{\textsf{Enigma-HW-2M}}, \textsf{VBI-Full}, and \textsf{Perfect TLB} normalized to the performance of \textsf{Native-2M}. We enhance the original design of 
Enigma~\cite{enigma} by replacing the OS system call handler for address translation on a CTC miss with a completely \mhp{hardware-managed} address translation, similar to \sys. For legibility, the figure shows results for only a subset of the applications. However, the chosen applications capture the behavior of all the applications, and the \onurvii{average} (and \onurvii{average} without \emph{mcf}) is calculated across all evaluated applications. We draw three observations from the figure. 

\onurvi{First, managing memory at 2~MB granularity improves the performance of applications compared to managing memory at 4~KB granularity.  This is}
because (1)~the larger page size lowers the average TLB miss count \onurvi{(e.g., 66\% lower for \textsf{Native-2M} compared to \textsf{Native}), and (2)~requires fewer page table accesses on average to serve TLB misses (e.g., 73\% fewer for \textsf{Native-2M} compared to \textsf{Native})}.

\onurvi{Second}, \onurii{\textsf{Enigma-HW-2M}} improves overall performance for programs running \emph{natively} on the system by 34\% compared to \textsf{Native-2M}, averaged across all benchmarks (including mcf). The performance gain is a direct result of
(1)~the very large CTC (16K~entries), which reduces the number of translation-related memory accesses by 89\% on average compared to \textsf {Native-2M}; and 
(2)~our hardware-managed address translation enhancement, which removes the costly system calls on each page walk request.

Third, \textsf{\sys-Full}, with all three of our optimizations in \pointer{\Cref{sec:mtl_optimizations}}, maps most VBs using direct mapping, thereby significantly reducing the number of TLB misses and translation-related memory accesses compared to \textsf{Native-2M} (on average by 79\% and 99\%, respectively). In addition, \textsf{\sys-Full} retains the benefits of \textsf{\sys-2}, which reduces the number of \emph{overall} DRAM accesses. \textsf{\sys-Full} reduces the \emph{total} number of DRAM accesses (including translation-related memory accesses) by 56\% on average compared to \textsf{Perfect TLB}. Consequently, \textsf{\sys-Full} outperforms all four comparison points including \textsf{Perfect TLB}. Specifically, \textsf{\sys-Full} improves performance by 77\% compared to \textsf{Native-2M}, 43\% compared to \onurii{\textsf{Enigma-HW-2M}} and 89\% compared to \textsf{Virtual-2M}.

We conclude that \onurii{by employing all of the optimizations that it enables, \sys} significantly outperforms all of our baselines in both native and virtualized environments.

\subsubsection{Multicore Evaluation}
\begin{figure}[t]
  \centering
  \includegraphics[width=0.93\linewidth]{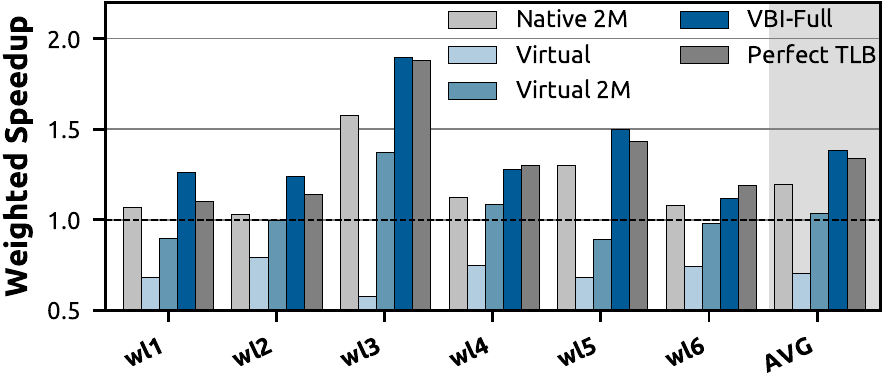}
  \caption{Multiprogrammed workload performance (normalized to \textsf{Native}).}
  \label{fig:mp}
\end{figure}

Figure~\ref{fig:mp} compares the weighted speedup of \textsf{VBI-Full} against four baselines in a quad-core system. We examine six different workload bundles, listed in Table~\ref{tbl:wl}, which consist of the applications studied in our single-core evaluations.  From the figure, we make two observations.  First, averaged across all bundles, \textsf{VBI-Full} improves performance by 38\% and 18\%, compared to \textsf{Native} and \textsf{Native-2M}, respectively. Second, \textsf{VBI-Full} outperforms \textsf{Virtual} and \textsf{Virtual-2M} by an average 67\% and 34\%, respectively. We conclude that the benefits of \sys persist even in the presence of higher memory load in multicore systems.

\begin{table}[b]
\scriptsize
\setlength\tabcolsep{3pt}
  \centering
    \setlength{\aboverulesep}{0pt}
    \setlength{\belowrulesep}{0pt}
  \begin{tabular}{ll|ll}
  \toprule
  \mhp{{\textbf wl1}} &  deepsjeng, omnetpp, bwaves, lbm & \mhp{{\textbf wl4}} &  milc, namd, GemsFDTD, bzip2\\
  \mhp{{\textbf wl2}} &  graph500, astar, img-dnn, moses & \mhp{{\textbf wl5}} &  bzip2, GemsFDTD, sjeng, mcf\\
  \mhp{{\textbf wl3}} &  mcf, GemsFDTD, astar, milc & \mhp{{\textbf wl6}} &  namd, bzip2, astar, sjeng\\
  \bottomrule
\end{tabular}
  \caption{Multiprogrammed workload bundles.}
  \label{tbl:wl}
\end{table}

\subsection{Use Case 2: Memory Heterogeneity}
\label{sec:tld}

As mentioned earlier, 
extracting the best performance from heterogeneous-latency DRAM \onurii{architectures}~\cite{charm, diva-dram, tldram, dynsub, chang.sigmetrics2016, chang2016low, kim2018solar,clrdram,seshadri2013rowclone} and hybrid memory architectures~\cite{yoon2012, refree, raoux2008, li2017utility, dhiman2009pdram, ramos11, het2, zhang2009exploring,chop,banshee} critically depends on \onurii{mapping data to the memory that suits the data requirements, and migrating data as its} 
\onurvii{requirements} change. We quantitatively show the performance benefits of \sys in exploiting heterogeneity by evaluating (1)~a PCM--DRAM hybrid memory~\cite{ramos11}; and (2)~TL-DRAM~\cite{tldram}, a \mhp{heterogeneous-latency} DRAM \onurii{architecture}. We evaluate five systems:
(1)~\textsf{\sys PCM--DRAM} and (2)~\textsf{\sys TL-DRAM}, in which \sys maps and migrates frequently-accessed \onurthird{VBs} to the low-latency memory (the fast memory region in the case of TL-DRAM);
(3)~\textsf{Hotness-Unaware PCM--DRAM} and (4)~\textsf{Hotness-Unaware TL-DRAM}, where the mapping mechanism is unaware of the hotness (i.e., the access frequency) of the data and therefore do not necessarily map the frequently-accessed data to the fast region; and
(5)~\textsf{IDEAL} in each plot refers to an unrealistic perfect mapping mechanism, which uses oracle knowledge to always map \mhp{frequently-accessed} data to the \mhp{fast portion of} memory. 

\begin{figure}[h]
  \centering
  \includegraphics[width=\linewidth]{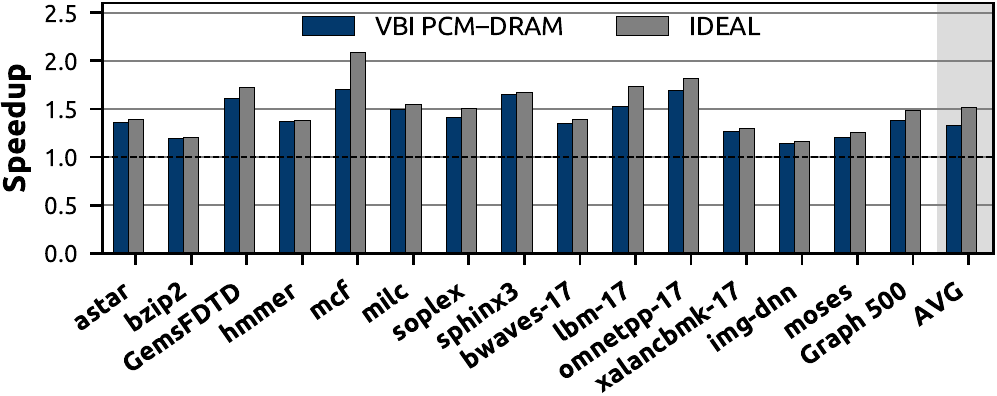}
  \caption{Performance of \sys PCM-DRAM (normalized to \mhp{data-hotness-unaware} mapping).}
  \label{fig:pcm}
\end{figure}

\begin{figure}[h]
  \centering
  \includegraphics[width=\linewidth]{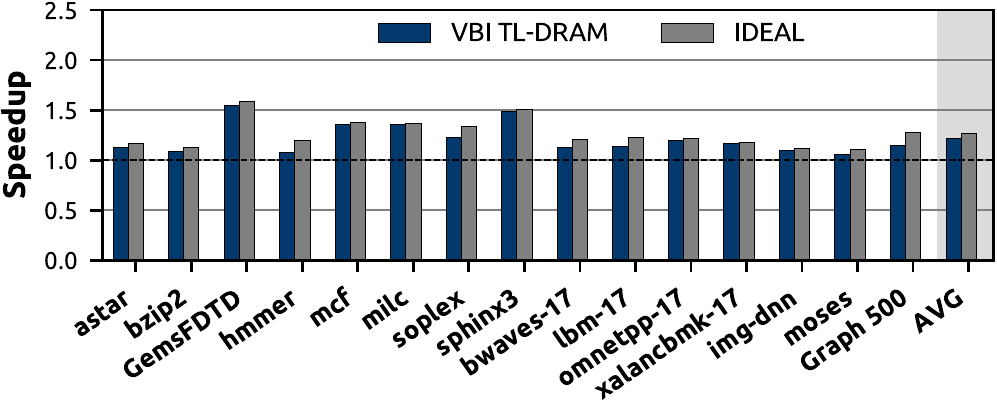}  
  \caption{Performance of \sys TL-DRAM (normalized to \mhp{data-hotness-unaware} mapping).}
  \label{fig:tldram}
\end{figure}

Figures~\ref{fig:pcm} and~\ref{fig:tldram} show the speedup obtained by \sys-enabled mapping over the hotness-unaware mapping in a PCM--DRAM hybrid memory and in TL-DRAM, respectively. We draw three observations from the figures. First, for PCM--DRAM, \textsf{\sys PCM--DRAM} improves performance by 33\% on average compared to the \textsf{Hotness-Unaware PCM--DRAM}, by accurately mapping the frequently-accessed data structures to the low-latency DRAM. Second, by mapping \mhp{frequently-accessed} data to the \mhp{fast DRAM regions}, \textsf{\sys TL-DRAM} takes \onurii{better} advantage of the benefits of TL-DRAM, with a performance improvement of 21\% on average compared to \textsf{Hotness-Unaware TL-DRAM}. Third, \textsf{\sys TL-DRAM} performs only 5.3\% slower than \textsf{IDEAL}, which is the upper bound of performance achieved by mapping hot data to the fast \mhp{regions} of DRAM. 

We conclude that \sys is effective for enabling efficient data mapping and migration in heterogeneous memory systems.
\section{Related Work}
\label{sec:related}


\onurthird{To our knowledge, \sys is the first virtual memory framework to fully delegate physical memory allocation and 
address translation to the hardware. This section compares \sys with other virtual memory designs and related works.}



\textbf{Virtual Memory in Modern Architectures.} 
\sg{Modern virtual memory architectures, such as those employed as part of modern instruction set architectures~\cite{intelx86manual, arm2013, powerpc2015, pa-risc}, have evolved into sophisticated systems.
These architectures have support for features such as large pages, multi-level page tables, hardware-managed TLBs, and variable-size memory segments, but require significant system software support to enable these features and to manage memory.
%
While system software support provides some flexibility to adapt to new ideas, it must communicate with hardware through a rigid \onurii{contract. Such rigid hardware/software communication introduces} costly overheads for many applications (e.g., high overheads with fixed-size \onurii{per-application virtual address spaces,} for applications that only need a small fraction of the space) and \onurii{prevents} the easy adoption of significantly different virtual memory architectures \onurii{or ideas that depend on large changes to the existing virtual memory framework}.
\sys is a completely different framework from existing virtual memory architectures. It supports the functionalities of existing virtual memory architectures, but can do much more by reducing translation overheads, \onurii{inherently and} seamlessly supporting virtual caches, and avoiding unnecessary physical memory allocation. These benefits come from enabling completely hardware-managed physical memory allocation and address translation, which no other virtual memory architecture does \onurtt{(including, for example, Multics~\cite{multics, multics2, multics3})}.}

\onurv{Several memory management frameworks~\cite{vijaykumar2016,etc,vm8,mask,pichai2014,vm9} are designed to minimize the virtual memory overhead in GPUs. 
Unlike \sys, these works provide optimizations \emph{within the existing virtual memory design}, so their benefits are constrained to the design of conventional virtual memory.}

\textbf{OS Support for Virtual Memory.}
There has been extensive work on how address spaces should be mapped to execution contexts~\cite{lindstrom1995}. Unix-like OSes provide a rigid one-to-one mapping between virtual address spaces and processes~\cite{ritchie1978, mckusick2014}.
SpaceJMP~\cite{vm17} proposes a design in which processes can jump from one virtual address space to another in order to access larger amounts of physical memory.
Single address space OSes \onurvii{rely on system-software-based mechanisms to expose a single global address space to processes}, to facilitate efficient data sharing between processes~\cite{heiser1998, opal, chase1992}. 
\onurvii{\sys makes use of a similar concept \onurii{as single address space OSes} \nas{with its single globally-visible VBI address space.} 
However, while existing single address space OS designs expose the single address space to processes, \sys does not do so, and instead has processes operate on CVT-relative virtual addresses.
This allows \sys to enjoy the same advantages as single address space OSes (e.g., synonym-/homonym-free VIVT caches), while providing further benefits (e.g., non-fixed addresses for shared libraries, hardware-based memory management). Additionally, \sys naturally supports single address space sharing between the host OS and guest OSes in virtualized environments.}


\textbf{User-Space Memory Management.}
Several OS designs propose user-space techniques to provide an application with more control over memory management\onurii{~\cite{engler1995,avm,vm19,barrelfish,hand1999,fos,ebbrt,sel4, kaashoek1997}}. For example,
\sg{the exokernel OS architecture~\cite{engler1995, kaashoek1997} allows applications to manage their own memory and
provides} memory protection via capabilities, thereby minimizing OS
involvement. Do-It-Yourself Virtual Memory Translation \sg{(DVMT)~\cite{vm19} decouples} memory translation from protection in the OS, and allows applications to handle their virtual-to-physical memory translation. 
\sg{These solutions (1)~increase application complexity and add non-trivial programmer burden to directly manage hardware resources, and (2)~do not expose \onurii{the rich} runtime information available in the hardware to memory managers. 
In contrast to these works, which continue to rely on software for physical memory management, \sys does not use \emph{any} part of the software stack for physical memory management.  \onurii{By partitioning the duties differently between software and hardware, and, importantly, performing physical memory management in the memory controller,} \sys provides similar flexibility benefits as user-space memory management without introducing additional programmer burden.}


\textbf{Reducing Address Translation Overhead.}
Several studies have characterized the overhead of virtual-to-physical address translation in \sg{modern} systems, which occurs primarily due to growing physical memory sizes, inflexible memory mappings, and virtualization~\cite{vm2, vm40, vm20, intelx86manual, merrifield2016, vm29}. 
Prior works try to ameliorate the address translation issue by: (1)~increasing the TLB reach to address a larger physical address space~\cite{vm6, pham2014, karakostas2015, vm36, vm12, vm37, vm42,vm8}; (2)~using TLB speculation to speed up address translation~\cite{vm38, pham2015tr, papadopoulou2015, Kavita1994}; (3)~introducing and optimizing page walk caches to store intermediate page table addresses~\cite{vm1, vm10, vm11, esteve14}; (4)~adding sharing and coherence between caching structures to share relevant address translation updates~\cite{vm33, esteve14, vm6, Bogdan2010, vm34, bharadwaj2018, kaxiras2013, latr}; (5)~allocating and using large contiguous regions of memory such as superpages~\cite{vm2, vm3, vm35, vm25, vm8, pham2015, yale2020}; (6)~improving memory virtualization with large, contiguous memory allocations and better paging structures~\cite{vm25, pham2015, pham2015tr, vm37, vm8, vm35}; (7)~prioritizing page walk data throughout the memory hierarchy~\cite{mask}; \onurtt{and (8)~reducing the overheads associated with address translation and maintaining TLB consistency in shared-memory multiprocessors~\cite{teller1990, teller1988, black1989, ritchie1985, wood1986}.} While all \sg{of} these works can mitigate the translation overhead, they build on top of \onurv{the} existing rigid \sg{virtual memory} \onurv{framework} and do not address the underlying \sg{overheads inherent to \onurv{the existing rigid framework} and to software-based memory management. Additionally, several prior works propose mechanisms for fine grain protection domains and sub-page sharing (e.g., \cite{tiwarixx, Greathouse12}), for example by exploiting the contiguity of fine grained permission rights across larger address ranges~\cite{mondrian, mondrianthesis}. However, these techniques do not address the other issues and overheads associated with the conventional virtual memory frameworks such as high address translation overhead. Unlike these works, \sys is a completely new framework for virtual memory, which eliminates several underlying sources of address translation overhead and enables many other benefits (e.g., efficient memory management in virtual machines, easy extensibility to heterogeneous memory systems). \sys can be combined with \onurii{some} of the above proposals to further optimize address translation.}

\section{Summary \onurtt{and Contributions}}
\label{sec:conclusion_vbi}

\onurii{We introduce the \sysfull (\sys), a new virtual memory framework to address the challenges in adapting conventional virtual memory to increasingly diverse system configurations \onurii{and workloads}.
The key idea \onurv{of} \sys is to delegate memory management to hardware in the memory controller.
\nasi{The memory-controller-based memory management in \sys} leads to many benefits not easily attainable in existing virtual memory, such as \nasi{inherently} virtual caches, \nasi{avoiding 2D page walks} in virtual machines, and delayed physical memory allocation.
We \onurv{experimentally} show that \sys (1)~reduces the overheads of address translation by reducing the number of translation requests and associated memory accesses, and
(2)~increases the effectiveness of managing heterogeneous main memory architectures.
We conclude that \sys is a promising new virtual memory framework that can enable several important optimizations and increased design flexibility for virtual memory.
We believe and hope that \sys will open up a new direction and many opportunities for future work \onurv{in novel virtual memory frameworks}.}\\

\noindent In this chapter, we make the following key contributions:\\
\begin{itemize}
    \item We propose the first virtual memory framework that relieves the OS of explicit physical memory management and delegates this duty to the \onurii{hardware, i.e., the} memory controller. 
    \item We propose VBI, a new virtual memory framework that efficiently enables \onurii{memory-controller-based} memory management by exposing a purely virtual memory interface to \onurii{applications, the OS, and the hardware caches}. \sys \onurv{naturally and seamlessly} supports several optimizations (e.g., low-cost page walks in virtual machines, purely virtual caches, delayed physical memory allocation), and \onurii{integrates well with a wide range of system designs}.
    \item We provide a detailed reference implementation of \sys, including required changes to the user applications, system software, ISA, and hardware.
    \item We quantitatively evaluate \sys using two concrete use cases: (1)~address translation improvements for native execution and virtual machines, and (2)~two different heterogeneous memory architectures. Our evaluations show that \sys significantly improves performance in both use cases.
\end{itemize}

\chapter{Conclusions and Future Work}
\label{sec:future}

\section{Conclusions}

The goal of this thesis is to enable efficient data handling in modern computing systems \onurt{via new} frameworks that \onurt{are developed based on a fundamental rethinking of the computing paradigm and key concepts and components in modern computing systems with the goal to make them data-centric and data-aware.}\\

In this thesis, we demonstrate that the overall performance and efficiency of the system can improve significantly using (1) data-centric architectures that minimize data movement and compute data in or near where the data resides, and (2) data-aware frameworks that understand what can be done with and to each piece of data and makes use of different properties of data (e.g., compressibility, approximability, locality, sparsity, access semantics) to improve performance, efficiency and other metrics. We propose two novel frameworks that follow these \onurt{fundamental guiding} principles.\\

First, we propose SIMDRAM, an end-to-end processing-using-DRAM framework that follows the data-centric approach and provides the programming interface, the ISA, and the hardware support for: (1) efficiently computing complex operations \onurt{inside DRAM chips, the predominant main memory technology}, and (2) providing the ability to implement arbitrary operations as required. SIMDRAM achieves this using an in-DRAM massively-parallel SIMD substrate that requires minimal changes to the DRAM architecture. We show that SIMDRAM significantly improves the performance and the energy efficiency of the system when computing a wide variety of complex operations and commonly-used real-world applications. \onurt{We conclude that SIMDRAM is a promising processing-using-memory framework that can enable significant performance and efficiency improvements in the system by efficiently computing complex and arbitrary operations in DRAM. We believe and hope that future work builds on our framework to \omiii{further} \omii{ease} \omiii{the} adoption \omiii{and improve the performance and efficiency} of processing-using-DRAM architectures and applications.}\\

Second, we introduce the Virtual Block Interface (VBI), an alternative virtual memory framework that follows the data-aware approach and (1) addresses the important challenges in adapting conventional virtual memory to increasingly large and diverse data demand in modern applications, (2) understands, conveys, and exploits the properties of different pieces of program data to enable more intelligent management of main memory, and (3) efficiently and flexibly supports increasingly diverse system configurations that are employed today to process the high data demand in modern applications. VBI achieves these while providing the key features of the conventional virtual memory frameworks. As two example use cases of the VBI framework, we show that (1) VBI significantly improves the overall system performance for both native execution and virtual machine environments, and (2) VBI significantly improves the effectiveness of heterogeneous main memory architectures. We conclude that VBI is a promising new virtual memory framework, that can enable several important optimizations, and increase the design flexibility for virtual memory to support efficient handling of data in modern computing systems. We believe and hope that VBI will open up a new direction and many opportunities for future work in novel virtual memory frameworks.\\

\section{Future Work}

This dissertation opens up \onurt{many} new research directions and opportunities. In this section, we discuss several \onurt{major high-level} future directions in which the ideas and approaches presented in this thesis can be extended to tackle other issues in modern computing systems regarding efficient handling of large amount of data in modern applications. \\

\subsection{Data-Aware Memory Architectures}
As we showed in Chapter~\ref{sec:data-aware}, conveying the properties of different pieces of program data to hardware can enable significant performance optimizations. While conveying data properties is difficult to implement on top of conventional virtual memory, VBI (Chapter~\ref{sec:data-aware}) is designed from the ground up to efficiently convey properties of program data to the hardware, including the memory.
We believe that data-aware memories, i.e., memory architectures that understand and exploit the properties of the data to make intelligent utilization decisions, can significantly transform the computing landscape, by exploiting information previously unavailable to the memory in an easy and flexible manner. The native support for conveying data properties to the hardware in VBI can enable a wide range of research and development in this area, serving as a platform to demonstrate numerous data-aware hardware optimizations in main memory management.\\

\subsection{Enabling Support for Designing New \onurt{Unconventional} Memory Subsystems}

As we discuss in Chapter~\ref{sec:data-aware}, conventional OS-based virtual memories are unable to efficiently adapt to and fully exploit today's diverse memory designs \onurt{(e.g., hybrid memory systems)}, as the OS has poor visibility into the physical memory architecture and lacks the rich fine-grained information on runtime memory behavior that is essential in managing the memory resources.
VBI (Chapter~\ref{sec:data-aware}) provides significant flexibility in enabling efficient virtual memory support for emerging memory architectures, without breaking the interfaces that programmers are used to, and without requiring bespoke virtual memory architectures for each different system design. In this thesis, we showed how VBI can provide efficient support for heterogeneous memory systems, which are becoming widely available (e.g., systems with Optane memory modules side-by-side with DRAM). Given its flexibility and customization opportunities, we believe that VBI can serve as a foundation for systems that incorporate non-traditional memory subsystems, and opens up \onurt{many} new research opportunities in using emerging memory technologies \onurt{(e.g., combining main memory and storage devices~\cite{meza2013})} and designing new unconventional memory subsystems \onurt{(e.g., potentially using neuromorphic hardware}) that can further enhance the efficiency of handling the large amount of data in modern applications.\\

\subsection{Virtual Memory Support for Processing-Using-Memory architectures}

Processing-using-memory reduces/eliminates the need to move data from the main memory to the processor for computation. SIMDRAM (Chapter~\ref{sec:data-centric}) eases the adoption of processing-using-DRAM architectures by enabling efficient implementation of complex operations and providing the ability to perform any arbitrary operations. However, full adoption of processing-using-memory solutions such as SIMDRAM requires support for some key virtual memory functionalities. More specifically, relying on the CPU to provide the processing-using-memory architectures with address translation, memory allocation, \onurt{and potentially memory security} can potentially nullify the very benefits of processing-using-memory approach which \onurt{aims to} reduce the interaction between the main memory and the CPU. Accordingly, designing efficient support in processing-using-memory architectures for critical virtual memory functionalities such as address translation, memory allocation, \onurt{and security mechanisms} is a promising research direction.

\chapter{Other Works of the Author}

\onurt{In addition to the works presented in this thesis, I have also contributed to several other research works done in collaboration with SAFARI Research Group members at CMU and ETH. In this section, I briefly overview these works.\\}

\onurt{\textbf{Expressive Memory (XMem)~\cite{xmem}:} Programs are traditionally conveyed to the hardware in the form of ISA instructions and a set of memory accesses to virtual addresses. This semantic gap leads to hardware treating all data as the same, thereby \onurt{being} unable to exploit data's semantics properties to employ more intelligent management or optimization policies. This work introduces a new cross-layer interface, called Expressive Memory (XMem), to communicate higher-level program semantics from the application to the system software and hardware architecture. By bridging the semantic gap, XMem provides two key benefits. First, it enables architectural/system-level techniques to leverage key program semantics that are challenging to predict or infer. Second, it improves the efficacy and portability of software optimizations by alleviating the need to tune code for specific hardware resources (e.g., cache space). This work was published in ISCA 2018~\cite{xmem}.\\}

\onurt{\textbf{CoNDA~\cite{boroumand2019conda, lazypim}:} Recent advances in memory technology have enabled near-data accelerators (NDAs), which are located off-chip, close to main memory. The lack of an efficient communication mechanism between CPUs and accelerators creates a significant overhead to synchronize data updates between the two. Accordingly, enforcing coherence with the rest of the system, which is already a major challenge for accelerators, becomes more difficult for NDAs. This work introduces CoNDA, a coherence mechanism that lets an NDA optimistically execute an NDA kernel, under the assumption that the NDA has all necessary coherence permissions. This optimistic execution allows CoNDA to gather information on the memory accesses performed by the NDA and by the rest of the system. CoNDA then exploits this information to avoid performing unnecessary coherence requests, and as a result, reduce the data movement for coherence significantly. CoNDA was published in ISCA 2019, and an earlier version of it, LazyPIM, in IEEE CAL~\cite{lazypim}.\\}

\onurt{\textbf{Demystifying Complex Workload-DRAM Interactions~\cite{saugata2019, saugata2019arx}:} With the increasingly diversifying application behavior and the wide array of available DRAM types, it has become very difficult to identify the best DRAM type for a given workload. Much of this difficulty lies in the complex interaction between memory access latency, bandwidth, parallelism, energy consumption, and application memory access patterns. Importantly, changes made by DRAM vendors in new DRAM types can significantly affect the behavior of an application in ways that are often difficult to intuitively and easily understand. This work identifies important families of workloads, as well as widely used types of DRAM chips, and comprehensively analyze the combined DRAM–workload behavior. We provide an experimental study of the interaction between nine different DRAM types and 115 modern applications and multi-programmed workloads. Furthermore, we perform a rigorous experimental characterization of system performance and DRAM energy consumption, and introduce new metrics to capture the sophisticated interactions between memory access patterns and the underlying hardware. The trends identified from the characterization performed in this work can drive optimizations in both hardware and software design. This work was published in SIGMETRICS 2019~\cite{saugata2019}.}\\

\onurtt{\textbf{AirLift~\cite{airlift}:} Genome sequencing is a technique that determines the DNA sequence of an organism. Modern genome sequencing machines~\cite{gs1, gs2, gs3, gs4, gs5, gs6, cali2020nanopore} extract small random fragments of the original DNA sequence, known as reads~\cite{cali2020genasm, mohammedprimer,mohammedsnake, mohammedshouji,hongyiseed,hongyihash,kim2018grim}. To adapt an existing genomic study (i.e., read sets from many samples) to a new reference genome, we need to remap the reads (i.e., update a read’s alignment location from the original (old) reference to another (new) reference). AirLift presents a methodology for quickly and comprehensively mapping a set of reads from one reference to another reference. AirLift is the first methodology and tool that leverages the similarity between two reference genomes to (1)~substantially reduce the time to remap a read set from an old (i.e., previously mapped to) reference genome to a new reference genome, (2)~comprehensively remap a read set, i.e., attempt to remap all reads in a read set, (3)~provide accurate remapping results, i.e., provide alignments with error rates below a specified acceptable error rate, and (4)~provide an end-to-end remapping solution on which downstream analysis (e.g., variant calling) can be immediately performed.}

%
%
%
%
%

\backmatter%
	\addtoToC{Bibliography}
	\bibliographystyle{unsrt}
	\bibliography{references}

\begin{appendices} 
\chapter{AIG-to-MIG Conversion}
\label{apdx:aoi-to-mig}


The conversion from AND/OR/NOT representation of an operation to its MAJ/NOT representation \juang{relies} on a set of \juang{transformation} rules that are derived \juang{from the characteristics of the MAJ operation.} 
Table~\ref{table:rules} lists the set of transformation rules that we use 
to synthesize a circuit for a desired operation with MAJ and NOT gates. 
We use full \omdefi{addition} as a \juang{running} example to describe the process of synthesizing a \juang{MAJ/NOT-based} circuit\omdefi{,} starting from an AND/OR/NOT representation of the circuit \omdefi{and} using the transformation rules. 
\juang{We obtain MAJ/NOT-based circuits for other \mech operations following the same method. 
In a later step (\cref{sec:mi-aap}), we translate \omdefi{a} MAJ/NOT-based circuit to sequences of \omdefi{\aaps} operations.}

\begin{table}[h]
    \caption{\omdefi{MAJ/NOT} transformation Rules~\cite{epflmaj}.}
    \label{table:rules}
    \centering
    \footnotesize
    \tempcommand{1.3}
    \renewcommand{\arraystretch}{0.8}
    \resizebox{\columnwidth}{!}{
    \begin{tabular}{ll}
    \toprule
    \textbf{Commutativity (C)} & $M(x, y, z) = M(y, x, z) = M(z, y, x)$\\
    \cmidrule(rl){1-2}
    \multirow{2}{*}{\textbf{Majority (M)}} & if$(x = y): M(x, y, z) = x = y$ \\
    & if$(x = \overline{y}): M(x, y, z) = z$ \\
    \cmidrule(rl){1-2}
    \textbf{Associativity (A)} & $M(x, u, M(y, u, z)) = M(z, u, M(y, u, x))$\\
    \cmidrule(rl){1-2}
    \textbf{Distributivity (D)} & $M(x, y, M(u, v, z)) = M(M(x, y, u), M(x, y, v), z)$\\
    \cmidrule(rl){1-2}
    \textbf{Inverter Propagation (I)} & $\overline{M}(x, y, z) = M(\overline{x}, \overline{y}, \overline{z})$\\
    \cmidrule(rl){1-2}
    \textbf{Relevance (R)} & $M(x, y, z) = M(x, y, \mathbf{z}^{}_{x/\overline{y}})$\\
    \cmidrule(rl){1-2}
    \textbf{Complementary Associativity (CA)} & $M(x, u, M(y, \overline{u}, z)) = M(x, u, M(y, x, z))$\\
    \bottomrule
    \end{tabular}
    }
\end{table}



\begin{figure*}[!b]
    \centering
    \includegraphics[width=\linewidth]{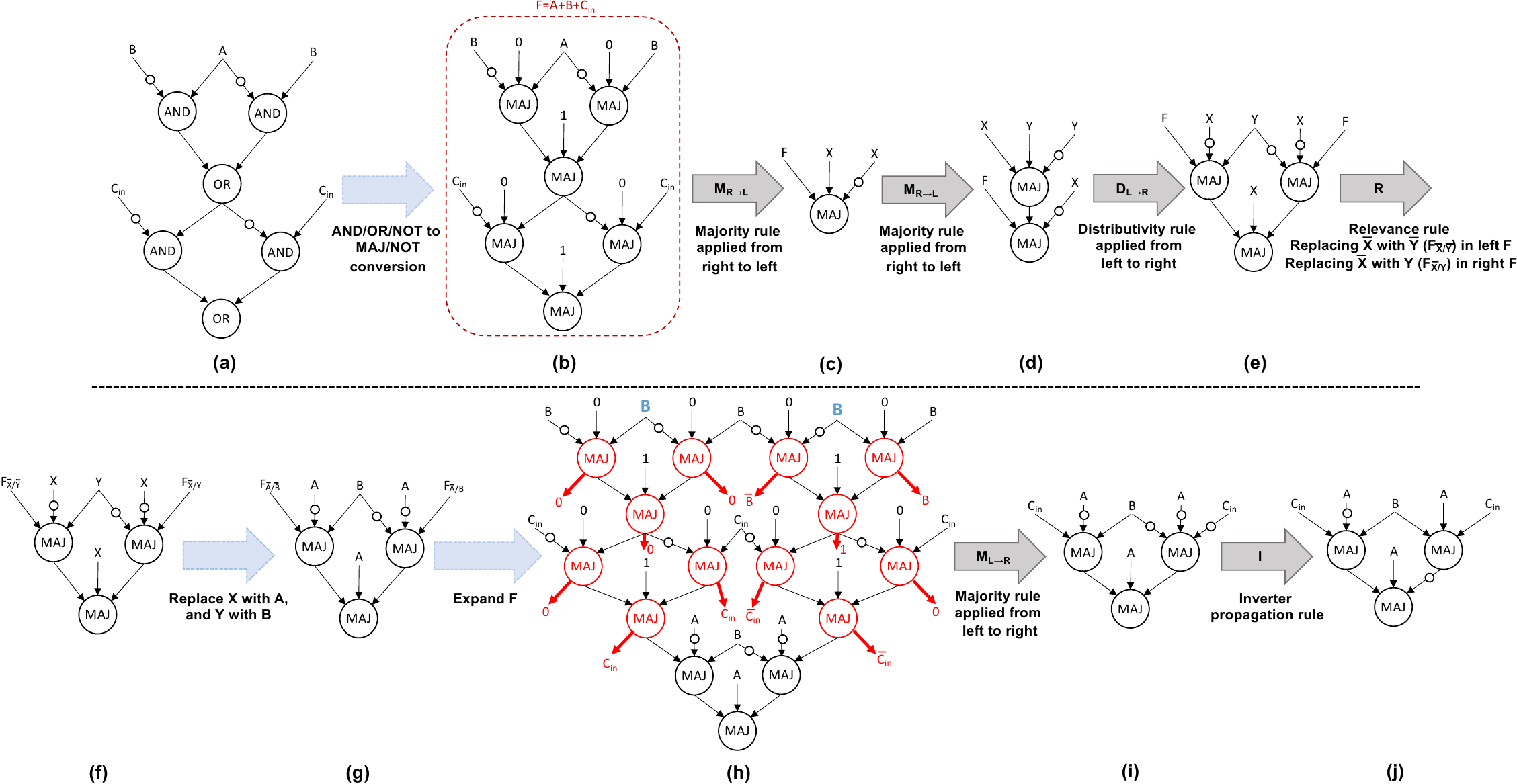}
    \caption{Synthesizing SIMDRAM circuit for a full addition.}
    \label{fig_add-maj}
\end{figure*}

\omdefi{Figure}~\ref{fig_add-maj}a shows the optimized AND/OR/\omdefi{Inverter} (i.e., AND/OR/NOT) \omdefi{Graph} (AOIG) representation of a full \omdefi{addition} (i.e., F = A + B + C$_{in}$). As shown in \omdefi{Figure}~\ref{fig_add-maj}b, the naive way to transform the AOIG to a \omdefi{Majority/Inverter} (i.e., MAJ/NOT) Graph (MIG) representation, is to replace every AND and OR \omdefi{primitive} with a \omdefi{three-input} MAJ \omdefi{primitive} where the third input is 0 or 1, respectively. The resulting MIG is in fact Ambit's~\cite{seshadri2017ambit} representation of the full \omdefi{addition}. 
While the AOIG in \omdefii{Figure~\ref{fig_add-maj}a} is optimized for AND/OR/NOT operations, the resulting MIG in Figure~\ref{fig_add-maj}b can be further \omdefi{optimized} by exploiting the 
\juang{transformation rules of the MAJ \omdefi{primitive} (Table~\ref{table:rules}\omdefii{, replicated from~\cite{epflmaj}})}. \omdefi{The MIG optimization is performed in two key steps: (1) node reduction, and (2) MIG reshaping.}

\omdefi{\textbf{Node reduction.} In order to optimize the MIG in Figure~\ref{fig_add-maj}b, the first step is to reduce the number of MAJ nodes in the MIG.} 
As shown in Table 1, rules \textbf{M} and \textbf{D} reduce the number of nodes in a MIG if applied from left to right (i.e., $\mathbf{M}^{}_{L\rightarrow{R}}$) and from right to left (i.e., $\mathbf{D}^{}_{R\rightarrow{L}}$), respectively. $\mathbf{M}^{}_{L\rightarrow{R}}$ replaces a MAJ \omdefi{node} with \omdefi{a single} \omdefi{value}, and $\mathbf{D}^{}_{R\rightarrow{L}}$ replaces three \omdefi{MAJ nodes} with two \omdefi{MAJ nodes in the MIG}. \omdefi{The node reduction step applies} $\mathbf{M}^{}_{L\rightarrow{R}}$ and $\mathbf{D}^{}_{R\rightarrow{L}}$ as many times as possible to reduce the the number of MAJ operations \omdefi{in the MIG. We can see in Figure~\ref{fig_add-maj}b that none of the two rules are applicable \juang{in the particular case of the full addition MIG}.} \omdefii{Therefore, Fig~\ref{fig_add-maj}b remains unchanged after applying node reduction.}

\omdefi{\textbf{MIG reshaping.}} When no further node reduction is possible, we \emph{reshape} the MIG in an effort to enable more node reduction opportunities by repeatedly using two sets of rules: 
(1)~rules $\mathbf{M}^{}_{R\rightarrow{L}}$, $\mathbf{D}^{}_{L\rightarrow{R}}$, and \textbf{R} to temporarily inflate the MIG and create more node reduction opportunities with the \omdefi{help of the} new nodes, and (2)~rules \textbf{A} and \revAMicro{\textbf{CA}}, to exchange variables between adjacent nodes.  \revAMicro{Note that in this step, rules \textbf{M} and \textbf{D} are \omdefi{applied} in the reverse direction compared to \omdefi{the} previous step \omdefi{(i.e., node reduction step)} which results in increasing the number of nodes in the MIG.} \omdefi{We now describe the MIG reshaping process for the full addition example (Figure~\ref{fig_add-maj}b). For simplicity,} we first assume that the entire MIG is represented as function \textbf{F} \omdefi{that computes the full addition of the input operands \textbf{A} and \textbf{B}. Then, we} apply rule $\mathbf{M}^{}_{R\rightarrow{L}}$ \omdefi{while} introducing variable \textbf{X} to the MIG \omdefi{(as \omdefii{$F = M (F, x, \overline{x})$)}} without impacting the functionality of the MIG (Figure \ref{fig_add-maj}c). We then apply \omdefi{the same rule again, and replace \textbf{X} with a new MAJ node while introducing variable \textbf{Y} (Figure~\ref{fig_add-maj}d). Next,}  
by applying rule $\mathbf{D}^{}_{L\rightarrow{R}}$, we introduce a new \omdefi{MAJ} node and distribute the function \textbf{F} across \omdefi{the} two MAJ nodes \omdefi{(Figure~\ref{fig_add-maj}e)}. \omdefii{Now, by applying rule \textbf{R} to the function \textbf{F} on the left, variable ${\overline{\textbf{X}}}$ is replaced with variable $\overline{\textbf{Y}}$ in the function \textbf{F} on the left. Similarly, by applying rule \textbf{R} to the function \textbf{F} on the right, variable ${\overline{\textbf{X}}}$ is replaced with variable ${\textbf{Y}}$ in the function \textbf{F} on the right \omdefi{(Figure~\ref{fig_add-maj}f)}.} 
At this point, since \omdefi{rule} $\mathbf{M}^{}_{R\rightarrow{L}}$ holds with any given two variables, we can safely replace \textbf{X} and \textbf{Y} with variables \textbf{A} and \textbf{B}, respectively (Figure~\ref{fig_add-maj}g). Next, we expand function \textbf{F} \omdefi{(Figure~\ref{fig_add-maj}h)} and the variables replaced as a result of the previous rule are highlighted in blue. As shown in Figure~\ref{fig_add-maj}h, the resulting graph after expanding function \textbf{F} has multiple node reduction opportunities using rule $\mathbf{M}^{}_{L\rightarrow{R}}$ and starting from the top of the graph. The nodes that can be eliminated using this rule are marked in red and the replacing value is indicated with a red arrow leaving the node. Figure~\ref{fig_add-maj}i shows the same MIG after resolving all the node reductions. We next use rule \textbf{I} to remove all three NOT \omdefi{primitives} in the rightmost MAJ node. The final optimized MIG that is shown in Figure~\ref{fig_add-maj}j \omdefi{requires only 3 MAJ \omdefi{primitives} to perform the full addition operation}\omdefii{, as opposed to the 6 we started with (in Figure~\ref{fig_add-maj}b)}. 

\omdefi{The node reduction step followed by the MIG reshaping step are repeated (for a predefined number of times) until we achieve an optimized MIG that requires minimal number of MAJ operations to perform the desired in-DRAM operation.} \revBMicro{The process of converting an operation to a MAJ-based implementation can be automated as suggested by prior work~\cite{epflmaj,soeken2016mig}.}

\chapter{Row-to-Operand Allocation}
 \label{apdx:row-to-op}


\nasrev{Algorithm~\ref{alg_mapping} describes SIMDRAM’s row-to-operand allocation procedure. \sgii{To enable in-DRAM computation, our \omdef{allocation} algorithm copies (i.e., maps) input operands for each MAJ node in the MIG from D-group rows (where the operands normally reside) into compute rows.  However, due to the limited number of compute rows, the \omdef{allocation} algorithm cannot \omdef{allocate DRAM rows to} all input operands from all MAJ nodes at once.} 
To address this issue, \juangg{the} \omdef{allocation} algorithm divides the \omdef{allocation} process into \emph{phases}. Each phase \omdef{allocates as many compute rows to} operands as possible. For example, \sgii{because no rows are \omdef{allocated} yet, the initial phase (Phase~0)} has all six compute rows \sgii{available for \omdef{allocation} (i.e., the rows are vacant),} and can \omdef{allocate up to six input operands to the compute rows}. 
\sgii{A phase} is considered finished \sgii{when either
(1)~there are not enough vacant compute rows to \omdef{allocate} all input operands for the next logic primitive that needs to be computed, or
(2)~there are no more MAJ primitives left to process in the MIG.} 
\omdef{The phase information is used when generating the \uprog{} for the MIG in Task 2 of Step 2 of SIMDRAM framework (\Cref{sec:framework:step2:generating}), where \uop{}s for all MAJ primitives in phase $i$ are generated prior to the MAJ primitives in phase $i+1$. Knowing that all the MAJ primitives in phase $i$ are performed before the next phase $i+1$ starts, \omi{the \omdef{allocation} algorithm can safely \omdef{\omdefii{reuse}}} \sgii{the compute rows for use} in phase $i+1$, without worrying about the output of a MAJ primitive being overwritten by a new row-to-operand allocation. 
}}

\nasrev{\sgii{We now describe the \omdef{row-to-operand allocation} algorithm in detail, using the MIG for full \omdef{addition} in \cmr{Figure}~\ref{fig_output_mapping}a as an example of a MIG being traversed by the algorithm.}
The \omdef{allocation} algorithm starts \sgii{at} Phase~0. \sgii{Throughout its execution, the algorithm maintains 
(1)~the list of free compute rows that are available for \omdef{allocation} (\emph{B\_rows} and \emph{B\_rows\_DCC} in Algorithm~\ref{alg_mapping}\omdefii{, initialized in lines 3--4}); and
(2)~}the list of \omdef{row-to-operand} \omdef{allocations} associated with each MAJ node, tagged with the phase number that the \omdef{allocations} were performed in (\emph{row\_operand\_allocation} in Algorithm~\ref{alg_mapping}). Once a \omdef{row-to-operand} \omdef{allocation} is performed, the algorithm removes the compute row used for the \omdef{allocation} from the list of the free compute rows, and adds the new \omdef{allocation} to the list of \omdef{row-to-operand allocations generated} in that phase for the corresponding MAJ node. The algorithm follows a simple procedure to \omdef{allocate compute rows to}  
the input operands of the MAJ nodes in the MIG. The algorithm does a topological traversal starting with the leftmost MAJ node in the highest level of the MIG (\sgii{e.g.}, Level 0 in \cmr{Figure}~\ref{fig_output_mapping}a), and traverses all the MAJ nodes in each level, before moving to the next \omdef{lower} level of the graph. }

\algnewcommand{\algorithmicgoto}{\textbf{go to}}%
\algnewcommand{\Goto}[1]{\algorithmicgoto~\ref{#1}}%
\begin{algorithm}[!ht]
  \caption{\juang{\mech's \omdef{Row-to-Operand Allocation Algorithm}.}}
  \label{alg_mapping}
  \tiny
  \begin{algorithmic}[1]

    \State \textbf{Input:} MIG \texttt{G} = (\texttt{V}, \texttt{E})   \algorithmiccomment{Majority-Inverter Graph \texttt{G} nodes <vertex, edge>}
    \State \textbf{Output:} \texttt{row\_operand\_allocation} \algorithmiccomment{Allocation map of rows to operands}
    
    \vspace{0.5em}
    \State \texttt{B\_rows} $\gets$ \texttt{\{T0, T1, T2, T3\}}
    \State \texttt{B\_rows\_DCC} $\gets$  \texttt{\{DCC0, DCC1\}} 
    \State \texttt{phase} $\gets$ 0
    \State \texttt{row\_operand\_allocation\_map} $\gets \emptyset$
    \vspace{0.5em}
  
    \ForEach{level in \texttt{G}}
        \ForEach{\texttt{V} in \texttt{G}[level]}
            \ForEach{\texttt{input edge} in \texttt{E[V]}}
                \State Search for \texttt{input edge}'s parent
    \If{input edge has no parents}\tikzmark{top}
                    \If{\texttt{input edge} is negated}
                    \State Allocate row in \texttt{B\_rows\_DCC} to \texttt{input edge}
                        \State Remove allocated row from \texttt{B\_rows\_DCC}
                    \Else
                        \State Allocate row in \texttt{B\_rows} to  \texttt{input edge}
                        \State Remove allocated row from \texttt{B\_rows} 
                    \EndIf \tikzmark{bottom}
                \Else \tikzmark{topblue}
                    \If{ \texttt{input edge} is negated}
                        \State Map allocated parent row in \texttt{B\_rows\_DCC} to  \texttt{input edge}
                    \Else
                        \State Map allocated parent row in \texttt{B\_rows} to  \texttt{input edge}
                    \EndIf
                \EndIf \tikzmark{bottomblue}

            \If{\texttt{B\_rows} and \texttt{B\_rows\_DCC} are empty} \tikzmark{topgreen}
                \State \texttt{phase} $\gets$ \texttt{phase + 1}
                \State \texttt{B\_rows} $\gets$ \texttt{\{T0, T1, T2, T3\}}
                \State \texttt{B\_rows\_DCC} $\gets$ \texttt{\{DCC0, DCC1\}} 
            \EndIf \tikzmark{bottomgreen}
            
             \State \texttt{row\_operand\_allocation} $\gets$ (\texttt{input edge}, allocated row, \texttt{phase}) \tikzmark{right}
            \EndFor 
        \EndFor
    \EndFor
    
  \end{algorithmic}
  \AddNoteRed{top}{bottom}{right}{\normalsize Case 1}
    \AddNoteBlue{topblue}{bottomblue}{right}{\normalsize Case 2}
    \AddNoteGreen{topgreen}{bottomgreen}{right}{\normalsize Case 3}
  
\end{algorithm}

\nasrev{For each of the three 
input edges (i.e., operands) of any given MAJ node, the algorithm checks for the following three possible cases and performs the \omdef{allocation} accordingly: }

\noindent\nasrev{\sgii{\textbf{Case~1:}} if the edge is not connected to another MAJ node in \sgii{a} \omdef{higher} level of the graph \omdef{(line 11 in Algorithm~\ref{alg_mapping})}, i.e., the edge does not have a parent (e.g., \sgii{the three edges entering} the blue node in \cmr{Figure}~\ref{fig_output_mapping}a), \sgii{and a compute row is available,} the 
input operand associated with the edge is considered \sgii{to be a source input, and is currently located} in the D-group rows of the subarray. As a result, the algorithm copies the input operand associated with the edge from \sgii{its D-group row} to the first available compute row. Note that if the edge \omdef{\emph{is}} complemented, i.e., the input operand is negated (e.g., the edge with operand A for the blue node in \cmr{Figure}~\ref{fig_output_mapping}a), the algorithm \omdef{allocates the first available compute row with dual contact cells (DCC0 or DCC1) to}  the input operand of the edge (lines \omdef{12--14} in Algorithm~\ref{alg_mapping}). If the edge is \omdef{\emph{not}} complemented (e.g., the edge with operand B for the blue node in \cmr{Figure}~\ref{fig_output_mapping}a), \omdef{a regular compute row is allocated to} the input operand 
(lines \omdef{15--17} in Algorithm~\ref{alg_mapping}). }

\noindent\nasrev{\sgii{\textbf{Case~2:}} if the edge is connected to another MAJ node in \sgii{a} higher level of the graph \omdef{(line 18 in Algorithm~\ref{alg_mapping})}, 
\omdef{the edge has a parent node and the value of the input operand associated with the edge equals the \omdefii{result} of the parent node, \omdefii{which}} is available in the compute rows that hold the result of the parent MAJ node. As a result, the algorithm maps the input operand of the edge to a compute row that holds the result of its parent node (lines \omdef{19--22} in Algorithm~\ref{alg_mapping}). 
}

\noindent\nasrev{\sgii{\textbf{Case~3:}} if there are no free compute rows available, the algorithm \omdef{considers} the phase as \emph{complete} and continues the \omdef{allocations} in the next phase (lines \omdef{23--26} in Algorithm~\ref{alg_mapping}). }

\nasrev{Once \omdef{DRAM rows are allocated to} all the edges connected to a MAJ node, the algorithm stores the \omdef{row-to-operand allocation} information of the three input operands of the MAJ node in \omdef{\emph{row\_operand\_allocation}} (line \omdef{27} in Algorithm~\ref{alg_mapping}) 
and associates this information with the MAJ node and the phase number that the \omdef{allocations} were performed in. The algorithm finishes once \omdef{DRAM rows are allocated to} \emph{all} the input operands of all the MAJ nodes in the MIG. \cmr{Figure}~\ref{fig_output_mapping}b shows these \omdef{allocations} as the output of Task 1 for the full \omdef{addition} example. \sgii{The resulting} \emph{row\_operand\_allocation} is then used in \omdef{Task 2 of Step 2 of the SIMDRAM framework} (\cmr{\Cref{sec:framework:step2:generating}}) to generate the \omiii{series} of \uop{}s to compute the operation that the MIG represents. } 

\chapter{Scalability of Operations}
 \label{apdx:op-class}

\revGeraldo{\nastaran{Table~\ref{table_aaps_operations} lists the semantics and the total number of AAP/APs required for each of the 16 \mech operations that we evaluate in this work (\Cref{sec_evaluation}) for input element(s) of size $n$.}} \omdef{Each operation is classified based on how the latency of the operation scales with respect to the element size $n$. Class 1, 2, and 3 operations scale linearly, logarithmically,and quadratically with $n$, respectively.}

\begin{table}[h]
\caption{\sgii{Evaluated \mech operations (for $n$-bit data)}.}
\label{table_aaps_operations}
\centering
\tempcommand{1.2}
\resizebox{\linewidth}{!}{
\begin{tabular}{|l|l|l|l|l|}
\hline
\multicolumn{1}{|l|}{\textbf{Type}} & \multicolumn{1}{l|}{\textbf{Operation}} & \multicolumn{1}{l|}{\textbf{\# AAPs/APs}} & \multicolumn{1}{l|}{\textbf{Class}} & \multicolumn{1}{l|}{\textbf{Semantics}} \\ 
\hline
\hline
\multirow{10}{*}{Arithmetic}& abs                       & \(10n - 2\)         & Linear              & \(dst = (src > 0) ?\ src : -(src)\) \\ \cline{2-5} 
                            & addition                  & \(8n + 1\)          & Linear              & $dst = src_1 + src_2 $                             \\ \cline{2-5} 
                            & \multirow{2}{*}{bitcount} &  $\Omega = 8n - 8\log_2 (n+1) $       &   \multirow{2}{*}{Linear}   & \multirow{2}{*}{ $\sum_{i=0}^{n} src(i)$}                        \\ \cline{3-3}
                            &                &  $O = 8n $             &                         &                    \\ \cline{2-5} 
                            & division       & $8n^{2} + 12n$         &  Quadratic  &            $dst = \frac{src_1}{src_2} $                \\ \cline{2-5} 
                            & max            & $10n + 2$              &  Linear     & $dst = (src_1 > src_2)?\ src_1 : src_2 $\\ \cline{2-5} 
                            & min            & $10n + 2   $           &  Linear     & $dst = (src_1 < src_2)?\ src_1 : src_2 $   \\ \cline{2-5} 
                            & multiplication & $ 11n^{2} - 5n - 1 $   &  Quadratic  & $dst = src_1 \times  src_2  $                      \\ \cline{2-5} 
                            & ReLU           & $3n + ((n-1)\ mod\ 2)$ &  Linear     & $dst = (src \geq 0)?\ src : 0  $ \\ \cline{2-5} 
                            & subtraction    & $8n + 1  $             &  Linear     & $dst = src_1 - src_2   $                        \\ \hline \hline
Predication                 & if\_else       & $7n$                   &  Linear     & $dst = (sel)?\ src_1 : src_2   $                 \\ \hline \hline
\hline
\multirow{3}{*}{Reduction}  & and\_reduction &  $5\floor{\frac{n}{2}} + 2$   &  Logarithmic & $Y = src(1)  \wedge src(2) \wedge src(3)  $                                  \\ \cline{2-5} 
                            & or\_reduction  & $5\floor{\frac{n}{2}}  + 2$   &  Logarithmic & $ Y = src(1)  \vee src(2) \vee src(3)$             \\ \cline{2-5} 
                            & xor\_reduction & $6\floor{\frac{n}{2}}  + 1$   &  Logarithmic & $Y = src(1)  \oplus src(2) \oplus src(3)$                                          \\ \hline \hline
\multirow{3}{*}{Relational} & equal          & $4n + 3$                      &  Linear      & $dst = (src_1 == src_2) $                       \\ \cline{2-5} 
                            & greater        & $3n + 2      $                &  Linear      & $dst = (src_1 > src_2)       $       \\ \cline{2-5} 
                            & greater\_equal & $3n + 2      $                & Linear       & $dst = (src_1 \geq src_2)    $       \\ \hline
\end{tabular}
}

\end{table}
\chapter{Evaluated Real-World Applications}
 \label{appendix}


\noindent{ \textbf{\omdefii{Convolutional} Neural Networks (CNNs).} \geraldo{CNNs~\omdefii{\cite{rastegari2016xnor,he2020sparse,krizhevsky2012imagenet}} are used in many classification tasks \omdefi{such as} image and handwriting classification. \nastaran{CNNs are often} \omdefii{computationally} \nastaran{intensive} as they use many general-matrix-multiplication (GEMM) operations \omdefii{using} \omdefii{floating}-point operations for each convolution. 
Prior works~\cite{rastegari2016xnor, lin2017towards, he2020sparse} demonstrate that \omdefi{instead of the costly \omdefii{floating-point} multiplication operations, convolutions can be performed} 
\nastaran{using a series of bitcount, addition, shift, and XNOR operations}. In this work, we use the XNOR-NET~\cite{rastegari2016xnor} implementations of VGG-13, VGG-16, and LeNET provided by \cite{he2020sparse}, \omdefi{to evaluate the functionality of SIMDRAM}.  We modify \omdefi{these} implementations to make use of SIMDRAM's bitcout, addition, shift, and XNOR operations. We evaluate \omdefi{all} three networks \revBMicro{for inference} using two different datasets: VGG-13 and VGG-16 (using CIFAR-10~\cite{krizhevsky2010convolutional}), and LeNet-5 (using MNIST~\cite{deng2012mnist}).}}

\geraldo{\noindent{\textbf{k-Nearest Neighbor Classifier (kNN).} 
\omdefii{We use a kNN classifier to solve the handwritten digits recognition problem~\cite{Lecun95learningalgorithms}. The kNN classifier finds a group of k objects in the input set using a simple distance algorithms such as Euclidean distance~\cite{friedman2001elements}.} 
\omdefi{In our evaluations, we} use SIMDRAM to implement the Euclidean distance algorithm \omdefi{entirely} in DRAM. We evaluate a kNN algorithm using the MNIST dataset~\omdefii{\cite{deng2012mnist}} with 3000 training images and 1000 testing images.  We quantize the inputs using an 8-bit representation. }}

\geraldo{\noindent{\textbf{Database.} We evaluate \omdefi{SIMDRAM using} two different database workloads. First, 
we evaluate a simple table scan query \texttt{‘select count(*) from T where c1 <= val <= c2'} using \omdefii{the} BitWeaving algorithm~\cite{li2013bitweaving}. 
Second, we evaluate the performance of the TPC-H~\cite{tpch} scheme using query 01, \omdefi{\omdefii{which} executes many arithmetic operations, including addition and multiplication}. \omdefi{For our evaluation,} we follow the column-based data layout employed in \cite{santos2017operand} and use a scale factor of 100. }}

\geraldo{\noindent{\textbf{Brightness.} We use a simple image brightness algorithm~\omdefii{\cite{foley1996computer}} to demonstrate the benefits of \omdefii{the} SIMDRAM predication operation. The algorithm evaluates if a given brightness value is larger than 0. If so, it increases the pixel value of the image by the brightness value. Before assigning the new brightness value to the pixel, the algorithm verifies if the new pixel value is between 0 and 255. In our SIMDRAM implementation, we use both addition and predication operations. }}

\end{appendices}
\end{document}